\documentclass[11pt]{combine}
\usepackage{amsmath,amsfonts,amssymb,amsthm,epsfig, graphicx, hyperref}
\usepackage[dvipsnames,table,dvipsnames*, svgnames*, hyperref]{xcolor}
\usepackage{natbib,amsmath,amssymb,amsthm,graphicx,setspace,paralist,booktabs,rotating,subcaption,float,color}
\usepackage{multirow}
\usepackage{fancyhdr}
\usepackage{bibunits}
\usepackage[small]{titlesec}

\newtheorem{theorem}{Theorem}
\newtheorem{corollary}{Corollary}
\newtheorem{proposition}{Proposition}

\newtheorem{lemma}{Lemma}
{
	\theoremstyle{definition}
	\newtheorem{definition}{Definition}
	\newtheorem{example}{Example}

}
\newcommand{\beq}{\begin{equation}}
\newcommand{\eeq}{\end{equation}}
\newcommand{\beas}{\begin{align*}}
\newcommand{\eeas}{\end{align*}}
\newcommand{\bea}{\begin{align}}
\newcommand{\eea}{\end{align}}
\newcommand{\bei}{\begin{itemize}}
	\newcommand{\eei}{\end{itemize}}
\newcommand{\ben}{\begin{enumerate}}
	\newcommand{\een}{\end{enumerate}}
\newcommand{\bet}{\begin{theorem}}
	\newcommand{\eet}{\end{theorem}}
\newcommand{\bel}{\begin{lemma}}
	\newcommand{\eel}{\end{lemma}}
\newcommand{\bep}{\begin{proposition}}
	\newcommand{\eep}{\end{proposition}}
\newcommand{\bed}{\begin{definition}}
	\newcommand{\eed}{\end{definition}}
\newcommand{\bec}{\begin{corollary}}
	\newcommand{\eec}{\end{corollary}}
\newcommand{\bex}{\begin{example}}
	\newcommand{\eex}{\end{example}}

\newcommand{\R}{\mathbb{R}}
\newcommand{\E}{\mathbb{E}}

\newcommand{\HH}{\mathcal{H}}

\newcommand{\supp}{\text{supp}}
\newcommand{\argmin}{\mathop{\rm arg\min}}

\def\limsup{\mathop{\overline{\rm lim}}}
\def\liminf{\mathop{\underline{\rm lim}}}

\numberwithin{equation}{section}

\numberwithin{remark}{section}

\def\limsup{\mathop{\overline{\rm lim}}}
\def\liminf{\mathop{\underline{\rm lim}}}

\begin{document}

\title{\scshape Statistical Inference for Genetic Relatedness Based on High-Dimensional Logistic Regression}
\author{Rong Ma$^1$, Zijian Guo$^2$, T. Tony Cai$^3$ and Hongzhe Li$^3$\\
	\\
	Stanford University$^1$\\
    Rutgers University$^2$\\
	University of Pennsylvania$^3$}
\date{}
\maketitle
\thispagestyle{empty}

\begin{abstract}
	This paper studies the problem of statistical inference for genetic relatedness between binary traits based on individual-level genome-wide association  data. Specifically, under the high-dimensional logistic regression models, we define parameters characterizing the cross-trait genetic correlation, the genetic covariance and the trait-specific genetic variance. A novel weighted debiasing method is developed for the logistic Lasso estimator  and  computationally efficient debiased estimators are proposed. The rates of convergence for these estimators are studied and their asymptotic normality is established under mild conditions. Moreover, we construct confidence intervals and statistical tests for these parameters, and provide theoretical justifications for the methods, including the coverage probability and expected length of the confidence intervals, as well as the size and power of the proposed tests. Numerical studies are conducted under both model generated data and simulated genetic data to show the superiority of the proposed methods. By analyzing a real data set on autoimmune diseases, we demonstrate its ability to obtain novel insights about the shared genetic architecture between ten pediatric autoimmune diseases.
	
	\bigskip
	
	\noindent\emph{KEY WORDS}: Confidence interval; Debiasing methods; Functional estimation; Genetic correlation;   Hypothesis testing.
\end{abstract}

\section{Introduction}

Genome-wide association studies (GWAS)  have identified thousands of  genetic variants or single nucleotide polymorphisms (SNPs) that are associated with various complex phenotypes. Among them, many variants were found to be associated with multiple complex traits, reflecting the pleiotropic action of genes or correlation between causal loci in two traits. Understanding the shared genetic  architecture among different traits can potentially lead to  further insights into  the biological etiology of diseases and inform therapeutic interventions \citep{van2019genetic}.  

Various definitions of genetic relatedness or correlation have been proposed in different contexts to characterize quantitatively the shared genetic associations between complex traits based on GWAS data.  Understanding of genetic relatedness between complex traits  help identify new trait-associated variants \citep{turley2018multi}, improve genetic risk prediction \citep{maier2015joint} and assist inference on causality \citep{o2018distinguishing}. Comparing to the traditional approaches from family studies, where measurements of both traits are required for the same individuals, methods based on GWAS enjoy the advantages of  increased sample sizes and reduced risk of confounding or ascertainment biases, and have greater potential for large-scale analysis involving multiple traits \citep{zhang2020comparison}. 

Bivariate linear mixed-effects models have been widely applied to estimate the genetic covariance and genetic correlation between two traits from individual-level GWAS data \citep{lee2011estimating,lee2012estimation,vattikuti2012heritability,lee2013genetic}. These models decompose the phenotypic variance into genetic and residual variance components, and define the genetic  correlation to be the correlation between the two trait-specific random generic effects. However, the mixed-effect model approach requires knowledge about  the genetic relationship matrix, which is commonly approximated by the genetic relationship across the set of all genotyped variants \citep{yang2010common}.  More recently, computationally efficient methods based on the cross-trait linkage disequilibrium (LD) score regression were developed \citep{bulik2015atlas,ning2020high} to estimate genetic correlation using GWAS summary statistics over a large set of SNPs. This approach relies on the classical asymptotics that does not take into account the high-dimensionality of the SNPs compared to the sample sizes, and might lead to inaccurate inference results \citep{zhao2019cross}. Some other approaches such as \cite{shi2017local}, \cite{lu2017powerful} and \cite{guo2021detecting} aim to explore differences in local genetic correlations between traits through genome partitioning based on genomic annotations. \cite{weissbrod2018estimating} noticed that many of the existing methods are primarily geared toward quantitative traits, and direct application of these methods to data sets with binary outcomes may suffer from reduced statistical power; they  proposed a mixed effects model to estimate the genetic correlation between binary traits.

In this study, we take a high-dimensional regression approach with fixed genetic effects for identifying trait-associated genetic variants and quantifying the genetic relatedness between two traits.  An important advantage of multiple regression over the simple univariate regression is its potential of identifying more trait-associated variants \citep{wu2009genome}. Existing studies of heritability or co-heritability within the high-dimensional regression framework include, for example, \cite{bonnet2015heritability,janson2017eigenprism,verzelen2018adaptive,guo2019optimal,zhao2019cross,zhao2019genetic,guo2021group}. Under the linear regression model, \cite{guo2019optimal} proposed bias-corrected estimators for the genetic covariance and correlation parameters based on individual-level GWAS data and  \cite{zhao2019cross} proposed  consistent estimators for polygenic risk score and genetic correlation based on GWAS summary statistics.   However, these papers focus on the genetic relatedness between continuous traits, and do not provide inference procedures such as statistical tests.

This paper aims to address the following two questions concerning binary traits.  How to define and study the genetic relatedness between two binary traits under the high-dimensional regression framework? How to perform valid statistical inference such as testing hypothesis or constructing confidence intervals for the genetic relatedness parameters? We address  these questions in a principled way with rigorous statistical justifications.

To that end,  for a pair of binary traits $(y,w)\in\{0,1\}^2$, we consider the following high-dimensional logistic regression models
\beq \label{m1}
y|X\sim \text{Bernoulli}(\pi_y(X)), \qquad \log \left\{\frac{\pi_y(X)}{1-\pi_y(X)}\right\}=\alpha+X^\top\beta,
\eeq
\beq \label{m2}
w|X\sim \text{Bernoulli}(\pi_w(X)), \qquad \log \left\{\frac{\pi_w(X)}{1-\pi_w(X)}\right\}=\zeta+X^\top\gamma,
\eeq
where $\pi_y(X)=P(y=1 | X)$, $\pi_w(X)=P(w=1|X)$, $X\in \R^p$ is a random vector of $p$ genetic variants with population covariance matrix $\Sigma\in\R^{p\times p}$,  $\beta,\gamma\in\R^p$ are the corresponding trait-specific regression coefficients, which are assumed to be sparse vectors, and $\alpha,\zeta\in\R$ are the trait-specific intercepts. The genetic covariance  between the two traits is defined as the covariance between the log-odds ratios associated to the two traits, i.e.,
$$\textup{genetic covariance}(y,w)={\rm Cov}\left(\log \left\{\frac{\pi_y(X)}{1-\pi_y(X)}\right\},\log \left\{\frac{\pi_w(X)}{1-\pi_w(X)}\right\}\right),$$
which, by definition, admits the following expressions 
$$
{\rm Cov}\left(\log \left\{\frac{\pi_y(X)}{1-\pi_y(X)}\right\},\log \left\{\frac{\pi_w(X)}{1-\pi_w(X)}\right\}\right)={\rm Cov}(X^\top\beta, X^\top \gamma)=\beta^{\top}\Sigma \gamma.
$$
Similarly, we define the genetic variance of the binary trait $y$ as  the variance of its associated log-odds, i.e.,
$$
\textup{genetic variance}(y)={\rm Var}\left(\log \left\{\frac{\pi_y(X)}{1-\pi_y(X)}\right\}\right),
$$
which satisfies ${\rm Var}\left(\log \left\{\frac{\pi_y(X)}{1-\pi_y(X)}\right\}\right)={\rm Var}(X^\top\beta)=\beta^{\top}\Sigma \beta$, and define the genetic variance of the trait $w$  as ${\rm Var}\left(\log \left\{\frac{\pi_w(X)}{1-\pi_w(X)}\right\}\right)={\rm Var}(X^\top\gamma)=\gamma^{\top}\Sigma \gamma$. Whenever both the genetic variances of $y$ and $w$ are nonzero,  we can define the genetic correlation $R(y,w)$ between the two traits as the correlation between the associated log-odds ratios, that is, $
{\rm Corr}\left(\log \left\{\frac{\pi_y(X)}{1-\pi_y(X)}\right\},\log \left\{\frac{\pi_w(X)}{1-\pi_w(X)}\right\}\right)
=\frac{\beta^{\top}\Sigma \gamma}{\sqrt{\beta^\top\Sigma\beta\gamma^\top\Sigma\gamma}}, $
and set $R(y,w)=0$ whenever $\beta^\top\Sigma\beta \cdot \gamma^\top\Sigma\gamma=0$. 

The concept of covariance or correlation between two log-odds ratios is  both statistically and empirically meaningful, and has been adopted by \cite{wei2013estimating} to account for the correlated outcomes in meta-analysis, and by \cite{bagos2012covariance} when the data are presented in the form of contingency tables. In our context, as parameters or functionals quantifying the conditional co-occurrence risk of two traits, the genetic covariance and correlation defined above characterize the shared effect size of the genetic variants by taking into account the true covariance structure of the variants.

This paper studies the problem of statistical inference for these genetic relatedness functionals based on individual-level  GWAS data with binary outcomes. By carefully analyzing the logistic Lasso estimator, we develop a novel weighted debiasing method and propose  computationally efficient debiased estimators for these functionals. We further  study their rates of convergence and obtain their asymptotic normality under mild theoretical conditions. Moreover,  confidence intervals and statistical tests for these functionals are constructed. We provide theoretical justifications for the methods, including the coverage probability and expected length of the confidence intervals, as well as the size and power of the proposed tests. 
Our results provide a rigorous statistical inference framework for studying the genetic relatedness between binary traits.

Throughout, 
for a symmetric matrix $A\in \R^{p\times p}$, $\lambda_i(A)$ stands for its $i$-th largest eigenvalue and $\lambda_{\max}(A)  = \lambda_1(A)$, $\lambda_{\min}(A) = \lambda_{p} (A)$. For a smooth function $f(x)$ defined on $\R$, we denote $\dot{f}(x) = d f(x)/dx$ and $\ddot{f}(x) = d^2 f(x)/dx^2$.  
For sequences $\{a_n\}$ and $\{b_n\}$, we write 
$a_n = o(b_n)$, $a_n\ll b_n$ or $b_n\gg a_n$ if $\lim_{n} a_n/b_n =0$, and write $a_n = O(b_n)$, $a_n\lesssim b_n$ or $b_n \gtrsim a_n$ if there exists a constant $C$ such that $a_n \le Cb_n$ for all $n$. We write $a_n\asymp b_n$ if $a_n \lesssim b_n$ and $a_n\gtrsim b_n$. 

\section{Estimation of Genetic Relatedness} \label{method.est.sec}

\subsection{Genetic  Relatedness under Various Settings of Data Availability}

We consider two types of data collection scenarios that are commonly adopted for studying genetic relatedness between two traits based on individual-level GWAS data. Data sets obtained from these two scenarios are widely available in current genetic research. In the first scenario, measurements of two traits along with the subject genotypes are obtained from different groups of unrelated individuals. In other words, there are two independent data sets, each containing measurements of a single trait and genotypes for a group of unrelated individuals. This scenario arise commonly when researchers attempt to conduct a cross-trait analysis based on multiple independent GWAS data. In the second scenario, measurements of multiple traits of interest along with the subject genotypes may be obtained from a same group of unrelated individuals. This type of data set is also widely available by virtue of many large-scale studies such as UK Biobank \citep{sudlow2015uk}. The above two scenarios are formally defined as follows.

{\bf Scenario (I): Data from independent samples.} The observations are $\{(y_i,X_{i\cdot})\}_{i=1}^{n_1}$ and  $\{(w_i,Z_{i\cdot})\}_{i=1}^{n_2}$, where $X_{i\cdot}$ and $Z_{i\cdot}$  are drawn independently from some probability measure $P_\theta$ on $\R^p$ with covariance matrix $\Sigma$, and $y_i$ and $w_i$ are generated based on (\ref{m1})  and (\ref{m2}), respectively.

{\bf Scenario (II): Data from overlapped samples.} The observations are $\{(y_i,X_{i\cdot})\}_{i=1}^{n_1}$ and  $\{(w_i,Z_{i\cdot})\}_{i=1}^{n_2}$, where $Z_{i\cdot}=X_{i\cdot}$ for $i\in\{1,2,...,m\}$, $1\le m\le \min\{n_1,n_2\}$. The samples in $\{Z_{i\cdot}\}_{i=1}^{m}$, $\{X_{i\cdot}\}_{i=m+1}^{n_1}$ and $\{Z_{i\cdot}\}_{i=m+1}^{n_2}$ are drawn independently from some probability measure $P_\theta$ on $\R^p$ with covariance matrix $\Sigma$, and $y_i$ and $w_i$ are generated  from (\ref{m1})  and (\ref{m2}), respectively.

Note that Scenario (I) corresponds to Scenario (II) with $m=0$.
In what follows, we introduce our main results by focusing on Scenario (I) to avoid unnecessary complications in the notation. The discussions under Scenario II are delayed to Section S5 of the Supplement \citep{ma2020supp} as our methods and results in this case are very similar.

\subsection{Weighted Bias Correction and the Proposed Estimators} \label{est.sec}

Estimation of the genetic correlation $R$ can be reduced to estimating the genetic covariance functional $\beta^\top\Sigma\gamma$ and the genetic variance functionals $\beta^\top\Sigma\beta$ and $\gamma^\top\Sigma\gamma$, respectively. The novel bias correction method developed here will lead to nearly unbiased estimators of these functionals of interest, and the construction of which can be summarized as the following two-step procedure. In the first step, an initial plug-in estimator of the functional is obtained based on the pooled sample covariance matrix $\widehat{\Sigma}=\frac{1}{n_1+n_2}\big[\sum_{i=1}^{n_1}X_{i\cdot}X_{i\cdot}^\top+\sum_{i=1}^{n_2}Z_{i\cdot}Z_{i\cdot}^\top\big]$, and the logistic Lasso estimators
\beq \label{lasso}
\begin{aligned}
	(\widehat{\alpha},\widehat{\beta}) = \argmin_{\alpha,\beta} \bigg[\frac{1}{n_1}\sum_{i=1}^{n_1} \bigg\{-y_i(\alpha+\beta^\top X_{i\cdot})+\log(1+e^{\alpha+\beta^\top X_{i\cdot}}) \bigg\}+\lambda( \|\beta\|_1+|\alpha|)\bigg],\\
	(\widehat{\zeta},\widehat{\gamma}) = \argmin_{\zeta,\gamma} \bigg[\frac{1}{n_2}\sum_{i=1}^{n_2} \bigg\{-w_i(\zeta+\gamma^\top Z_{i\cdot})+\log(1+e^{\zeta+\gamma^\top Z_{i\cdot}}) \bigg\}+\lambda( \|\gamma\|_1+|\zeta|)\bigg],
\end{aligned}
\eeq
with $\lambda=C \sqrt{\log p/n}$ for some constant $C>0$. In the second step, the final estimator is obtained by modifying the initial estimator with a carefully designed bias correction term.

We begin with genetic covariance functional $\beta^\top\Sigma\gamma$. With the logistic Lasso estimators (\ref{lasso}) and $\widehat{\Sigma}$, the corresponding plug-in estimator is defined as $\widehat{\beta}^{\top} \widehat{\Sigma}\widehat{\gamma}$, whose error can be decomposed as
$
\widehat{\beta}^{\top} \widehat{\Sigma}\widehat{\gamma}-\beta^{\top}\Sigma \gamma=\widehat{\gamma}^\top\Sigma(\widehat{\beta}-\beta)+\widehat{\beta}^\top\Sigma(\widehat{\gamma}-\gamma)-(\widehat{\beta}-\beta)^\top\Sigma(\widehat{\gamma}-\gamma)+\widehat{\beta}^\top(\widehat{\Sigma}-\Sigma)\widehat{\gamma}.
$
It turns out that the term $\widehat{\beta}^\top(\widehat{\Sigma}-\Sigma)\widehat{\gamma}$ only contributes to the variance of the plug-in estimator, the terms $\widehat{\gamma}^\top\Sigma(\widehat{\beta}-\beta)$ and $\widehat{\beta}^\top\Sigma(\widehat{\gamma}-\gamma)$ contribute to the leading order bias of the plug-in estimator, whereas the contribution from $(\widehat{\beta}-\beta)^\top\Sigma(\widehat{\gamma}-\gamma)$ is negligible. Therefore, the bias of the plug-in estimator can be further reduced by estimating $\widehat{\gamma}^\top\Sigma(\widehat{\beta}-\beta)$ and $\widehat{\beta}^\top\Sigma(\widehat{\gamma}-\gamma)$ directly. To accomplish this, set $h(u)=\frac{e^u}{1+e^u}$, then by Taylor's expansion, 
$
h(\widehat{\alpha}+X_{i\cdot}^\top\widehat{\beta})-h(\alpha+X_{i\cdot}^\top\beta)=\frac{e^{\widehat{\alpha}+X_{i\cdot}^{\top}\widehat{\beta}}X^{\top}_{i\cdot}(\widehat{\beta}-\beta) }{(1+e^{\widehat{\alpha}+X_{i\cdot}^{\top}\widehat{\beta}})^2}+\frac{e^{\widehat{\alpha}+X_{i\cdot}^{\top}\widehat{\beta}} \left(\widehat{\alpha}-\alpha\right)}{(1+e^{\widehat{\alpha}+X_{i\cdot}^{\top}\widehat{\beta}})^2}+\Delta_i,
$
where $\Delta_i=\ddot{h}[{X'_{i\cdot}}^\top\{t\beta'+(1-t)\widehat{\beta'}\}]\{{X'_{i\cdot}}^\top(\widehat{\beta}'-\beta')\}^2$ for some $t\in(0,1)$, ${\beta}'=({\alpha},{\beta}^\top)^\top$, $\widehat{\beta}'=(\widehat{\alpha},\widehat{\beta}^\top)^\top$ and $X'_{i\cdot}=(1,X_{i\cdot}^\top)^\top$.
Furthermore, if we define $\epsilon_i=y_i-h({\alpha}+X_{i\cdot}^\top{\beta})$, 
\begin{align*}
	&\quad\{h(\widehat{\alpha}+X_{i\cdot}^\top\widehat{\beta})-y_i\}X_{i\cdot}\\
	&=\bigg\{\frac{e^{\widehat{\alpha}+X_{i\cdot}^{\top}\widehat{\beta}}}{(1+e^{\widehat{\alpha}+X_{i\cdot}^{\top}\widehat{\beta}})^2}X^{\top}_{i\cdot}(\widehat{\beta}-\beta)+\frac{e^{\widehat{\alpha}+X_{i\cdot}^{\top}\widehat{\beta}}}{(1+e^{\widehat{\alpha}+X_{i\cdot}^{\top}\widehat{\beta}})^2}\left(\widehat{\alpha}-\alpha\right)+\Delta_i-\epsilon_i\bigg\}X_{i\cdot}\\
	&=\frac{e^{\widehat{\alpha}+X_{i\cdot}^{\top}\widehat{\beta}}}{(1+e^{\widehat{\alpha}+X_{i\cdot}^{\top}\widehat{\beta}})^2}X_{i\cdot}X^{\top}_{i\cdot}(\widehat{\beta}-\beta)+(\Delta_i-\epsilon_i)X_{i\cdot}+\frac{e^{\widehat{\alpha}+X_{i\cdot}^{\top}\widehat{\beta}}}{(1+e^{\widehat{\alpha}+X_{i\cdot}^{\top}\widehat{\beta}})^2}\left(\widehat{\alpha}-\alpha\right)X_{i\cdot}.
\end{align*}
In order to construct a good estimator of $\Sigma(\widehat{\beta}-\beta)$, we rescale each item $\{h(\widehat{\alpha}+X_{i\cdot}^\top\widehat{\beta})-y_i\}X_{i\cdot}$ by a sample-specific weight $\frac{(1+e^{\widehat{\alpha}+X_{i\cdot}^{\top}\widehat{\beta}})^2}{e^{\widehat{\alpha}+X_{i\cdot}^{\top}\widehat{\beta}}}$ so that
\begin{align*}
	&\sum_{i=1}^{n_1}\frac{(1+e^{\widehat{\alpha}+X_{i\cdot}^{\top}\widehat{\beta}})^2}{e^{\widehat{\alpha}+X_{i\cdot}^{\top}\widehat{\beta}}}\{h(\widehat{\alpha}+X_{i\cdot}^\top\widehat{\beta})-y_i\}X_{i\cdot}\\
	&=\bigg(\sum_{i=1}^{n_1}X_{i\cdot}X^{\top}_{i\cdot}\bigg)(\widehat{\beta}-\beta)+\sum_{i=1}^{n_1}\frac{(1+e^{\widehat{\alpha}+X_{i\cdot}^{\top}\widehat{\beta}})^2}{e^{\widehat{\alpha}+X_{i\cdot}^{\top}\widehat{\beta}}}(\Delta_i-\epsilon_i)X_{i\cdot}+(\widehat{\alpha}-\alpha)\sum_{i=1}^{n_1}X_{i\cdot}.
\end{align*}
Consequently, as long as the last two terms in the above equation are negligible comparing to the leading term $\big(\sum_{i=1}^{n_1}X_{i\cdot}X^{\top}_{i\cdot}\big)(\widehat{\beta}-\beta)$, an estimator of $\widehat{\gamma}^\top\Sigma(\widehat{\beta}-\beta)$ can be constructed as
\beq \label{bc1}
\widehat{\gamma}^{\top}\frac{1}{n_1}\sum_{i=1}^{n_1}\frac{(1+e^{\widehat{\alpha}+X_{i\cdot}^{\top}\widehat{\beta}})^2}{e^{\widehat{\alpha}+X_{i\cdot}^{\top}\widehat{\beta}}}\{h(\widehat{\alpha}+X_{i\cdot}^\top\widehat{\beta})-y_i\}X_{i\cdot}.
\eeq
Similarly, we can estimate the error term $\widehat{\beta}^\top\Sigma(\widehat{\gamma}-\gamma)$ by
\beq \label{bc2}
\widehat{\beta}^{\top}\frac{1}{n_2}\sum_{i=1}^{n_2}\frac{(1+e^{\widehat{\zeta}+Z_{i\cdot}^{\top}\widehat{\gamma}})^2}{e^{\widehat{\zeta}+Z_{i\cdot}^{\top}\widehat{\gamma}}}\{h(\widehat{\zeta}+Z_{i\cdot}^\top\widehat{\gamma})-w_i\}Z_{i\cdot}.
\eeq
As a result, in light of the error decomposition, a bias-corrected estimator for $\beta^\top\Sigma\gamma$ is defined as
\beq\label{eq: weighted FDE}
\begin{aligned}
	\widehat{\beta^{\top}\Sigma\gamma}&=\widehat{\beta}^{\top} \widehat{\Sigma}\widehat{\gamma}-\widehat{\gamma}^{\top}\frac{1}{n_1}\sum_{i=1}^{n_1}\frac{(1+e^{\widehat{\alpha}+X_{i\cdot}^{\top}\widehat{\beta}})^2}{e^{\widehat{\alpha}+X_{i\cdot}^{\top}\widehat{\beta}}}\{h(\widehat{\alpha}+X_{i\cdot}^\top\widehat{\beta})-y_i\}X_{i\cdot}\\
	&\quad-\widehat{\beta}^{\top}\frac{1}{n_2}\sum_{i=1}^{n_2}\frac{(1+e^{\widehat{\zeta}+Z_{i\cdot}^{\top}\widehat{\gamma}})^2}{e^{\widehat{\zeta}+Z_{i\cdot}^{\top}\widehat{\gamma}}}\{h(\widehat{\zeta}+Z_{i\cdot}^\top\widehat{\gamma})-w_i\}Z_{i\cdot}.
\end{aligned} 
\eeq
The above  estimator modifies the simple plug-in estimator by adding a carefully constructed bias-correction term accounting for the leading order bias of the plug-in estimator. The bias-correction terms (\ref{bc1}) and (\ref{bc2}) are weighted averages, where the weights, originated from the nonlinearity of the link function, reflect each sample's contribution to the overall bias. 

In the same vein of our construction of  the estimator $\widehat{\beta^\top\Sigma\gamma}$,  bias-corrected estimators for the genetic variances can be defined similarly as 
\begin{align}
	&\widehat{\beta^{\top}\Sigma\beta}=\widehat{\beta}^{\top} \widehat{\Sigma}\widehat{\beta}-2\widehat{\beta}^{\top}\frac{1}{n_1}\sum_{i=1}^{n_1}\frac{(1+e^{\widehat{\alpha}+X_{i\cdot}^{\top}\widehat{\beta}})^2}{e^{\widehat{\alpha}+X_{i\cdot}^{\top}\widehat{\beta}}}\{h(\widehat{\alpha}+X_{i\cdot}^\top\widehat{\beta})-y_i\}X_{i\cdot}, \label{q.est.1}\\
	&\widehat{\gamma^{\top}\Sigma\gamma}=\widehat{\gamma}^{\top} \widehat{\Sigma}\widehat{\gamma}-2\widehat{\gamma}^{\top}\frac{1}{n_2}\sum_{i=1}^{n_2}\frac{(1+e^{\widehat{\zeta}+Z_{i\cdot}^{\top}\widehat{\gamma}})^2}{e^{\widehat{\zeta}+Z_{i\cdot}^{\top}\widehat{\gamma}}}\{h(\widehat{\zeta}+Z_{i\cdot}^\top\widehat{\gamma})-w_i\}Z_{i\cdot} \label{q.est.2}
\end{align}
Based on the above genetic variance and covariance estimators, a natural estimator of the genetic correlation is $\bar{R}=	\frac{\widehat{\beta^\top\Sigma\gamma}}{\sqrt{\widehat{\beta^\top\Sigma\beta}\widehat{\gamma^\top\Sigma\gamma}}}.$
Taking into account the actual range of $R$, we propose its final estimator as
\beq \label{R.hat}
\widehat{R} = \left\{ \begin{array}{ll}
	\bar{R}, & \textrm{if $(\widehat{\beta^\top\Sigma\gamma})^2<{\widehat{\beta^\top\Sigma\beta}\widehat{\gamma^\top\Sigma\gamma}}$}\\
	0, & \textrm{if ${\widehat{\beta^\top\Sigma\beta}\widehat{\gamma^\top\Sigma\gamma}}=0$}\\
	\text{sign}(\bar{R}),& \textrm{otherwise}
\end{array} \right. .
\eeq
Compared to the existing methods for constructing debiased estimators in high-dimensional regression \citep{zhang2014confidence,javanmard2014confidence,javanmard2014hypothesis,van2014asymptotically,cai2017confidence,guo2019optimal,ma2020global,cai2018semi,cai2020statistical,guo2021inference}, our proposed method has two distinct advantages.  Firstly, the proposed estimators can be directly obtained from their explicit expressions as in (\ref{eq: weighted FDE}) to (\ref{R.hat}), which only rely on the initial logistic Lasso estimator, and simple plug-in procedures. Its main computational task is to solve for the initial Lasso estimator, which can be efficiently done with a standard tuning process (Section \ref{simu.sec}), and therefore is more scalable to the large data sets in genetic studies. This is very different from the existing methods, which, in addition to the initial estimator, involve solving other high-dimensional optimization problems for bias correction, which are computationally challenging, time-consuming, and subject to difficult tuning processes.  Secondly, with our carefully constructed weighted bias-correction method, many commonly used but stringent technical conditions 
can be avoided. This significantly expands the range of applicability of our proposed methods; see also the discussions after Theorems \ref{est.thm} and \ref{asym.thm1}. 

\section{Confidence Intervals and Statistical Tests} \label{inf.sec}

As an important consequence, it can be shown that each of the above proposed estimators is asymptotically normally distributed.  This can be used to construct confidence intervals and statistical tests for the functionals.

Specifically,  it can be shown that the estimator  $\widehat{\beta^\top\Sigma\gamma}$ has variance
$$
v^2=\frac{n_1+n_2}{n_1}E\{ \eta_i^{(X)}(\widehat{\gamma}^\top X_{i\cdot})^2\}+\frac{n_1+n_2}{n_2}E\{ \eta_i^{(Z)}(\widehat{\beta}^\top Z_{i\cdot})^2\}+E \big\{ \widehat{\beta}^\top (X_{i\cdot}X_{i\cdot}^\top -\Sigma)\widehat{\gamma}\big\}^2,
$$
where $\eta_i^{(X)}=\frac{(1+e^{\widehat{\alpha}+X_{i\cdot}^{\top}\widehat{\beta}})^4e^{{\alpha}+X_{i\cdot}^{\top}{\beta}}}{(1+e^{\alpha+X_{i\cdot}^{\top}{\beta}})^2e^{2\widehat{\alpha}+2X_{i\cdot}^{\top}\widehat{\beta}}}$ and $\eta_i^{(Z)}=\frac{(1+e^{\widehat{\zeta}+Z_{i\cdot}^{\top}\widehat{\gamma}})^4e^{{\zeta}+Z_{i\cdot}^{\top}{\gamma}}}{(1+e^{{\zeta}+Z_{i\cdot}^{\top}{\gamma}})^2e^{2\widehat{\zeta}+2Z_{i\cdot}^{\top}\widehat{\gamma}}}.$
Intuitively,  the parameters $\beta$ and $\gamma$ in the above expressions can be estimated by their initial Lasso estimators, so that a moment estimator of the asymptotic variance can be defined as
$
\widehat{v}^2=\frac{n_1+n_2}{n^2_1}\sum_{i=1}^{n_1} \frac{(1+e^{\widehat{\alpha}+X_{i\cdot}^\top\widehat{\beta}})^2}{e^{\widehat{\alpha}+X_{i\cdot}^\top\widehat{\beta}}} (\widehat{\gamma}^\top X_{i\cdot})^2+\frac{n_1+n_2}{n^2_2}\sum_{i=1}^{n_2} \frac{(1+e^{\widehat{\eta}+Z_{i\cdot}^\top\widehat{\gamma}})^2}{e^{\widehat{\eta}+Z_{i\cdot}^\top\widehat{\gamma}}} (\widehat{\beta}^\top Z_{i\cdot})^2
+\frac{1}{n_1+n_2}\big\{\sum_{i=1}^{n_1}(\widehat{\beta}X_{i\cdot}X_{i\cdot}^\top\widehat{\gamma}-\widehat{\beta}\widehat{\Sigma}\widehat{\gamma})^2+\sum_{i=1}^{n_2}(\widehat{\beta}Z_{i\cdot}Z_{i\cdot}^\top\widehat{\gamma}-\widehat{\beta}\widehat{\Sigma}\widehat{\gamma})^2\big\}. 
$
Hence, an $(1-\alpha)$-level CI for the genetic covariance is
$
{\text{CI}_{\alpha}(\beta^{\top}\Sigma\gamma,\mathcal{D})}=\big[ \widehat{\beta^{\top}\Sigma\gamma}-\widehat{\rho},  \widehat{\beta^{\top}\Sigma\gamma}+\widehat{\rho}\big],
$
where $\widehat{\rho}=\frac{z_{\alpha/2}\widehat{v}}{\sqrt{n_1+n_2}}$ and $z_{\alpha/2}=\Phi^{-1}(1-\alpha/2)$ is the upper $\alpha/2$-quantile of the standard normal distribution. 
Similarly,  the asymptotic variance of the genetic variance estimator $\widehat{\beta^\top\Sigma\beta}$ can be derived as
$
v^2_\beta=\frac{4(n_1+n_2)}{n_1}E\{ \eta_i^{(X)} (\widehat{\beta}^\top X_{i\cdot})^2\}+E\{\widehat{\beta}^\top(X_{i\cdot}X_{i\cdot}^\top-\Sigma)\widehat{\beta}\}^2,
$
which can be estimated by
$
\widehat{v}^2_\beta=\frac{4(n_1+n_2)}{n^2_1}\sum_{i=1}^{n_1} \frac{(1+e^{\widehat{\alpha}+X_{i\cdot}^\top\widehat{\beta}})^2}{e^{\widehat{\alpha}+X_{i\cdot}^\top\widehat{\beta}}} (\widehat{\beta}^\top X_{i\cdot})^2
+\frac{1}{n_1+n_2}\bigg\{\sum_{i=1}^{n_1}(\widehat{\beta}X_{i\cdot}X_{i\cdot}^\top\widehat{\beta}-\widehat{\beta}\widehat{\Sigma}\widehat{\beta})^2+\sum_{i=1}^{n_2}(\widehat{\beta}Z_{i\cdot}Z_{i\cdot}^\top\widehat{\beta}-\widehat{\beta}\widehat{\Sigma}\widehat{\beta})^2\bigg\}.
$
Then, an $(1-\alpha)$-level confidence interval for $\beta^\top\Sigma\beta$ is
$
{\text{CI}_{\alpha}(\beta^{\top}\Sigma\beta,\mathcal{D})}=\big[ \widehat{\beta^{\top}\Sigma\beta}-\widehat{\rho}_\beta,  \widehat{\beta^{\top}\Sigma\beta}+\widehat{\rho}_\beta\big],
$
where $\widehat{\rho}_\beta=\frac{z_{\alpha/2}\widehat{v}_\beta}{\sqrt{n_1+n_2}}$.  The confidence interval ${\text{CI}_{\alpha}(\gamma^{\top}\Sigma\gamma,\mathcal{D})}$  can be obtained by symmetry.

The confidence interval for the genetic correlation $R$ is a direct consequence of the Slutsky's theorem. Specifically, for the estimator $\widehat{R}$ defined in (\ref{R.hat}), whenever $\widehat{\beta^\top\Sigma\beta}\widehat{\gamma^\top\Sigma\gamma}\ne 0$, we can estimate its asymptotic variance by
$
\widehat{v}_R^2 = \frac{\widehat{v}^2}{\widehat{\beta^\top\Sigma\beta}\widehat{\gamma^\top\Sigma\gamma}}, 
$
and define the corresponding $(1-\alpha)$-level confidence interval as
$
{\text{CI}_{\alpha}(R,\mathcal{D})}=\big[ \widehat{R}-\widehat{\rho}_R,  \widehat{R}+\widehat{\rho}_R\big]\cap[-1,1],
$
where $\widehat{\rho}_R=\frac{z_{\alpha/2}\widehat{v}_R}{\sqrt{n_1+n_2}}$.

Converting the above CIs, we obtain statistical tests for each of the null hypotheses $H_{0,1}: \beta^{\top}\Sigma\gamma=B_0$,$H_{0,2}: \beta^{\top}\Sigma\beta=Q_0$, and $H_{0,3}: R=R_0$, 
for some $B_0\in\R$, $Q_0\ge0$ and $R_0\in[-1,1]$.
Specifically, we define test statistics 
$T_1=\frac{\sqrt{n_1+n_2}(\widehat{\beta^{\top}\Sigma\gamma}-B_0)}{\widehat{v}}$, $T_2=\frac{\sqrt{n_1+n_2}(\widehat{\beta^{\top}\Sigma\beta}-Q_0)}{\widehat{v}_\beta}$ and $T_3=\frac{\sqrt{n_1+n_2}(\widehat{R}-R_0)}{\widehat{v}_R},$
so that for each $\ell\in\{1,2,3\}$, to obtain an $\alpha$-level test, we reject the null hypothesis $H_{0,\ell}$ whenever $|T_{\ell}|>z_{\alpha/2}$.

\section{Theoretical Properties} \label{theory.sec}

\subsection{Rates of Convergence and Optimality}

The random covariates are characterized by the following conditions.

\noindent {\bf(A1)}  For each $1\le i\le n_1$ and $1\le j\le n_2$, $X_{i\cdot}$ and $Z_{j\cdot}$ are centred $i.i.d.$ sub-Gaussian random vectors where $\Sigma=E(X_{i\cdot}X_{i\cdot}^\top)\in\R^{p\times p}$ satisfies $ M^{-1}\le \lambda_{\min}(\Sigma)\le \lambda_{\max}(\Sigma)\le M$ for some constant $M>1$.

\noindent{\bf(A2)} There exists a positive constant $c_0$ such that $E\left(  \frac{\beta^\top X_{i\cdot}X_{i\cdot}^\top \gamma}{\beta^\top\Sigma\gamma}-1\right)^2>c_0$.

About the regression coefficients, we denote $k=\max\{\|\beta\|_0,\|\gamma\|_0\}$, $U(\beta,\gamma)=\max\{\|\beta\|_2,\|\gamma\|_2\}$ and $L(\beta,\gamma)=\min\{\|\beta\|_2,\|\gamma\|_2\}$. We assume

\noindent {\bf(A3)} $\max\{|\alpha|,|\zeta|\}\le C$ and $U(\beta,\gamma)\le C$ for some constant $C>0$. 

Intuitively, assumptions (A1) and (A3) imply that the marginal case probabilities $P(y_i=1)$ and $P(w_i=1)$ are balanced, or bounded away from 0 and 1, whereas (A2)  ensures the asymptotic variances does not diminish.

For technical reasons, for each trait we split the corresponding samples into halves so that the initial Lasso estimation step and the rest of the steps such as covariance estimation and bias-correction are conducted on independent data sets. Without loss of generality, we assume under Scenario I  there are $2(n_1+n_2)$ samples in $\mathcal{D}$, divided into two disjoint subsets $\mathcal{D}_1$ and $\mathcal{D}_2$, each containing $n_1+n_2$ independent samples, with $n_1$ samples corresponding to trait $y_i$ and $n_2$ samples corresponding to trait $w_i$. The initial Lasso estimators  are obtained from $\mathcal{D}_1$, whereas the sample covariance, the bias-correction terms and the asymptotic variance estimators are based on $\mathcal{D}_2$ and the initial Lasso estimators. 
We emphasize that the sample splitting procedure is only used to facilitate the theoretical analysis, and is not needed in practice. We demonstrate this point numerically in Section \ref{simu.sec}; see also Section \ref{diss.sec} for more discussions.

The following theorem concerns the rate of convergence of the bias-corrected estimators $\widehat{\beta^\top\Sigma\gamma}$ and $\widehat{\beta^\top\Sigma\beta}$; the results for $\widehat{\gamma^\top\Sigma\gamma}$ are similar.

\bet[Rates of Convergence] \label{est.thm}
Suppose (A1) and (A3) hold, $n_1\asymp n_2\asymp n$ and $k\lesssim \frac{n}{\log p{\log n}}$. Then, for sufficiently large $(n,p)$ and any $t>0$, 
\begin{align}
	|\widehat{\beta^\top\Sigma\gamma}-\beta^\top\Sigma\gamma|&\lesssim \frac{tU(\beta,\gamma)}{\sqrt{n}}+\{1+U(\beta,\gamma)\sqrt{\log n}\}\frac{k\log p}{n},\\
	|\widehat{\beta^\top\Sigma\beta}-\beta^\top\Sigma\beta|&\lesssim \frac{t\|\beta\|_2}{\sqrt{n}}+(1+\|\beta\|_2\sqrt{\log n})\frac{k\log p}{n},
\end{align}
with probability at least $1-p^{-c}-n^{-c}-t^{-2}$ for some constant $c>0$.
\eet

In Theorem \ref{est.thm}, in addition to the mild sparsity condition, the consistency of the proposed estimators only require the balanced marginal case probabilities through (A1) and (A3), and the general sub-Gaussian design with a regular covariance matrix, which includes many important cases such as Gaussian, bounded, and binary designs, or any combinations of them. It makes the proposed methods widely applicable to various practical settings.

To establish the optimality of the proposed genetic covariance estimator, our next result concerns the minimax lower bound for estimating  ${\beta^\top\Sigma\gamma}$.
To this end, we define the parameter space for $\theta=(\beta,\gamma,\Sigma)$ as
\[
\Theta(k,L_n)=\left\{(\beta,\gamma,\Sigma): \begin{aligned} &\max\{\|\beta\|_0,\|\gamma\|_0\}\le k,U(\beta,\gamma)\le L_n \\
	&M^{-1}\le \lambda_{\min}(\Sigma)\le \lambda_{\max}(\Sigma)\le M \end{aligned}\right\}
\]
for some constant $M>1$, and denote $\xi=\beta^\top\Sigma\gamma$.

\bet[Minimax Lower Bound] \label{lower.est.thm}
Suppose $X_i$ and $Z_i\stackrel{i.i.d.}{\sim} N(0,\Sigma)$ for $i=1,...,n$, and $k \lesssim \min\big\{ p^\nu, \frac{n}{\log p} \big\}$ for some $0< \nu <1/2$. Then
\beq
\inf_{\widehat{\xi}}\sup_{\theta\in\Theta(k,L_n)} P_\theta\bigg(|\widehat{\xi}-\xi|\gtrsim \frac{L_n^2}{\sqrt{n}}+ \min\bigg\{\frac{L_n}{\sqrt{n}}+k\frac{\log p}{n}, L_n^2 \bigg\}\bigg)\ge c
\eeq
for some constant $c>0$.
\eet

By Theorem \ref{est.thm}, a uniform upper bound over the parameter space $\Theta(k,L_n)$ can be obtained as
$
\sup_{\theta\in\Theta(k,L_n)} P_\theta\big(|\widehat{\beta^\top\Sigma\gamma}-\beta^\top\Sigma\gamma|\lesssim \frac{tL_n}{\sqrt{n}}+(1+L_n\sqrt{\log n})\frac{k\log p}{n}\big)\ge 1-p^{-c}-n^{-c}-t^{-2}.
$
Combining this with the  lower bound from Theorem \ref{lower.est.thm}, we conclude that,  for all $k\lesssim \min\{\frac{n}{\log p\log n},p^{\nu}\}$ with any $\nu\in(0,1/2)$, and $\sqrt{\frac{k\log p}{n}}\lesssim L_n\lesssim 1$, our genetic covariance estimator $\widehat{\beta^\top\Sigma\gamma}$ is minimax rate-optimal over $\Theta(k,L_n)$, up to  a $\sqrt{\log n}$ factor. 
In particular, in this case, the exact rate optimality of  $\widehat{\beta^\top\Sigma\gamma}$ is guaranteed over the ultra-sparse region $k\lesssim \frac{\sqrt{n}}{\log p\sqrt{\log n}}$, or the weak signal regime $L_n\lesssim (\log n)^{-1/2}$, over which the minimax rate is $\frac{L_n}{\sqrt{n}}+\frac{k\log p}{n}$. Moreover, this suggests that the uncertainty due to the covariance estimation $\widehat\beta^\top(\widehat\Sigma-\Sigma)\widehat\gamma$ in the plug-in estimator is fundamental and may not be removed as for the leading order biases.



\bet[Rate of Convergence] \label{R.est.thm}
Suppose (A1) (A2) and (A3) hold, $n_1\asymp n_2\asymp n$,  $k\ll \frac{n}{\log p\log n}$ and $L(\beta,\gamma)\gg \sqrt{{k\log p}/{n}}$. Then $|\widehat{R}-R|\to 0$ in probability. In particular, for sufficiently large $(n,p)$ and any constant $t>\sqrt{2}$, with probability at least $1-2t^{-2}$, it holds that
\beq
|\widehat{R}-R|\lesssim \frac{t\{U(\beta,\gamma)+U^2(\beta,\gamma)\}}{L^2(\beta,\gamma)\sqrt{n}}+\frac{1+U(\beta,\gamma)\sqrt{\log n}}{L^2(\beta,\gamma)}\cdot\frac{k\log p}{n}.
\eeq
\eet

Comparing to Theorem \ref{est.thm}, the consistency of $\widehat{R}$ requires an additional condition (A2) and a lower bound on the minimal effect size. These conditions are necessary to ensure the true genetic variances are bounded away from zero and the genetic correlation is well-defined.

\subsection{Theoretical Properties of Inference Procedures}

We establish the asymptotic normality of the proposed bias-corrected estimators and provide theoretical justifications of the CIs and the statistical tests. We start with a theorem that  provides a refined analysis of the estimation errors and consequently the asymptotic normality of the estimators.

\bet[Asymptotic Normality] \label{asym.thm0}
Suppose (A1) (A2) (A3) hold, $n_1\asymp n_2\asymp n$, $k\lesssim \frac{n}{\log p{\log n}}$ and $L(\beta,\gamma)\gg\sqrt{{k\log p}/{n}}$. Then
\begin{enumerate}
	\item It holds that $\widehat{\beta^\top\Sigma\gamma}-\beta^\top\Sigma\gamma=A_{n}+B_{n}$,
	where 
	$
	P\big\{A_{n}\lesssim \{U(\beta,\gamma)\sqrt{\log n}+1\}\frac{k\log p}{n}\big\} \ge 1-p^{-c}-n^{-c},
	$
	and $\frac{\sqrt{n_1+n_2}B_{n}}{v}\big| \mathcal{D}_1\to_d N(0,1)$ as $(n,p)\to\infty$. Additionally, if $k\ll  \frac{U(\beta,\gamma)\sqrt{n}}{\{1+U(\beta,\gamma)\sqrt{\log n}\}\log p}$, we establish the asymptotic normality
	$
	\frac{\sqrt{n_1+n_2}(\widehat{\beta^\top\Sigma\gamma}-\beta^\top\Sigma\gamma)}{v}\big| \mathcal{D}_1\to_d N(0,1).
	$
	\item It holds that $\widehat{\beta^\top\Sigma\beta}-\beta^\top\Sigma\beta=A'_{n}+B'_{n}$,
	where 
	$
	P\big\{A'_{n}\lesssim (\|\beta\|_2\sqrt{\log n}+1)\frac{k\log p}{n}\big\} \ge 1-p^{-c}-n^{-c},
	$
	and $\frac{\sqrt{n_1+n_2}B'_{n}}{v_\beta}\big| \mathcal{D}_1\to_d N(0,1)$ as $(n,p)\to\infty$. Additionally, if $k\ll  \frac{\|\beta\|_2\sqrt{n}}{[1+\|\beta\|_2\sqrt{\log n}]\log p}$, we establish the asymptotic normality
	$
	\frac{\sqrt{n_1+n_2}(\widehat{\beta^\top\Sigma\beta}-\beta^\top\Sigma\beta)}{v_\beta}\big| \mathcal{D}_1\to_d N(0,1).
	$
\end{enumerate}
\eet

The second part of the theorem applies to the estimator $\widehat{\gamma^\top\Sigma \gamma}$ by symmetry. A direct consequence of Theorems \ref{est.thm} and \ref{asym.thm0}, in combination with Slutsky's theorem, is the following theorem concerning the asymptotic normality of the  genetic correlation estimator $\bar{R}$ in Section \ref{est.sec}.

\bet[Asymptotic Normality] \label{asym.thm1}
Under the conditions of Theorem \ref{asym.thm0}, if in addition we have $k\ll \min\{\frac{n}{\log p\log n}, \frac{U(\beta,\gamma)\sqrt{n}}{\{1+U(\beta,\gamma)\sqrt{\log n}\}\log p}  \}$, then we have $\frac{\sqrt{n_1+n_2}(\bar{R}-R)}{v_R}\big| \mathcal{D}_1\to_d N(0,1)$ as $(n,p)\to\infty$.
\eet

Some remarks about the technical innovations leading to the above theorems are in order. Firstly, distinct from the existing works on the statistical inference in high-dimensional logistic regression, the proposed methods do not require the commonly assumed but stringent theoretical conditions such as the bounded individual probability condition \citep{van2008high,van2014asymptotically,ning2017general,ma2020global,guo2021inference} where $P(y_i=1|X_{i\cdot})\in(\delta,1-\delta)$ for all $1\le i\le n$ and some $\delta\in(0,1/2)$, the sparse inverse population Hessian condition \citep{van2014asymptotically,belloni2016post,ning2017general,jankova2018biased} or the sparse precision condition \citep{ma2020global}. Secondly, from a practical point of view, the removal of these technical assumptions significantly expands the range of applicability of the proposed methods. For example, as was argued by \cite{cai2020statistical} and \cite{xia2020revisit}, in practice, the bounded individual probability and the sparse inverse population Hessian conditions are seldom satisfied or can be verified from the data. In contrast, the balanced marginal case probability condition holds  easily and can be  checked based on the observed outcomes.

Built upon Theorems \ref{asym.thm0} and \ref{asym.thm1}, theoretical justifications such as the asymptotic coverage probability and the expected length of the proposed CIs ${\text{CI}_{\alpha}(\beta^\top\Sigma\gamma,\mathcal{D})}$, ${\text{CI}_{\alpha}(\beta^\top\Sigma\beta,\mathcal{D})}$ and ${\text{CI}_{\alpha}(R,\mathcal{D})}$ can be obtained.

\bet[Confidence Intervals] \label{ci.thm1}
Under the conditions of Theorem \ref{asym.thm0}, for any constant $0<\alpha<1$, if $k\ll \min\{\frac{n}{\log p\log n}, \frac{U(\beta,\gamma)\sqrt{n}}{\{1+U(\beta,\gamma)\sqrt{\log n}\}\log p}  \}$, then
\begin{enumerate}
	\item(Coverage) 
	$
	{\liminf_{n,p\to\infty}}P_\theta\{ \beta^\top\Sigma\gamma\in {\textup{CI}_{\alpha}(\beta^\top\Sigma\gamma,\mathcal{D})}\}\ge 1-\alpha,
	$
	$
	{\liminf_{n,p\to\infty}  }P_\theta\{ \beta^\top\Sigma\beta\in {\textup{CI}_{\alpha}(\beta^\top\Sigma\beta,\mathcal{D})}\}\ge 1-\alpha,
	$
	and
	$
	{\liminf_{n,p\to\infty}}P_\theta\{ R\in {\textup{CI}_{\alpha}(R,\mathcal{D})}\}\ge 1-\alpha;
	$
	\item(Length) if we denote $L\{\textup{CI}_\alpha(\cdot,\mathcal{D})\}$ as the length of  $\textup{CI}_\alpha(\cdot,\mathcal{D})$, then with probability at least $1-p^{-c}$, we have
	$
	L\{\textup{CI}_{\alpha}(\beta^\top\Sigma\gamma,\mathcal{D})\}\asymp \frac{U(\beta,\gamma)}{\sqrt{n}}$, $L\{\textup{CI}_{\alpha}(\beta^\top\Sigma\beta,\mathcal{D})\} \asymp \frac{\|\beta\|_2}{\sqrt{n}}.
	$  and 
	$
	L\{\textup{CI}_{\alpha}(R,\mathcal{D})\}\asymp\frac{1}{L(\beta,\gamma)\sqrt{n}}.
	$
\end{enumerate}
\eet

This theorem implies that the statistical tests proposed in Section \ref{inf.sec} have the following theoretical properties concerning their sizes and powers under certain local alternatives.

\bec[Hypotheses Testing] \label{hy.thm}
Under the conditions of Theorem \ref{ci.thm1},
\begin{enumerate}
	\item(Size) for each $\ell\in\{1,2,3\}$, for any constant $0<\alpha<1$, under the null hypothesis $H_{0,\ell}$, we have $\limsup_{n,p\to\infty}P_{\theta}(|T_{\ell}|>z_{\alpha/2})\le \alpha$;
	\item(Power) for any $0<\delta<1$, there exists some $c>0$ such that, for any $|\beta^\top\Sigma\gamma-B_0|\ge cU(\beta,\gamma)n^{-1/2}$, $\liminf_{n,p\to\infty}P_{\theta}(|T_1|>z_{\alpha/2})\ge 1-\delta$; for any $|\beta^\top\Sigma\beta-Q_0|\ge c\|\beta\|_2n^{-1/2}$,  $\liminf_{n,p\to\infty}P_{\theta}(|T_2|>z_{\alpha/2})\ge 1-\delta$; and for any $|R-R_0|\ge cL^{-1}(\beta,\gamma)n^{-1/2}$, $\liminf_{n,p\to\infty}P_{\theta}(|T_{3}|>z_{\alpha/2})\ge 1-\delta$.
\end{enumerate}
\eec

\section{Simulations} \label{simu.sec}

\subsection{Evaluations with Simulated Genetic Data}

To justify our proposed methods for analyzing real genetic data sets, we carried out  numerical experiments under the settings where the covariates were simulated genotypes with possible LD structures that resembled those of the human genome, and the inferences were made at a chromosomal basis. Specifically, focusing on the Scenario I with $n_1=n_2=n$, for given choices of $p$ and $n$,  using the \texttt{R} package \texttt{sim1000G} \citep{dimitromanolakis2019sim1000g}, 
we generated genotypes of $2n$ unrelated individuals containing $p$ SNPs based on the sequencing data over a region (GrCH37: bp 40,900 to bp 2,000,000) on chromosome 9 of 503 European samples from the 1000 Genomes Project Phase 3 \citep{10002015global}, and a comprehensive haplotype map integrated over 1,184 reference individuals \citep{international2010integrating}; see Section S4 of the  Supplement for the resulting correlation matrix among the generated SNPs. The true effect sizes for the two binary traits were generated such that for each trait there were 25 associated SNPs with 12 of them shared by both traits. The effect sizes of the associated SNPs were uniformly drawn from $[-1,1]$. For reasons of practical interest, we mainly focused on the estimation, confidence intervals and hypothesis testing about the genetic correlation parameter.  The results about the genetic covariance and variance can be found in Section \ref{ci.simu} below and Section S4 of the Supplement. 

For parameter estimation,  in addition to our proposed estimators (``pro"), we also considered (i) the simple plug-in 
(``plg") estimators $\hat{\beta}^\top\hat{\Sigma}\hat{\gamma}$, $\hat{\beta}^\top\hat{\Sigma}\hat{\beta}$ and $\hat{R}_{plg}=\frac{\hat{\beta}^\top\hat{\Sigma}\hat{\gamma}}{\sqrt{\hat{\beta}^\top\hat{\Sigma}\hat{\beta}\hat{\gamma}^\top\hat{\Sigma}\hat{\gamma}}}$; (ii) the component-wise projected Lasso (``lpj")  estimators $\breve{\beta}^\top\hat{\Sigma}\breve{\gamma}$, $\breve{\beta}^\top\hat{\Sigma}\breve{\beta}$ and $\hat{R}_{lpj}=\frac{\breve{\beta}^\top\hat{\Sigma}\breve{\gamma}}{\sqrt{\breve{\beta}^\top\hat{\Sigma}\breve{\beta}\breve{\gamma}^\top\hat{\Sigma}\breve{\gamma}}}$ where each component of $\breve{\beta}$ and $\breve{\gamma}$ was the debiased Lasso estimator implemented by the function \texttt{lasso.proj} in the R package \texttt{hdi} under default setting; and (iii) the component-wise projected Ridge (``rpj") estimators  $\check{\beta}^\top\hat{\Sigma}\check{\gamma}$, $\check{\beta}^\top\hat{\Sigma}\check{\beta}$ and $\hat{R}_{rpj}=\frac{\check{\beta}^\top\hat{\Sigma}\check{\gamma}}{\sqrt{\check{\beta}^\top\hat{\Sigma}\check{\beta}\check{\gamma}^\top\hat{\Sigma}\check{\gamma}}}$ where each component of $\check{\beta}$ and $\check{\gamma}$ were obtained from the function \texttt{ridge.proj} in the R package \texttt{hdi} under the default setting. For the proposed method, we used cross-validation to determine the tuning parameter (see Section S4.1  for details).
Table \ref{table:test} contains the empirical estimation errors (square-roots of the empirical mean-squared errors) for the genetic correlation estimators, which demonstrates the superior performance of the proposed method.

For confidence intervals, we compared our proposed CIs (``pro") with an alternative bootstrap CIs. Specifically, the bootstrap CIs are based on the plg estimators calculated from 100 observations sampled from the original data set, so that the final CIs are constructed based on the empirical distributions of 500 bootstrap estimators.  Table \ref{table:t5} contains the averaged coverage probabilities and lengths of the proposed and the plg-based bootstrap CIs, denoted as ``boot," with 500 rounds of simulation for each setting. Our results suggest the desirable coverage and shorter length of the proposed CIs. 
Finally, for hypotheses testing, we evaluated the empirical type I errors and the statistical powers of both our proposed tests and the plg-based bootstrap tests under the setup where the effect sizes were generated with an additional constraint $|\beta^\top \Sigma\gamma|>3$. Table \ref{table:t6} contains the empirical type I errors and statistical powers of the proposed tests over different settings, each based on 500 rounds of simulations. Our results suggest empirical validity of the proposed test, and its advantage over the bootstrap tests. Although in Tables \ref{table:t5} and \ref{table:t6}, the proposed method became a little  conservative when $n$ increased from 200 to 400, which was likely due to the limitation of our empirically determined tuning parameter, we still observed greater power for the test and shorter lengths for the CIs with larger $n$, and in both cases the advantage over the alternative methods.
For more simulations under a slightly different setting of association structure, see Section S4.5 of the Supplement (Table S8).

\begin{table}
	\centering
	\caption{ Estimation errors for the genetic correlation under simulated genetic data with $k=25$. pro: proposed estimators; plg: simple plug-in estimators; lpj:  component-wise projected Lasso estimators; rpj:  the component-wise projected Ridge estimators.} 
		\begin{tabular}{c|cccc|cccc|cccc}
			\hline
			\multirow{2}{1em}{$p$}&\multicolumn{4}{c}{$n=200$} &\multicolumn{4}{c}{$300$} &\multicolumn{4}{c}{$400$}\\
			\cline{2-13}
			&   pro & plg & lpj &  rpj & pro & plg &  lpj &  rpj &   pro & plg & lpj &  rpj \\  
			\hline
			700& 0.09 & 0.12  & 0.15 & 0.16  &0.09 & 0.11 &0.14  &0.13  &0.08 &0.11  &0.13 &0.12\\
			800 & 0.08 & 0.10  & 0.15 & 0.14  & 0.08&0.11 & 0.15 & 0.11 &0.09  &0.11  &0.15 &0.12\\
			900&  0.09 & 0.13 &0.16  &0.15  & 0.11 & 0.12 & 0.15 & 0.13 &0.07  &0.11 &0.14  &0.11\\
			1000 & 0.10 & 0.12 & 0.14  &0.15 & 0.09 & 0.11 &  0.14 & 0.12  &0.08 &0.09  &0.14 &0.09\\
			\hline
		\end{tabular}
		\label{table:test}
	\end{table}
	
	\begin{table}
		\caption{ Coverage  and length of the CIs for the genetic correlation under simulated genetic data with $\alpha=0.05$.\label{table:t5}} 
		\centering
		\begin{tabular}{c|cccc|cccc|cccc}
			\hline
			\multirow{3}{1em}{$p$}&\multicolumn{4}{c}{$n=200$} &\multicolumn{4}{c}{$300$} &\multicolumn{4}{c}{$400$}\\
			\cline{2-13}
			& \multicolumn{2}{c|}{coverage} &\multicolumn{2}{c|}{length} &\multicolumn{2}{c|}{ coverage} &\multicolumn{2}{c|}{length} &\multicolumn{2}{c|}{coverage} &\multicolumn{2}{c}{length} \\
			\cline{2-13}
			&   pro & boot &pro & boot& pro & boot  & pro & boot&   pro & boot& pro & boot\\  
			\hline
			700 &   96.4 & 82.4 &  0.30 & 0.37 & 97.6&  85.8  & 0.26 & 0.39 & 97.0 & 82.6 &0.27& 0.41 \\
			800&  97.0 & 85.4 &  0.29 &0.37 & 98.0 & 82.5 & 0.27 & 0.39 & 98.2 & 85.2 &0.26 & 0.39 \\
			900 &  96.6 & 84.2  &  0.31 & 0.36&    96.8& 86.2 &0.26&0.38 & 97.6 & 84.0 & 0.25& 0.39\\
			1000 &  97.5 & 86.0  & 0.30 & 0.34 & 97.6& 80.0 & 0.26& 0.36 &97.8 & 84.9  &0.26 & 0.41 \\
			\hline
		\end{tabular}
	\end{table}
	
	\begin{table}
		\caption{Type I errors and powers for testing the genetic correlation under simulated genetic data with $\alpha=0.05$.	\label{table:t6}}
		\centering
		\begin{tabular}{c|cccc|cccc|cccc}
			\hline
			\multirow{3}{1em}{${p}$}&\multicolumn{4}{c}{$n=200$} &\multicolumn{4}{c}{$300$} &\multicolumn{4}{c}{$400$}\\
			\cline{2-13}
			& \multicolumn{2}{c|}{ type I error} &\multicolumn{2}{c|}{power} &\multicolumn{2}{c|}{type I error} &\multicolumn{2}{c|}{power} &\multicolumn{2}{c|}{ type I error} &\multicolumn{2}{c}{power} \\
			\cline{2-13}
			&   pro & boot &pro & boot&  pro & boot &   pro & boot & pro & boot&   pro & boot \\  
			\hline
			700&   0.04 & 0.41  & 0.47 & 0.72 &  0.04& 0.35 &0.63& 0.68& 0.02 & 0.34 &0.69& 0.65 \\
			800 &   0.04 & 0.42 &0.46& 0.74 & 0.03& 0.37  &0.59& 0.71 &0.03& 0.34 &0.70& 0.66\\
			900&  0.04& 0.42 &0.45& 0.70& 0.03&  0.35 &0.64& 0.66  & 0.02& 0.32&0.69&0.73  \\
			1000 &  0.06&  0.41 &0.42& 0.71 & 0.02&  0.36 &0.63& 0.70 & 0.02& 0.36& 0.68&0.70\\
			\hline
		\end{tabular}
	\end{table}

	\subsection{Evaluation with Model-Generated Data} \label{ci.simu}
	
	We consider the high-dimensional setting $p> n$, and set the sparsity level as $k=25$. For the true regression coefficients, given the support $\mathcal{S}$ such that $|\mathcal{S}|=k$, we generated $\beta_j$ and $\gamma_j$ uniformly from $[-1,1]$ for all $j\in \mathcal{S}$. For the design covariates, we focused on Scenario I, where $n_1=n_2=n$ and the covariates were generated from a multivariate Gaussian distribution with covariance matrix as either $\Sigma=\Sigma_B$, where $\Sigma_B$ is a $p\times p$ blockwise diagonal matrix of $10$ identical  unit diagonal Toeplitz matrices whose off-diagonal entries descend from 0.3 to 0 (see Section S4.1 of the Supplement for its explicit form), or $\Sigma=\Sigma_E$ where $\Sigma_E$ is an exchangeable covariance matrix with unit diagonals and off-diagonals being 0.2. The numerical result on each setting was based on 500 rounds of simulations.
	
	For parameter estimation, we evaluated the proposed method and the three alternative methods defined in the previous section. The results, due to space limit, were put in Section S4.2 of the Supplement (Tables S1, S2), which demonstrated the superiority of each of the proposed estimators over the alternatives. 
	Under the same simulation setups, we evaluated and compared different method for constructing $95\%$ CIs for the parameters.  Specifically, we compared our proposed CIs (``pro") with two alternative bootstrap CIs, 
	based on 500 plg estimators or rpj estimators calculated from 100 observations sampled from the original data set.  Table \ref{table:t2} contains the averaged coverage probabilities and lengths of the proposed and the plg-based bootstrap CIs  (``boot") under the blockwise diagonal covariant matrix. For reason of space, the results under the exchangeable covariance, and the results for the rpj-based bootstrap CIs, {whose coverage were in general poorer than the plg-based CIs for $\beta^\top\Sigma\gamma$ and $\beta^\top\Sigma\beta$, and only slightly better than the plg-based CIs for $R$,} are delayed to Section S4.3 of the Supplement (Tables S3-S5). In general our proposed CIs achieve the $95\%$ nominal confidence levels whereas the bootstrap CIs are off-target or biased. In particular, for the genetic correlation $R$, the proposed CI has better coverage and smaller length. In addition, our proposed methods were computationally more efficient than the bootstrap CIs as the averaged running time (MacBook Pro with 2.2 GHz 6-Core Intel Core i7) for the proposed CIs is only about 1 second whereas the bootstrap CIs takes more than 1.6 mins for the plg-based CIs and 1 hour for the rpj-based CIs on average. When the sample size increased from 300 to 500, the empirical coverage of the proposed CIs for $\beta^\top\Sigma\gamma$ and $R$ seemed to inflate slightly, which again was likely due to our empirically determined tuning parameter. Nevertheless, the proposed CIs had shorter length for larger $n$, and its advantage over the alternative method was notable.
	
	For hypothesis testing, we also compared the empirical type I errors and statistical powers of our proposed tests and the plg-based bootstrap tests, demonstrating the empirical superiority of the proposed method. For reason of space, we relegate our simulation results  to Section S4.4 (Tables S6 and S7) of the Supplement.

	\begin{table}
		\centering
		\caption{ Coverage and length of the CIs with $\Sigma=\Sigma_B$, $\alpha=0$.$05$ and sparsity $k=25$. pro: proposed estimators; boot: the plg-based bootstrap confidence intervals.} 	
		\begin{tabular}{c|cccc|cccc|cccc}
			\hline
			\multirow{3}{1em}{$p$}&\multicolumn{4}{c|}{$\beta^\top\Sigma\gamma$} &\multicolumn{4}{c|}{$\beta^\top\Sigma\beta$} &\multicolumn{4}{c}{$R$}\\
			\cline{2-13}
			& \multicolumn{2}{c|}{ pro} &\multicolumn{2}{c|}{boot} &\multicolumn{2}{c|}{ pro} &\multicolumn{2}{c|}{boot} &\multicolumn{2}{c|}{ pro} &\multicolumn{2}{c}{boot} \\
			\cline{2-13}
			&   cov & len &cov& len&  cov& len  &   cov& len &   cov& len&   cov& len \\  
			\hline
			&\multicolumn{12}{c}{$n=300$}\\
			700 &   94.8 &  6.24 &46.4& 2.05& 94.4&  7.61 &13.5& 2.42  & 96.6 & 0.35 &76.0&0.37\\
			800 &  97.4 &  7.72 & 47.8& 1.91& 92.4&  7.89 &13.2& 2.30& 95.0 & 0.37& 76.4&0.36  \\
			900 &  93.6& 5.59& 50.2 & 1.85& 93.8&  6.71  & 14.6& 2.27& 96.4 & 0.34 & 73.6&0.35\\
			1000 &  93.2 & 5.85& 42.6& 1.93& 92.6& 7.88 &7.2& 2.39 & 93.0 & 0.32  & 76.4& 0.36\\
			&\multicolumn{12}{c}{$n=400$}\\
			700&   96.0 & 6.11& 56.6 & 2.30 & 92.0 &  7.85  &30.0 & 2.96  & 96.6 & 0.32& 76.6& 0.37\\
			800&  97.4 &  5.91 & 55.4& 2.20 & 92.4& 7.47  &22.8& 2.63  & 96.2& 0.32 &74.4& 0.37 \\
			900 &  96.6&  5.81& 51.0 & 2.19 &90.6 &  7.32   & 21.6& 2.69 & 96.6& 0.31& 73.0 & 0.37 \\
			1000 &  93.8 & 5.65& 47.8& 2.07 &90.4&  7.11  &19.8 & 2.58  & 93.4& 0.31& 72.6& 0.36 \\
			&\multicolumn{12}{c}{$n=500$}\\
			700 &   99.0 & 5.71 &61.0& 2.40 & 95.2 & 6.93   &43.2 & 2.92 & 98.6 & 0.30& 73.4& 0.37\\
			800&  98.6 &  5.70& 60.6& 2.38 & 93.4& 7.07 & 41.2& 2.83 & 97.2& 0.29 & 78.0& 0.37 \\
			900 &  99.2&  5.92& 58.0& 2.32 & 92.6 &  7.36  & 31.2& 2.88 & 98.4& 0.30 & 76.6& 0.36\\
			1000 &  98.6 & 5.44 &57.8& 2.18 & 90.4&  6.70 &30.0& 2.73  & 98.2& 0.29 &76.6&0.36 \\
			\hline
		\end{tabular}
		\label{table:t2}
	\end{table}

	\section{Analysis of Ten Pediatric Autoimmune Diseases} \label{real.sec}
	
	We  investigate the genetic correlations between each pair of ten pediatric autoimmune diseases, including autoimmune thyroiditis (THY), psoriasis (PSOR), juvenile idiopathic arthritis (JIA), ankylosing spondylitis (AS), common variable immunodeficiency (CVID), celiac disease (CEL), Crohn's disease (CD), ulcerative colitis (UC),  type 1 diabetes (T1D) and systemic lupus erythematosus (SLE). The diseased  subjects and controls were identified either directly from previous studies or from de-identified samples and associated electronic medical records in the genomics biorepository at The Children's Hospital of Philadelphia \citep{li2015meta}.  The data set includes  10,718 normal controls, 97 THY cases, 107 AS cases, 100 PSOR cases, 173 CEL cases, 254 SLE cases, 308 CVID cases, 865 UC cases, 1086 T1D cases, 1123 JIA cases, and 1922 CD cases. Specifically, for each pair of the ten diseases, we evaluated their chromosome-specific genetic relatedness by estimating and performing hypotheses testing about the genetic correlation parameter on each of the 22 autosomes. By focusing on the chromosome-specific genetic correlations, we are able to make better inference with limited sample sizes for many diseases, and to obtain insights on the genomic regions that relate the two diseases of interest.
	
	For each subject, after removing the SNPs with minor allele frequency less than 0.05, a total  of 475,324 SNPs were obtained across 22 autosomes (see Supplement for details).  To apply our proposed methods, for each pair of diseases, we randomly split the controls into two groups of equal size, combined them with each of the cases and fitted two high-dimensional logistic regressions between the disease outcomes and the SNPs to obtain the initial logistic Lasso estimators for each disease. Then the bias-corrected estimators were obtained, where the sample covariance matrix were calculated based on all the samples. Moreover, using our  proposed method, we tested the individual null hypothesis that the chromosome-specific genetic correlation is zero between each pair of diseases in order to identify i) the diseases  that are genetically associated and ii) the specific chromosome where the diseases have shared genetic architecture. 
	
\begin{figure}
		\centering
		\includegraphics[angle=0,width=13cm]{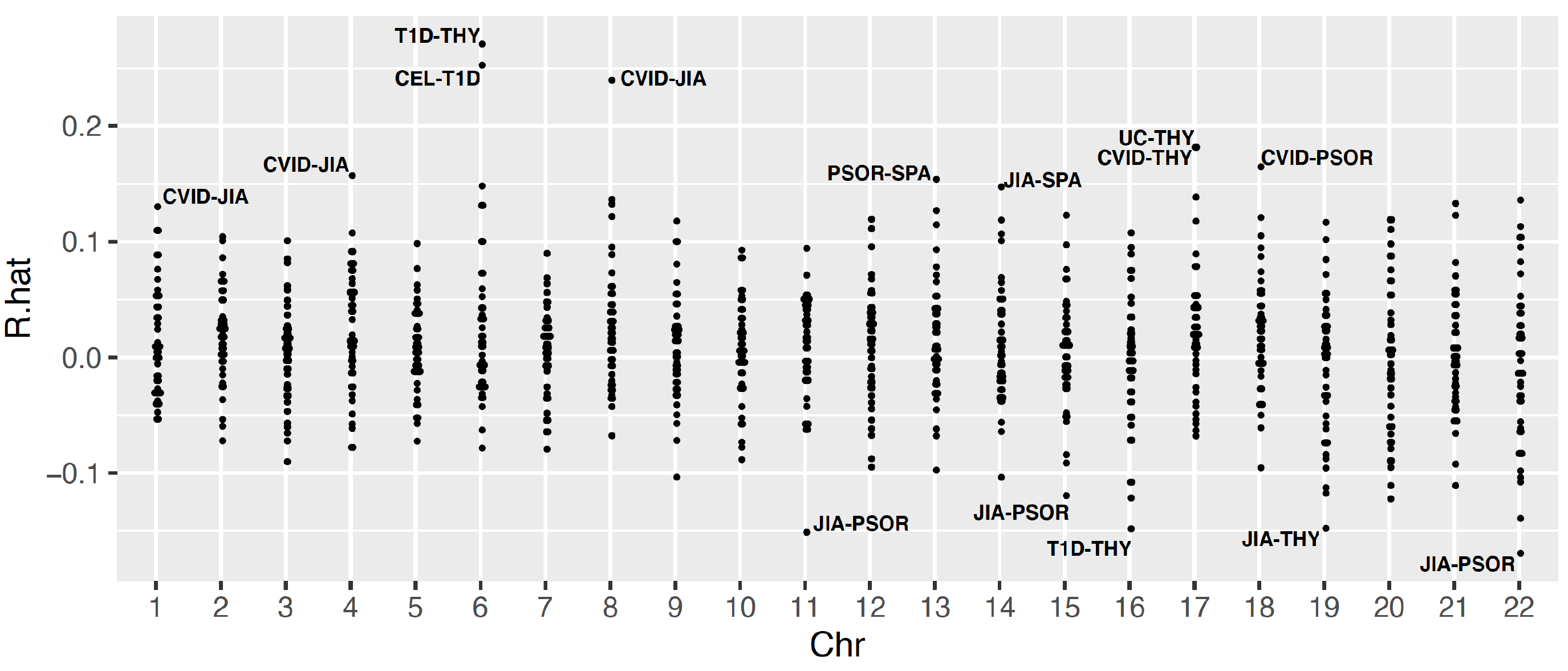}
		\includegraphics[angle=0,width=13cm]{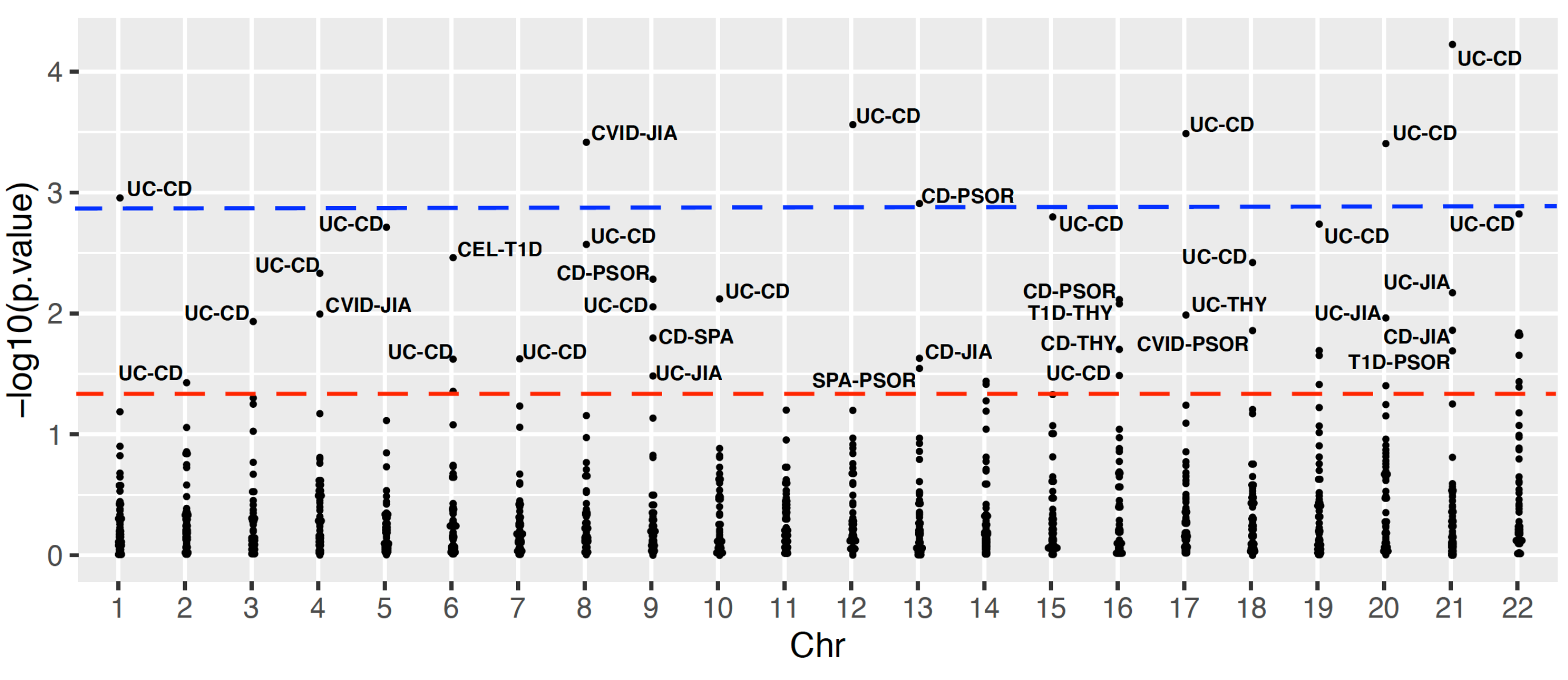}
		\caption{Analysis of genetic sharing between 10 autoimmune diseases. Top panel: estimated genetic correlations between each pair of diseases on each autosome. Bottom panel: negative log-transformed p-values for each pair of diseases, based on proposed method. 
			The red and blue dashed lines represent the original and Bonferroni-adjusted significance level at 0.05.} 
		\label{real.fig}
	\end{figure}

	The results are summarized in Figure \ref{real.fig}. The top panel shows the estimated chromosome-specific genetic correlations between each pair of diseases, where the disease pairs having larger absolute values were annotated. The bottom panel shows the negative log-transformed p-values for each pair of diseases.  Our  tests suggest strong genetic sharing between UC and CD on chromosomes 1, 12,  17, 20 and 21, CVID and JIA on chromosome 8, and CD and PSOR on chromosome 13. Many pairs of these diseases showed genetic relatedness at the nominal p-value of 0.05, however, due to small sample sizes, they did not reach the statistical significance after the Bonferroni adjustment of multiple comparisons. Note that the pairs UC and CD, and  CVID and JIA were also found to be statistically significant by \cite{li2015meta} using different measures of genetic sharing, yet our proposed methods were able to additionally locate the genetic sharing to specific chromosomes and provide theoretically valid uncertainty quantifications.

	%
	%

	\section{Discussion} \label{diss.sec}

	In this paper, a statistical inference framework for studying the genetic relatedness between two  binary traits was introduced under  the high-dimensional logistic regression models. Our model allows the number of SNPs to far exceed the sample sizes while producing efficient and valid statistical inference under mild conditions on  sparsity and effect size of the true associations, and the covariance structure or linkage disequilibrium of the variants. 
	Many efforts have been made to improve the speed of optimization and operation for genome-scale and ultrahigh-dimensional data sets. For example, in \cite{qian2019fast}, a new computational framework was proposed so that scalable Lasso solutions can be obtained for very large Biobank data set involving about 300,000 individuals and 800,000 genetic  variants. We expect that these new computational methods will increase the utility of the proposed methods in genetic correlation analysis at whole genome sequencing scale.

\bibliographystyle{chicago}
\bibliography{reference}

\newpage

	\title{Supplement to ``Statistical Inference for Genetic Relatedness Based on High-Dimensional Logistic Regression"}
\author{Rong Ma$^1$, Zijian Guo$^2$, T. Tony Cai$^3$ and Hongzhe Li$^3$\\
	\\
	Stanford University$^1$\\
	Rutgers University$^2$\\
	University of Pennsylvania$^3$}
\date{}
\maketitle
\thispagestyle{empty}

\setcounter{section}{0}
\setcounter{equation}{0}
\def\theequation{S\arabic{section}.\arabic{equation}}
\def\thesection{S\arabic{section}}
\def\thefigure{S\arabic{section}.\arabic{figure}}
\def\thetable{S\arabic{section}.\arabic{table}}

\section{Proofs of Upper Bound Results}

\subsection{Proof of Theorem 4}

The proof is separated into two parts, with the first part concerning  $\beta^\top\Sigma\gamma$ and the second part concerning  $\beta^\top\Sigma\beta$. We first state a useful lemma about the estimation bounds for the logistic Lasso estimators.

\bel[\citealt{cai2020statistical}] \label{lasso.bnd.lem}
Under the conditions of Theorem 1, with probability at least $1-p^{-c}$,  we have $		\|\widehat{\beta}-\beta\|_2\lesssim \sqrt{\frac{k\log p}{n}}$, $\|\widehat{\beta}-\beta\|_1\lesssim k\sqrt{\frac{\log p}{n}}$ and $\frac{1}{n}\|X(\widehat{\beta}-\beta)\|_2^2\lesssim \frac{k\log p}{n}$.
\eel

\paragraph{Part I.} The following decomposition holds
\begin{align*}
	\widehat{\beta}^\top\widehat{\Sigma}\widehat{\gamma}-\beta^\top\Sigma\gamma&=(\widehat{\beta}-\beta)^\top\Sigma\gamma-\widehat{\beta}^\top\Sigma\gamma+	\widehat{\beta}^\top\widehat{\Sigma}\widehat{\gamma}\\
	&=-(\widehat{\beta}-\beta)^\top\Sigma(\widehat{\gamma}-\gamma)+(\widehat{\beta}-\beta)^\top\Sigma\widehat{\gamma}-\widehat{\beta}^\top\Sigma\gamma+\widehat{\beta}^\top\widehat{\Sigma}\widehat{\gamma}\\
	&=-(\widehat{\beta}-\beta)^\top\Sigma(\widehat{\gamma}-\gamma)+(\widehat{\beta}-\beta)^\top\Sigma\widehat{\gamma}+\widehat{\beta}^\top\Sigma(\widehat{\gamma}-\gamma)-\widehat{\beta}^\top\Sigma\widehat{\gamma}+\widehat{\beta}^\top\widehat{\Sigma}\widehat{\gamma}\\
	&=-(\widehat{\beta}-\beta)^\top\Sigma(\widehat{\gamma}-\gamma)+(\widehat{\beta}-\beta)^\top\Sigma\widehat{\gamma}+\widehat{\beta}^\top\Sigma(\widehat{\gamma}-\gamma)+\widehat{\beta}^\top(\widehat{\Sigma}-\Sigma)\widehat{\gamma}.
\end{align*}
Although in our high-dimensional setting, the sample covariance matrix $\widehat\Sigma$ is no longer  consistent under the operator norm \citep{vershynin2010introduction}, what we needed in our analysis was the entrywise unbiasedness of $\widehat\Sigma$ and therefore the asymptotic normality of the quadratic form $\widehat\beta^\top (\widehat\Sigma-\Sigma)\widehat\gamma$ conditional on $(\widehat\beta,\widehat\gamma)$, established below. In particular,  the error due to estimating $\Sigma$ by $\widehat\Sigma$ contributed to a portion of uncertainty in the proposed debiased estimators, which was reflected in their associated asymptotic variance (for example, the last term $\E \big\{ \widehat{\beta}^\top (X_{i\cdot}X_{i\cdot}^\top -\Sigma)\widehat{\gamma}\big\}^2$ in $v^2$ defined in Section 3). 
Define $h(X_{i\cdot},\alpha,\beta)=\frac{\exp\left(\alpha+X_{i\cdot}^{\top}\beta\right)}{1+\exp\left(\alpha+X_{i\cdot}^{\top}\beta\right)}$. Since $\dot{h}(X_{i\cdot},\alpha,\beta)=\frac{\partial h(X_{i\cdot},\alpha,\beta)}{\partial \beta'}=\frac{\exp\left(\alpha+X_{i\cdot}^{\top}\beta\right)}{\left(1+\exp\left(\alpha+X_{i\cdot}^{\top}\beta\right)\right)^2}X'_{i\cdot}$ where $X'_{i\cdot}=(1,X_{i\cdot}^\top)^\top$ and $\beta'=(\alpha,\beta^\top)^\top$, by Taylor's expansion, 
$
h(X_{i\cdot},\alpha,{\beta})-h(X_{i\cdot},\widehat{\alpha},\widehat{\beta})=-\frac{\exp\left(\widehat{\alpha}+X_{i\cdot}^{\top}\widehat{\beta}\right)}{\left(1+\exp\left(\widehat{\alpha}+X_{i\cdot}^{\top}\beta\right)\right)^2}X^{\top}_{i\cdot}\left(\widehat{\beta}-\beta\right)-\frac{\exp\left(\widehat{\alpha}+X_{i\cdot}^{\top}\widehat{\beta}\right)}{\left(1+\exp\left(\widehat{\alpha}+X_{i\cdot}^{\top}\beta\right)\right)^2}\left(\widehat{\alpha}-\alpha\right)+\Delta_i,
$
where 
$
\Delta_i=\ddot{h}({X'_{i\cdot}}^\top[t\beta'+(1-t)\widehat{\beta'}])[{X'_{i\cdot}}^\top(\widehat{\beta}'-\beta')]^2,
$
for some $t\in(0,1)$ and $\widehat{\beta}'=(\widehat{\alpha},\widehat{\beta}^\top)^\top$.
Then, if we define $\epsilon_i=y_i-h(X_{i\cdot},\alpha,\beta)$, we have
\begin{equation}
	\begin{aligned}
		&\bigg(\frac{\exp(\widehat{\alpha}+X_{i\cdot}^{\top}\widehat{\beta})}{1+\exp(\widehat{\alpha}+X_{i\cdot}^{\top}\widehat{\beta})}-y_i\bigg)X_{i\cdot}\\
		&=\bigg(\frac{\exp(\widehat{\alpha}+X_{i\cdot}^{\top}\widehat{\beta})}{(1+\exp(\widehat{\alpha}+X_{i\cdot}^{\top}\widehat{\beta}))^2}X^{\top}_{i\cdot}(\widehat{\beta}-\beta)+\frac{\exp(\widehat{\alpha}+X_{i\cdot}^{\top}\widehat{\beta})}{(1+\exp(\widehat{\alpha}+X_{i\cdot}^{\top}\widehat{\beta}))^2}(\widehat{\alpha}-\alpha)+\Delta_i+\epsilon_i\bigg)X_{i\cdot}\\
		&=\bigg(\frac{\exp(\widehat{\alpha}+X_{i\cdot}^{\top}\widehat{\beta})}{(1+\exp(\widehat{\alpha}+X_{i\cdot}^{\top}\widehat{\beta}))^2}X_{i\cdot}X^{\top}_{i\cdot}\bigg)(\widehat{\beta}-\beta)+(\Delta_i+\epsilon_i)X_{i\cdot}+\frac{\exp(\widehat{\alpha}+X_{i\cdot}^{\top}\widehat{\beta})}{(1+\exp(\widehat{\alpha}+X_{i\cdot}^{\top}\widehat{\beta}))^2}(\widehat{\alpha}-\alpha)X_{i\cdot}.
	\end{aligned}
\end{equation}
By re-scaling each item $\big(\frac{\exp\left(\widehat{\alpha}+X_{i\cdot}^{\top}\widehat{\beta}\right)}{1+\exp\left(\widehat{\alpha}+X_{i\cdot}^{\top}\widehat{\beta}\right)}-y_i\big)X_{i\cdot}$, we have 
\begin{equation}
	\begin{aligned}
		&\sum_{i=1}^{n_1}\bigg(\frac{\exp(\widehat{\alpha}+X_{i\cdot}^{\top}\widehat{\beta})}{(1+\exp(\widehat{\alpha}+X_{i\cdot}^{\top}\widehat{\beta}))^2}\bigg)^{-1}\bigg(\frac{\exp(\widehat{\alpha}+X_{i\cdot}^{\top}\widehat{\beta})}{1+\exp(\widehat{\alpha}+X_{i\cdot}^{\top}\widehat{\beta})}-y_i\bigg)X_{i\cdot}\\
		&=\bigg(\sum_{i=1}^{n_1}X_{i\cdot}X^{\top}_{i\cdot}\bigg)(\widehat{\beta}-\beta)+\sum_{i=1}^{n_1}\bigg(\frac{\exp(\widehat{\alpha}+X_{i\cdot}^{\top}\widehat{\beta})}{(1+\exp(\widehat{\alpha}+X_{i\cdot}^{\top}\widehat{\beta}))^2}\bigg)^{-1}(\Delta_i+\epsilon_i)X_{i\cdot}+(\widehat{\alpha}-\alpha)\sum_{i=1}^{n_1}X_{i\cdot}
	\end{aligned}.
\end{equation}
Similarly, we have
\begin{equation}
	\begin{aligned}
		&\sum_{i=1}^{n_2}\bigg(\frac{\exp(\widehat{\zeta}+Z_{i\cdot}^{\top}\widehat{\gamma})}{(1+\exp(\widehat{\zeta}+Z_{i\cdot}^{\top}\widehat{\gamma}))^2}\bigg)^{-1}\bigg(\frac{\exp(\widehat{\zeta}+Z_{i\cdot}^{\top}\widehat{\gamma})}{1+\exp(\widehat{\zeta}+Z_{i\cdot}^{\top}\widehat{\gamma})}-w_i\bigg)Z_{i\cdot}\\
		&=\bigg(\sum_{i=1}^{n_2}Z_{i\cdot}Z^{\top}_{i\cdot}\bigg)\left(\widehat{\gamma}-\gamma\right)+\sum_{i=1}^{n_2}\bigg(\frac{\exp(\widehat{\zeta}+Z_{i\cdot}^{\top}\widehat{\gamma})}{(1+\exp(\widehat{\zeta}+Z_{i\cdot}^{\top}\widehat{\gamma}))^2}\bigg)^{-1}\left(\Lambda_i+\delta_i\right)Z_{i\cdot}+(\widehat{\zeta}-\zeta)\sum_{i=1}^{n_2}Z_{i\cdot},
	\end{aligned}
\end{equation}
where
\begin{equation}
	\Lambda_i=\ddot{h}({X'_{i\cdot}}^\top[t\gamma'+(1-t)\hat{\gamma}'])[{Z'_{i\cdot}}^\top(\hat{\gamma}'-\gamma')]^2,\;\; \text{and} \;\; \delta_i=w_i-\frac{\exp\left(\zeta+Z_{i\cdot}^{\top}\gamma\right)}{1+\exp\left(\zeta+Z_{i\cdot}^{\top}\gamma\right)},
\end{equation}
with $\gamma'=(\zeta,\gamma^\top)^\top$ and $\widehat{\gamma}'=(\widehat{\zeta},\widehat{\gamma}^\top)^\top.$
Then, if we denote $\Sigma_X = n_1^{-1}\sum_{i=1}^{n_1}X_{i\cdot}X_{i\cdot}^\top$ and $\Sigma_Z = n_2^{-1}\sum_{i=1}^{n_2}Z_{i\cdot}Z_{i\cdot}^\top$, the error of the proposed estimator $\widehat{\beta^{\top}\Sigma\gamma}-{\beta^{\top}\Sigma\gamma}$ can be decomposed as follows,
\begin{align*}
	& \frac{1}{n_1}\widehat{\gamma}^\top\sum_{i=1}^{n_1}\bigg(\frac{\exp(\widehat{\alpha}+X_{i\cdot}^{\top}\widehat{\beta})}{(1+\exp(\widehat{\alpha}+X_{i\cdot}^{\top}\widehat{\beta}))^2}\bigg)^{-1}\epsilon_iX_{i\cdot}+\frac{1}{n_2} \widehat{\beta}^\top\sum_{i=1}^{n_2}\bigg(\frac{\exp(\widehat{\zeta}+Z_{i\cdot}^{\top}\widehat{\gamma})}{(1+\exp(\widehat{\zeta}+Z_{i\cdot}^{\top}\widehat{\gamma}))^2}\bigg)^{-1}\delta_i Z_{i\cdot}+\widehat{\beta}^{\top}\left(\widehat{\Sigma}-\Sigma\right)\widehat{\gamma}\\
	& \quad +\frac{1}{n_1}\widehat{\gamma}^\top\sum_{i=1}^{n_1}\bigg(\frac{\exp(\widehat{\alpha}+X_{i\cdot}^{\top}\widehat{\beta})}{(1+\exp(\widehat{\alpha}+X_{i\cdot}^{\top}\widehat{\beta}))^2}\bigg)^{-1}\Delta_iX_{i\cdot}+\frac{1}{n_2}\widehat{\beta}^{\top}\sum_{i=1}^{n_2}\bigg(\frac{\exp(\widehat{\zeta}+Z_{i\cdot}^{\top}\widehat{\gamma})}{(1+\exp(\widehat{\zeta}+Z_{i\cdot}^{\top}\widehat{\gamma}))^2}\bigg)^{-1}\Lambda_i Z_{i\cdot}\\
	&\quad -(\widehat{\beta}-\beta)^{\top}{\Sigma}\left(\widehat{\gamma}-\gamma\right)+(\widehat{\beta}-\beta)^\top(\Sigma-\widehat{\Sigma}_X)\widehat{\gamma}+\widehat{\beta}^\top(\Sigma-\widehat{\Sigma}_Z)(\widehat{\gamma}-\gamma)\\
	&\quad-\frac{1}{n_1}(\widehat{\alpha}-\alpha)\sum_{i=1}^{n_1}\widehat{\gamma}^\top X_{i\cdot}-\frac{1}{n_2}(\widehat{\zeta}-\zeta)\sum_{i=1}^{n_2}\widehat{\beta}^\top Z_{i\cdot} 
\end{align*}
Now we denote
\begin{align}
	T_1 &= \frac{1}{n_1}\widehat{\gamma}^\top\sum_{i=1}^{n_1}\frac{(1+\exp({X'_{i\cdot}}^{\top}\widehat{\beta}'))^2}{\exp({X'_{i\cdot}}^{\top}\widehat{\beta}')}\epsilon_iX_{i\cdot}+\frac{1}{n_2} \widehat{\beta}^\top\sum_{i=1}^{n_2}\frac{(1+\exp({Z'_{i\cdot}}^{\top}\widehat{\gamma}'))^2}{\exp({Z'_{i\cdot}}^{\top}\widehat{\gamma}')}\delta_i Z_{i\cdot}+\widehat{\beta}^{\top}(\widehat{\Sigma}-\Sigma)\widehat{\gamma}\label{t1},\\
	T_2 &=\frac{1}{n_1} \widehat{\gamma}^\top\sum_{i=1}^{n_1}\frac{(1+\exp({X'_{i\cdot}}^{\top}\widehat{\beta}'))^2}{\exp({X'_{i\cdot}}^{\top}\widehat{\beta}')}\Delta_iX_{i\cdot}+\frac{1}{n_2}\widehat{\beta}^{\top}\sum_{i=1}^{n_2}\frac{(1+\exp({Z'_{i\cdot}}^{\top}\widehat{\gamma}'))^2}{\exp({Z'_{i\cdot}}^{\top}\widehat{\gamma}')}\Lambda_i Z_{i\cdot},\\
	T_3&= -(\widehat{\beta}-\beta)^{\top}{\Sigma}\left(\widehat{\gamma}-\gamma\right)+(\widehat{\beta}-\beta)^\top(\Sigma-\widehat{\Sigma}_X)\widehat{\gamma}+\widehat{\beta}^\top(\Sigma-\widehat{\Sigma}_Z)(\widehat{\gamma}-\gamma), \\
	T_4&=\frac{1}{n_1}(\widehat{\alpha}-\alpha)\sum_{i=1}^{n_1}\widehat{\gamma}^\top X_{i\cdot}+\frac{1}{n_2}(\widehat{\zeta}-\zeta)\sum_{i=1}^{n_2}\widehat{\beta}^\top Z_{i\cdot} \label{t4},
\end{align}
so that $\widehat{\beta^{\top}\Sigma\gamma}-{\beta^{\top}\Sigma\gamma}=T_1+T_2+T_3+T_4$.
In the following, we will analyze the above four terms individually.

\underline{Asymptotic normality of $T_1$.} Let $T_1=N_1+N_2$ where $N_1=\frac{1}{n_1}\sum_{i=1}^{n_1}\xi_i$ and $N_2=\frac{1}{n_2}\sum_{i=1}^{n_2}\mu_i$, such that
\[
\xi_i=\frac{(1+\exp({X'_{i\cdot}}^{\top}\widehat{\beta}'))^2}{\exp({X'_{i\cdot}}^{\top}\widehat{\beta}')}\epsilon_i \widehat{\gamma}^\top X_{i\cdot}+\frac{n_1}{n_1+n_2}\widehat{\beta}^\top (X_{i\cdot}X_{i\cdot}^\top -\Sigma)\widehat{\gamma},
\]
\[
\mu_i=\frac{(1+\exp({Z'_{i\cdot}}^{\top}\widehat{\gamma}'))^2}{\exp({Z'_{i\cdot}}^{\top}\widehat{\gamma}')}\delta_i \widehat{\beta}^\top Z_{i\cdot}+\frac{n_2}{n_1+n_2}\widehat{\beta}^\top (Z_{i\cdot}Z_{i\cdot}^\top -\Sigma)\widehat{\gamma}.
\]
Note that, conditional on $\mathcal{D}_1$, $\{\xi_i\}_{i=1}^{n_1}$ and $\{\mu_i\}_{i=1}^{n_2}$ are mutually independent. In the following, we will show that, there exists some $v_1,v_2>0$ such that, 
\beq \label{asym.norm}
\frac{\sqrt{n_1}N_1}{v_1}\bigg|\mathcal{D}_1\to_d N(0,1),\qquad 	\frac{\sqrt{n_2}N_2}{v_2}\bigg|\mathcal{D}_1\to_d N(0,1).
\eeq
Then, by the independence of $N_1$ and $N_2$, we have
\beq \label{clt.eq}
\frac{\sqrt{n_1+n_2}T_1}{v}\bigg|\mathcal{D}_1\to_d N(0,1),
\eeq
where
$
v^2=\frac{n_1+n_2}{n_1}v_1^2+\frac{n_1+n_2}{n_2}v_2^2.
$
The following arguments are all conditional on $\mathcal{D}_1$.
By definition, we have $\E \xi_i=0$, and by the fact that $\E [\epsilon_i|X_{i\cdot}]=0$, we have
$
\E_{\mathcal{D}_2}\xi_i^2=\E_{\mathcal{D}_2}\eta_i(X,\widehat{\alpha},\widehat{\beta})(\widehat{\gamma}^\top X_{i\cdot})^2+\frac{n_1^2}{(n_1+n_2)^2}\E_{\mathcal{D}_2} [ \widehat{\beta}^\top (X_{i\cdot}X_{i\cdot}^\top -\Sigma)\widehat{\gamma}]^2
$
where $\eta_i(X,\widehat{\alpha},\widehat{\beta})=\frac{(1+\exp(\widehat{\alpha}+X_{i\cdot}^{\top}\widehat{\beta}))^4\exp({\alpha}+X_{i\cdot}^{\top}{\beta})}{(1+\exp(\alpha+X_{i\cdot}^{\top}{\beta}))^2\exp(2\widehat{\alpha}+2X_{i\cdot}^{\top}\widehat{\beta})}$ and the expectation $\E_{\mathcal{D}_2}$ is with respect to the data set $\mathcal{D}_2$. Let $v^2_1=\E_{\mathcal{D}_2}\xi_i^2$. Since $\xi_i/v_1$ are $i.i.d.$ random variables with mean zero and variance 1, the asymptotic normality of $\sqrt{n_1}N_1/v_1$ follows from the classical CLT. Similarly, we can obtain asymptotic normality of $\sqrt{n_2}N_2/v_2$ where $v_2^2=\E_{\mathcal{D}_2}\mu_i^2$. This completes the proof of (\ref{asym.norm}) and hence (\ref{clt.eq}).

The following lemma concerns the variance components.

\bel \label{v.bnd.lem}
Under the conditions of Theorem 4.1, with probability at least $1-p^{-c}$, it holds that $\E_{\mathcal{D}_2} \eta_i(X,\widehat{\alpha},\widehat{\beta})(\widehat{\gamma}^\top X_{i\cdot})^2\asymp \|\gamma\|_2^2$, $\E_{\mathcal{D}_2} \eta_i(Z,\widehat{\eta},\widehat{\gamma})(\widehat{\beta}^\top X_{i\cdot})^2\asymp \|\beta\|_2^2$ and $\E_{\mathcal{D}_2}[ \widehat{\beta}^\top(X_{i\cdot}X_{i\cdot}^\top-\Sigma)\widehat{\gamma}]^2\asymp \|\beta\|_2^2\|\gamma\|_2^2$.
\eel

The lemma implies that
$
v^2\asymp \|\gamma\|_2^2+\|\beta\|_2^2+\|\beta\|_2^2\|\gamma\|_2^2.
$

\underline{Upper bound of $T_2$.} By Cauchy-Schwartz inequality, it holds that
\[
\bigg|\frac{1}{n_1} \widehat{\gamma}^\top\sum_{i=1}^{n_1}\frac{(1+\exp({X'_{i\cdot}}^{\top}\widehat{\beta}'))^2}{\exp({X'_{i\cdot}}^{\top}\widehat{\beta}')}\Delta_iX_{i\cdot}\bigg|\le \max_{1\le i\le n_1}|\widehat{\gamma}^\top X_{i\cdot}|\cdot \frac{1}{n_1}\sum_{i=1}^{n_1}\bigg|\frac{(1+\exp({X'_{i\cdot}}^{\top}\widehat{\beta}'))^2}{\exp({X'_{i\cdot}}^{\top}\widehat{\beta}')}\Delta_i\bigg|.
\]
On the one hand, since by Lemma \ref{lasso.bnd.lem}, with probability at least $1-p^{-c'}$,
\beq \label{gamma.hat}
\|\widehat{\gamma}\|_2\le \|\gamma-\widehat{\gamma}\|_2+\|\gamma\|_2\le \sqrt{\frac{k\log p}{n}}+\|\gamma\|_2,
\eeq
for $k\lesssim \frac{n}{\log p}$. Then by the sub-Gausssian property, it holds that
\[
P\bigg(\max_{1\le i\le n_1}|\widehat{\gamma}^\top X_{i\cdot}|\lesssim \bigg(\|\gamma\|_2+\sqrt{\frac{k\log p}{n}}\bigg)\sqrt{\log n}\bigg)\ge 1-n^{-c}-p^{-c}.
\]
On the other hand, 
\begin{align*}
	&\frac{1}{n_1}\sum_{i=1}^{n_1}\bigg|\frac{(1+\exp({X'_{i\cdot}}^{\top}\widehat{\beta}'))^2}{\exp({X'_{i\cdot}}^{\top}\widehat{\beta}')}\Delta_i\bigg|\\
	&\le \frac{1}{n_1}\sum_{i=1}^{n_1}\frac{(1+\exp({X'_{i\cdot}}^{\top}\widehat{\beta}'))^2h''({X'_{i\cdot}}^\top[t\beta+(1-t)\widehat{\beta}])}{\exp({X'_{i\cdot}}^{\top}\widehat{\beta}')} [{X'_{i\cdot}}^\top(\widehat{\beta}'-\beta')]^2\\
	&=\frac{1}{n_1}\sum_{i=1}^{n_1}\frac{(1+\exp({X'_{i\cdot}}^{\top}\widehat{\beta}'))^2[\exp({X'_{i\cdot}}^\top(t\beta'+(1-t)\widehat{\beta}'))-1]\exp({X'_{i\cdot}}^\top(t\beta'+(1-t)\widehat{\beta}'))}{\exp({X'_{i\cdot}}^{\top}\widehat{\beta}')[\exp({X'_{i\cdot}}^\top(t\beta'+(1-t)\widehat{\beta}'))+1]^3} [{X'_{i\cdot}}^\top(\widehat{\beta}'-\beta')]^2.
\end{align*}
Consider the function
$
g(x,w)=\frac{(1+e^x)^2 (e^{x+w}-1)e^{x+w}}{e^x(1+e^{x+w})^3},$ where  $x,w\in \R.
$
It can be checked that there exists some constant $T>0$ such that
\[
\frac{\partial}{\partial x}\bigg[\frac{(e^{x}-1)e^{x}}{(1+e^{x})^3}\bigg]<0,\quad \text{for all $|x|\ge T$.}
\]
For some fixed constant $c>2T$, we consider the following cases:
\begin{enumerate}
	\item If $|x|<c$ and $|w|<c/2$, then it follows directly that $g(x,w)<C$;
	\item If $|x|>c$ and $|w|<c/2$, then we have $|x+w|\ge |x|-|w|\ge |x|-c/2$ and
	$
	g(x,w)\le 4e^{|x|}\frac{(e^{|x+w|}-1)e^{|x+w|}}{(1+e^{|x+w|})^3}\lesssim \frac{e^{2|x|}(e^{|x|}-C_1)}{(C_2+e^{|x|})^3}.
	$
	Note that the right hand side of the above inequality satisfies
	$
	\lim_{x\to+\infty} \frac{e^{2|x|}(e^{|x|}-C_1)}{(C_2+e^{|x|})^3}=1.
	$
	It then follows that $g(x,w)<C$ at the region of interest.
\end{enumerate}
This indicates that, there exists some constant $c_0>0$ such that $g(x,w)<C$ for all $x\in \R$ and $|w|<c_0$. Now since with probability at least $1-n^{-c}-p^{-c}$,
\[
|{X'_{i\cdot}}^\top(\beta'-\widehat{\beta}')|\lesssim \|\beta'-\widehat{\beta}'\|_2\cdot\sqrt{\log n}\lesssim \sqrt{\frac{k\log p{\log n}}{n}},
\]
as long as { $k\lesssim \frac{n}{\log p {\log n}}$}, we have
\[
\frac{(1+\exp({X'_{i\cdot}}^{\top}\widehat{\beta}'))^2[\exp({X'_{i\cdot}}^\top(t\beta'+(1-t)\widehat{\beta}'))-1]\exp({X'_{i\cdot}}^\top(t\beta'+(1-t)\widehat{\beta}'))}{\exp({X'_{i\cdot}}^{\top}\widehat{\beta}')[\exp({X'_{i\cdot}}^\top(t\beta'+(1-t)\widehat{\beta}'))+1]^3}=O(1),
\]
with probability at least $1-n^{-c}-p^{-c}$. Hence, 
$
\frac{1}{n_1}\sum_{i=1}^{n_1}\bigg|\frac{(1+\exp({X'_{i\cdot}}^{\top}\widehat{\beta}'))^2}{\exp({X'_{i\cdot}}^{\top}\widehat{\beta}')}\Delta_i\bigg|\lesssim \frac{1}{n_1}\sum_{i=1}^{n_1} [{X'_{i\cdot}}^\top(\widehat{\beta}'-\beta')]^2\lesssim \frac{k\log p}{n},
$
where the last inequality follows from Lemma \ref{lasso.bnd.lem}. The proof of the second term in $T_2$ follows the same idea. In sum, we obtain
$
P\big(T_2\le C(\|\beta\|_2+\|\gamma\|_2+\sqrt{k \log p/n})\frac{k\log p\sqrt{\log n}}{n}\big)\ge 1-n^{-c}-p^{-c}.
$

\underline{Upper bound of $T_3$.} The first term can be control by 
$
|(\widehat{\beta}-\beta)^{\top}{\Sigma}\left(\widehat{\gamma}-\gamma\right)|\lesssim \|\widehat{\beta}-\beta\|_2 \|\widehat{\gamma}-\gamma\|_2.
$
It then follows that
$
|(\widehat{\beta}-\beta)^{\top}{\Sigma}\left(\widehat{\gamma}-\gamma\right)|\lesssim \frac{k\log p}{n},
$
with probability at least $1-p^{-c}$. On the other hand, 
$
(\widehat{\beta}-\beta)^\top(\Sigma-\widehat{\Sigma}_X)\widehat{\gamma}\le \|\widehat{\beta}-\beta\|_1\|(\Sigma-\widehat{\Sigma}_X)\widehat{\gamma}\|_\infty,
$
where by Lemma \ref{lasso.bnd.lem}, with probability at least $1-p^{-c}$
$
\|\widehat{\beta}-\beta\|_1\lesssim k\sqrt{\frac{\log p}{n}},
$
and by the concentration inequality of sub-exponential random variables, conditional on $\widehat{\gamma}$,
\[
P\bigg(\|(\Sigma-\widehat{\Sigma}_X)\widehat{\gamma}\|_\infty \le C\|\widehat{\gamma}\|_2\sqrt{\frac{\log p}{n}}\bigg| \widehat{\gamma}\bigg)\ge 1-p^{-c}.
\]
By (\ref{gamma.hat}), we have
$
P\big(\|(\Sigma-\widehat{\Sigma}_X)\widehat{\gamma}\|_\infty \le C_1\|{\gamma}\|_2\sqrt{\frac{\log p}{n}}+C_2\frac{k^{1/2}\log p}{n}\big)\ge 1-p^{-c},
$
so that, with probability at least $1-p^{-c}$, we have
$
(\widehat{\beta}-\beta)^\top(\Sigma-\widehat{\Sigma}_X)\widehat{\gamma}\lesssim \big(\|\gamma\|_2+\sqrt{\frac{k\log p}{n}}\big) \frac{k\log p}{n}
$
In sum, we have
\beq
P\bigg( T_3\lesssim(1+\|\beta\|_2+\|\gamma\|_2) \frac{k\log p}{n} \bigg)\ge 1-p^{-c}.
\eeq
\underline{Upper bound of $T_4$.} Since
$
\frac{1}{n}(\widehat{\alpha}-\alpha)\sum_{i=1}^n\widehat{\gamma}^\top X_{i\cdot}\le |\widehat{\alpha}-\alpha|\cdot \bigg|  \frac{1}{n}\sum_{i=1}^n\widehat{\gamma}^\top X_{i\cdot}\bigg|,
$
conditional on $\widehat{\beta}'$ and $\widehat{\gamma}'$, by concentration inequality for sub-Gaussian random variables, we have
$
\bigg|  \frac{1}{n}\sum_{i=1}^n\widehat{\gamma}^\top X_{i\cdot}\bigg|\lesssim \|\widehat{\gamma}\|_2 \sqrt{\frac{\log p}{n}}.
$
Again, by Lemma \ref{lasso.bnd.lem}, it follows that, with probability at least $1-p^{-c}$,
$
\frac{1}{n}(\widehat{\alpha}-\alpha)\sum_{i=1}^n\widehat{\gamma}^\top X_{i\cdot}\lesssim (\|\gamma\|_2+\sqrt{k\log p/n})\frac{k\log p}{n},
$
which is dominated by the error bound for $T_3$.

To sum up, by setting $A_n=T_2+T_3+T_4$, we have
\beq \label{ub.eq}
P\bigg(A_n\lesssim \bigg(\sqrt{\frac{k\log p}{n}}+\|\beta\|_2+\|\gamma\|_2\bigg)\frac{k\log p\sqrt{\log n}}{n}+(1+\|\beta\|_2+\|\gamma\|_2)\frac{k\log p}{n}\bigg)\ge 1-n^{-c}-p^{-c},
\eeq
\beq  \label{T2-T4}
P\bigg(A_n\lesssim [1+(\|\beta\|_2+\|\gamma\|_2)\sqrt{\log n}]\frac{k\log p}{n}\bigg)\ge 1-n^{-c}-p^{-c},
\eeq
for some constant $c>0$. This proves the first statement of the theorem. 

As for asymptotic normality, it suffices to set $B_n=T_1$ and show that
\beq \label{v>b}
\bigg|\frac{\sqrt{n}A_n}{v}\bigg|=o_P(1),
\eeq
and the asymptotic normality  follows from Slutsky's theorem.
Note that it has been shown that with high probability 
$
|A_n|\lesssim [1+(\|\beta\|_2+\|\gamma\|_2)\sqrt{\log n}]\frac{k\log p}{n}.
$
To ensure (\ref{v>b}), we need
\beq \label{cond.v>b}
[1+(\|\beta\|_2+\|\gamma\|_2)\sqrt{\log n}]\frac{k\log p}{n}\ll \frac{v}{\sqrt{n}}\asymp \frac{\|\gamma\|_2+\|\beta\|_2+\|\beta\|_2\|\gamma\|_2}{\sqrt{n}}.
\eeq
In other words,  a sufficient condition for (\ref{v>b}) is 
$
k\ll \frac{(\|\gamma\|_2+\|\beta\|_2+\|\beta\|_2\|\gamma\|_2)\sqrt{n}}{[1+(\|\beta\|_2+\|\gamma\|_2)\sqrt{\log n}]\log p}.
$

\paragraph{Part II.} In the following, we consider the cases without the intercepts as the  analysis is identical to the one with the intercepts up to an extra term $T_4$ in the error bound, which is dominated by $T_3$. Recall that
$
\widehat{\beta^{\top}\Sigma\beta}=\widehat{\beta}^{\top} \widehat{\Sigma}\widehat{\beta}-2\widehat{\beta}^{\top}\frac{1}{n_1}\sum_{i=1}^{n_1}\big(\frac{\exp(X_{i\cdot}^{\top}\widehat{\beta})}{(1+\exp(X_{i\cdot}^{\top}\widehat{\beta}))^2}\big)^{-1}\big(\frac{\exp(X_{i\cdot}^{\top}\widehat{\beta})}{1+\exp(X_{i\cdot}^{\top}\widehat{\beta})}-y_i\big)X_{i\cdot},
$
and
\begin{align*}
	\widehat{\beta}^\top\widehat{\Sigma}\widehat{\beta}-\beta^\top\Sigma\beta&=(\widehat{\beta}-\beta)^\top\Sigma\beta-\widehat{\beta}^\top\Sigma\beta+	\widehat{\beta}^\top\widehat{\Sigma}\widehat{\beta}\\
	&=-(\widehat{\beta}-\beta)^\top\Sigma(\widehat{\beta}-\beta)+(\widehat{\beta}-\beta)^\top\Sigma\widehat{\beta}-\widehat{\beta}^\top\Sigma\beta+\widehat{\beta}^\top\widehat{\Sigma}\widehat{\beta}\\
	&=-(\widehat{\beta}-\beta)^\top\Sigma(\widehat{\beta}-\beta)+2(\widehat{\beta}-\beta)^\top\Sigma\widehat{\beta}-\widehat{\beta}^\top\Sigma\widehat{\beta}+\widehat{\beta}^\top\widehat{\Sigma}\widehat{\beta}\\
	&=-(\widehat{\beta}-\beta)^\top\Sigma(\widehat{\beta}-\beta)+2(\widehat{\beta}-\beta)^\top\Sigma\widehat{\beta}+\widehat{\beta}^\top(\widehat{\Sigma}-\Sigma)\widehat{\beta}.
\end{align*}
We have the following decomposition for $\widehat{\beta^{\top}\Sigma\beta}-\beta^{\top}\Sigma\beta$
\begin{align*}
	&\quad \frac{2}{n_1}\widehat{\beta}^\top\sum_{i=1}^{n_1}\bigg(\frac{\exp(X_{i\cdot}^{\top}\widehat{\beta})}{(1+\exp(X_{i\cdot}^{\top}\widehat{\beta}))^2}\bigg)^{-1}\epsilon_iX_{i\cdot}+\frac{2}{n_1}\widehat{\beta}^\top\sum_{i=1}^{n_1}\bigg(\frac{\exp(X_{i\cdot}^{\top}\widehat{\beta})}{(1+\exp(X_{i\cdot}^{\top}\widehat{\beta}))^2}\bigg)^{-1}\Delta_iX_{i\cdot}+\widehat{\beta}^{\top}\left(\widehat{\Sigma}-\Sigma\right)\widehat{\beta}\\
	& \quad  +2(\widehat{\beta}-\beta)^\top(\Sigma-\widehat{\Sigma})\widehat{\beta}-(\widehat{\beta}-\beta)^{\top}{\Sigma}(\widehat{\beta}-\beta).
\end{align*}
Similar to the proofs in Part I, we define
\begin{align}
	T_1&=\frac{2}{n_1} \sum_{i=1}^{n_1}\bigg(\frac{\exp(X_{i\cdot}^{\top}\widehat{\beta})}{(1+\exp(X_{i\cdot}^{\top}\widehat{\beta}))^2}\bigg)^{-1}\epsilon_i\widehat{\beta}^\top X_{i\cdot}+\widehat{\beta}^\top(\widehat{\Sigma}-\Sigma)\widehat{\beta},\\
	T_2&=\frac{2}{n_1}\widehat{\beta}^\top\sum_{i=1}^{n_1}\bigg(\frac{\exp(X_{i\cdot}^{\top}\widehat{\beta})}{(1+\exp(X_{i\cdot}^{\top}\widehat{\beta}))^2}\bigg)^{-1}\Delta_iX_{i\cdot},\\
	T_3&=2(\widehat{\beta}-\beta)^\top(\Sigma-\widehat{\Sigma})\widehat{\beta}-(\widehat{\beta}-\beta)\widehat{\Sigma}(\widehat{\beta}-\beta).
\end{align}
Let $T_1=N_1+N_2$ where $N_1=\frac{1}{n_1}\sum_{i=1}^{n_1}\xi_i$ and $N_2=\frac{1}{n_2}\sum_{i=1}^{n_2}\mu_i$, such that
$
\xi_i=\frac{2(1+\exp({X_{i\cdot}}^{\top}\widehat{\beta}))^2}{\exp({X_{i\cdot}}^{\top}\widehat{\beta})}\epsilon_i \widehat{\beta}^\top X_{i\cdot}+\frac{n_1}{n_1+n_2}\widehat{\beta}^\top (X_{i\cdot}X_{i\cdot}^\top -\Sigma)\widehat{\beta},
$
$
\mu_i=\frac{n_2}{n_1+n_2}\widehat{\beta}^\top (Z_{i\cdot}Z_{i\cdot}^\top -\Sigma)\widehat{\beta}.
$
Note that, conditional on $\mathcal{D}_1$, $\{\xi_i\}_{i=1}^{n_1}$ and $\{\mu_i\}_{i=1}^{n_2}$ are mutually independent. In the following, we will show that, there exist $v_1,v_2>0$, with
\beq \label{asym.norm2}
\frac{\sqrt{n_1}N_1}{v_1}\bigg| \mathcal{D}_1\to_d N(0,1),\qquad 	\frac{\sqrt{n_2}N_2}{v_2}\bigg| \mathcal{D}_1\to_d N(0,1).
\eeq
Then, by the independence of $N_1$ and $N_2$, we have
$
\frac{\sqrt{n_1+n_2}T_1}{v_\beta}\bigg| \mathcal{D}_1\to_d N(0,1),
$
where
$
v_\beta^2=\frac{n_1+n_2}{n_1}v_1^2+\frac{n_1+n_2}{n_2}v_2^2.
$
Specifically, by CLT, we have (\ref{asym.norm2}) with 
$
v^2_1=4\E \eta_i(X,\widehat{\beta}) (X_{i\cdot}^\top\widehat{\beta})^2+\frac{n^2_1}{(n_1+n_2)^2}\E[\widehat{\beta}^\top(X_{i\cdot}X_{i\cdot}^\top-\Sigma)\widehat{\beta}]^2,
$
and
$
v^2_2=\frac{n^2_2}{(n_1+n_2)^2}\E[\widehat{\beta}^\top(Z_{i\cdot}Z_{i\cdot}^\top-\Sigma)\widehat{\beta}]^2.
$
In other words,
$
v_{\beta}^2=\frac{4(n_1+n_2)}{n_1}\E \eta_i(X,\widehat{\beta}) (X_{i\cdot}^\top\widehat{\beta})^2+\E[\widehat{\beta}^\top(X_{i\cdot}X_{i\cdot}^\top-\Sigma)\widehat{\beta}]^2.
$
On the other hand, by similar arguments in Part I, we have
\[
P\bigg( T_2\le C\bigg( \sqrt{\frac{k\log p}{n}}+\|\beta\|_2\bigg)\frac{k\log p\sqrt{\log n}}{n} \bigg)\ge 1-n^{-c}-p^{-c},
\]
\[
P\bigg(T_3\le C(1+\|\beta\|_2)\frac{k\log p}{n}\bigg)\ge 1-p^{-c}.
\]
The statements in the theorem then follows from the same arguments as in Part I and by setting $A'_n=T_2+T_3$ and $B'_n=T_1$.

\subsection{Proof of Theorem 1}

As in the proof of Theorem 4, we start with the decomposition
$
|\widehat{\beta^\top\Sigma\gamma}-\beta^\top\Sigma\gamma|=T_1+T_2+T_3+T_4,
$
where $T_1$ to $T_4$ are defined in (\ref{t1}) - (\ref{t4}). 
Note that
$
\E \frac{T_1^2}{v^2}=\frac{1}{n_1+n_2}.
$
By Markov's inequality, for any $t>0$
$
P(|T_1|\le tv)\ge 1-\frac{1}{t^2(n_1+n_2)},
$
or, by Lemma \ref{v.bnd.lem},
$
P\big(|T_1|\le C\frac{t(\|\beta\|_2+\|\gamma\|_2)}{\sqrt{n}}\big)\ge 1-\frac{1}{t^2}.
$
Combining this with  (\ref{T2-T4}) in the proof of Theorem 4, we obtain
$
P\big(|\widehat{\beta^\top\Sigma\gamma}-\beta^\top\Sigma\gamma|\lesssim \frac{t(\|\beta\|_2+\|\gamma\|_2)}{\sqrt{n}}+ [1+(\|\beta\|_2+\|\gamma\|_2)\sqrt{\log n}]\frac{k\log p}{n}\big)\ge 1-n^{-c}-p^{-c}-1/t^2.
$
Similar result can be obtained for $\widehat{\beta^\top\Sigma\beta}$.

\subsection{Proof of Theorem 5}

By definition, we have
$
v_R^2=\frac{v^2}{\beta^\top\Sigma\beta \gamma^\top\Sigma\gamma}\asymp \frac{v^2}{\|\beta\|_2^2\|\gamma\|_2^2}.
$
By Lemma \ref{v.bnd.lem}, it holds that with probability at least $1-p^{-c}$,
$
v_R^2\asymp \frac{\|\beta\|^2_2+\|\gamma\|_2^2+\|\beta\|_2^2\|\gamma\|_2^2}{\|\beta\|_2^2\|\gamma\|_2^2}\asymp \frac{1}{\|\beta\|_2^2}+\frac{1}{\|\gamma\|_2^2}+1.
$
Now since
$
\frac{\sqrt{n_1+n_2}\widehat{\beta^\top\Sigma\gamma}}{v_R\sqrt{\widehat{\beta^\top\Sigma\beta}\widehat{\gamma^\top\Sigma\gamma}}}=\frac{\sqrt{n_1+n_2}\widehat{\beta^\top\Sigma\gamma}}{v}\cdot \frac{\sqrt{{\beta^\top\Sigma\beta}{\gamma^\top\Sigma\gamma}}}{\sqrt{\widehat{\beta^\top\Sigma\beta}\widehat{\gamma^\top\Sigma\gamma}}},
$
if we can show that, for any constant $t>0$, with probability at least $1-t^{-2}$
\beq \label{demon.conv}
\bigg|\frac{{\widehat{\beta^\top\Sigma\beta}\widehat{\gamma^\top\Sigma\gamma}}}{{\beta^\top\Sigma\beta}{\gamma^\top\Sigma\gamma}}-1\bigg|=o(1),
\eeq
we can apply Slutsky's theorem to obtain the asymptotic normality. To show (\ref{demon.conv}), it suffices to prove
\beq \label{ratio}
\frac{|{{\widehat{\beta^\top\Sigma\beta}\widehat{\gamma^\top\Sigma\gamma}}}-{{\beta^\top\Sigma\beta}{\gamma^\top\Sigma\gamma}}|}{{{\beta^\top\Sigma\beta}{\gamma^\top\Sigma\gamma}}}\asymp 		\frac{|{{\widehat{\beta^\top\Sigma\beta}\widehat{\gamma^\top\Sigma\gamma}}}-{{\beta^\top\Sigma\beta}{\gamma^\top\Sigma\gamma}}|}{ \|\beta\|_2^2\|\gamma\|_2^2}=o(1).
\eeq
Since
\begin{align*}
	&\quad |{{\widehat{\beta^\top\Sigma\beta}\widehat{\gamma^\top\Sigma\gamma}}}-{{\beta^\top\Sigma\beta}{\gamma^\top\Sigma\gamma}}|\\
	&\le |(\widehat{\beta^\top\Sigma\beta}-{\beta^\top\Sigma\beta}){\gamma^\top\Sigma\gamma}|+|(\widehat{\gamma^\top\Sigma\gamma}-{\gamma^\top\Sigma\gamma}){\beta^\top\Sigma\beta}|+|(\widehat{\beta^\top\Sigma\beta}-{\beta^\top\Sigma\beta})(\widehat{\gamma^\top\Sigma\gamma}-{\gamma^\top\Sigma\gamma})|,
\end{align*}
under the conditions of Theorem 1, as long as 
$
\frac{t(\|\beta\|_2+\|\beta\|_2^2)}{\sqrt{n}}+(1+\|\beta\|_2\sqrt{\log n})\frac{k\log p}{n}\to 0,
$
and
$
\frac{t(\|\gamma\|_2+\|\gamma\|_2^2)}{\sqrt{n}}+(1+\|\gamma\|_2\sqrt{\log n})\frac{k\log p}{n}\to 0,
$
or
\beq \label{cond.1}
\|\gamma\|_2+\|\beta\|_2\ll n^{1/4},\qquad k\ll \frac{n}{[1+(\|\beta\|_2+\|\gamma\|_2)\sqrt{\log n}]\log p},
\eeq
with probability at least $1-p^{-c}-n^{-c}-t^{-2}$,
\begin{align}
	|{{\widehat{\beta^\top\Sigma\beta}\widehat{\gamma^\top\Sigma\gamma}}}-{{\beta^\top\Sigma\beta}{\gamma^\top\Sigma\gamma}}|&\lesssim \frac{t\|\gamma\|_2^2(\|\beta\|_2+\|\beta\|_2^2)}{\sqrt{n}}+\|\gamma\|_2^2(1+\|\beta\|_2\sqrt{\log n})\frac{k\log p}{n}\nonumber \\
	&\quad+\frac{t\|\beta\|_2^2(\|\gamma\|_2+\|\gamma\|_2^2)}{\sqrt{n}}+\|\beta\|_2^2(1+\|\gamma\|_2\sqrt{\log n})\frac{k\log p}{n}. \label{ratio.2}
\end{align}
and thus
\begin{align*}
	\frac{|{{\widehat{\beta^\top\Sigma\beta}\widehat{\gamma^\top\Sigma\gamma}}}-{{\beta^\top\Sigma\beta}{\gamma^\top\Sigma\gamma}}|}{  \|\beta\|_2^2\|\gamma\|_2^2}&\lesssim\frac{t(\|\beta\|_2+\|\beta\|_2^2)}{\|\beta\|_2^2\sqrt{n}}+\frac{1+\|\beta\|_2\sqrt{\log n}}{\|\beta\|_2^2}\frac{k\log p}{n}\\
	&\quad+\frac{t(\|\gamma\|_2+\|\gamma\|_2^2)}{\|\gamma\|_2^2\sqrt{n}}+\frac{1+\|\gamma\|_2\sqrt{\log n}}{\|\gamma\|_2^2}\frac{k\log p}{n}.
\end{align*}
As long as
\beq \label{cond.2}
\min\{\|\beta\|_2,\|\gamma\|_2 \}\gg \sqrt{\frac{k\log p}{n}},\qquad k\ll \frac{n}{\log p\log n},
\eeq
we have (\ref{ratio}) in probability.
Thus we complete  the proof as (\ref{cond.1}) and (\ref{cond.2}) are implied by the conditions $	k\ll \min\{\frac{n}{\log p\log n}, \frac{(\|\gamma\|_2+\|\beta\|_2+\|\beta\|_2\|\gamma\|_2)\sqrt{n}}{[1+(\|\beta\|_2+\|\gamma\|_2)\sqrt{\log n}]\log p}  \}$ and $\sqrt{\frac{k\log p}{n}}\ll \|\beta\|_2,\|\gamma\|_2\ll 1$ in the theorem.

\subsection{Proof of Theorem 3}
Since $R(\beta,\gamma,\Sigma)\in(-1,1)$, we have
$
|\widehat{R}-R(\beta,\gamma,\Sigma)|\le \bigg|\frac{\widehat{\beta^\top\Sigma\gamma}}{\sqrt{\widehat{\beta^\top\Sigma\beta}\widehat{\gamma^\top\Sigma\gamma}}}-R(\beta,\gamma,\Sigma)\bigg|.
$
Note that
\begin{align*}
	\bigg|\frac{\widehat{\beta^\top\Sigma\gamma}}{\sqrt{\widehat{\beta^\top\Sigma\beta}\widehat{\gamma^\top\Sigma\gamma}}}-R(\beta,\gamma,\Sigma)\bigg|\le \bigg| \frac{\widehat{\beta^\top\Sigma\gamma}-\beta^\top\Sigma\gamma}{\sqrt{\widehat{\beta^\top\Sigma\beta}\widehat{\gamma^\top\Sigma\gamma}}}\bigg|+|\beta^\top\Sigma\gamma|\bigg| \frac{1}{\sqrt{\widehat{\beta^\top\Sigma\beta}\widehat{\gamma^\top\Sigma\gamma}}}- \frac{1}{\sqrt{{\beta^\top\Sigma\beta}{\gamma^\top\Sigma\gamma}}}\bigg|.
\end{align*}
In the proof of Theorem 5, we have shown that under the conditions $k\ll \min\{\frac{n}{\log p\log n}, \frac{n}{(1+(\|\beta\|_2+\|\gamma\|_2)\sqrt{\log n})\log p}\}$ and $\sqrt{\frac{k\log p}{n}}\ll \|\beta\|_2,\|\gamma\|_2\ll 1$, (\ref{demon.conv}) holds. Thus,   for any constant $t>0$, with probability at least $1-t^{-2}$
\begin{align*}
	\bigg|\frac{\widehat{\beta^\top\Sigma\gamma}}{\sqrt{\widehat{\beta^\top\Sigma\beta}\widehat{\gamma^\top\Sigma\gamma}}}-R(\beta,\gamma,\Sigma)\bigg|&\lesssim \bigg| \frac{\widehat{\beta^\top\Sigma\gamma}-\beta^\top\Sigma\gamma}{\sqrt{{\beta^\top\Sigma\beta}{\gamma^\top\Sigma\gamma}}}\bigg|+\frac{|\widehat{\beta^\top\Sigma\gamma}|}{\sqrt{{\beta^\top\Sigma\beta}{\gamma^\top\Sigma\gamma}}}\bigg|\frac{\sqrt{{\beta^\top\Sigma\beta}{\gamma^\top\Sigma\gamma}}}{\sqrt{\widehat{\beta^\top\Sigma\beta}\widehat{\gamma^\top\Sigma\gamma}}}-1\bigg|\\
	&\lesssim \bigg| \frac{\widehat{\beta^\top\Sigma\gamma}-\beta^\top\Sigma\gamma}{\sqrt{{\beta^\top\Sigma\beta}{\gamma^\top\Sigma\gamma}}}\bigg|+\bigg|\frac{\sqrt{{\beta^\top\Sigma\beta}{\gamma^\top\Sigma\gamma}}}{\sqrt{\widehat{\beta^\top\Sigma\beta}\widehat{\gamma^\top\Sigma\gamma}}}-1\bigg|
\end{align*}
Note that, for any constant $t>0$, with probability at least $1-t^{-2}$,
\[
\bigg| \frac{\widehat{\beta^\top\Sigma\gamma}-\beta^\top\Sigma\gamma}{\sqrt{{\beta^\top\Sigma\beta}{\gamma^\top\Sigma\gamma}}}\bigg|\lesssim \frac{t(\|\beta\|_2+\|\gamma\|_2+\|\beta\|_2\|\gamma\|_2)}{\|\beta\|_2\|\gamma\|_2\sqrt{n}}+\frac{(1+(\|\beta\|_2+\|\gamma\|_2)\sqrt{\log n})}{\|\beta\|_2\|\gamma\|_2}\frac{k\log p}{n},
\]
and by (\ref{demon.conv}),
\begin{align*}
	\bigg|\frac{\sqrt{{\beta^\top\Sigma\beta}{\gamma^\top\Sigma\gamma}}}{\sqrt{\widehat{\beta^\top\Sigma\beta}\widehat{\gamma^\top\Sigma\gamma}}}-1\bigg|&=\bigg|\frac{{{\beta^\top\Sigma\beta}{\gamma^\top\Sigma\gamma}}}{{\widehat{\beta^\top\Sigma\beta}\widehat{\gamma^\top\Sigma\gamma}}}-1\bigg|\cdot\bigg[\frac{\sqrt{{\beta^\top\Sigma\beta}{\gamma^\top\Sigma\gamma}}}{\sqrt{\widehat{\beta^\top\Sigma\beta}\widehat{\gamma^\top\Sigma\gamma}}}+1\bigg]^{-1}\\
	&\lesssim\frac{t(\|\beta\|_2+\|\beta\|_2^2)}{\|\beta\|_2^2\sqrt{n}}+\frac{1+\|\beta\|_2\sqrt{\log n}}{\|\beta\|_2^2}\frac{k\log p}{n}\\
	&\quad+\frac{t(\|\gamma\|_2+\|\gamma\|_2^2)}{\|\gamma\|_2^2\sqrt{n}}+\frac{1+\|\gamma\|_2\sqrt{\log n}}{\|\gamma\|_2^2}\frac{k\log p}{n},
\end{align*}
where the last inequality follows from (\ref{ratio.2}).
Hence, for any constant $t>0$, with probability at least $1-t^{-2}$,
\begin{align*}
	\bigg|\frac{\widehat{\beta^\top\Sigma\gamma}}{\sqrt{\widehat{\beta^\top\Sigma\beta}\widehat{\gamma^\top\Sigma\gamma}}}-R(\beta,\gamma,\Sigma)\bigg|&\lesssim \frac{t(\|\beta\|_2+\|\gamma\|_2+\|\beta\|_2\|\gamma\|_2)}{\|\beta\|_2\|\gamma\|_2\sqrt{n}}+\frac{(1+(\|\beta\|_2+\|\gamma\|_2)\sqrt{\log n})}{\|\beta\|_2\|\gamma\|_2}\frac{k\log p}{n}\\
	&\quad+\frac{t(\|\beta\|_2+\|\beta\|_2^2)}{\|\beta\|_2^2\sqrt{n}}+\frac{1+\|\beta\|_2\sqrt{\log n}}{\|\beta\|_2^2}\frac{k\log p}{n}\\
	&\quad+\frac{t(\|\gamma\|_2+\|\gamma\|_2^2)}{\|\gamma\|_2^2\sqrt{n}}+\frac{1+\|\gamma\|_2\sqrt{\log n}}{\|\gamma\|_2^2}\frac{k\log p}{n},
\end{align*}
as long as the conditions of Theorem 5 holds.

\subsection{Proof of Theorem 6}

The proof of this theorem is divided into two parts, with the first part corresponding to $\widehat{\beta^\top\Sigma\gamma}$ and $\widehat{\beta^\top\Sigma\beta}$, and the second part corresponding to $\widehat{R}$, respectively. Throughout, the expectation is with respect to  $\mathcal{D}_2$.

\paragraph{Part I.} We only focus on the proof related to $\widehat{\beta^\top\Sigma\gamma}$ as the proof for  $\widehat{\beta^\top\Sigma\beta}$ is similar.
First, we show that 
\beq \label{v.consistent}
\bigg|\frac{\hat{v}^2}{v^2}-1\bigg|\to_P 0.
\eeq
Recall that
$
v^2=\frac{n_1+n_2}{n_1}\E \eta_i(X,\widehat{\beta}')(\widehat{\gamma}^\top X_{i\cdot})^2+\frac{n_1+n_2}{n_2}\E \eta_i(Z,\widehat{\gamma}')(\widehat{\beta}^\top X_{i\cdot})^2+\E [ \widehat{\beta}^\top (X_{i\cdot}X_{i\cdot}^\top -\Sigma)\widehat{\gamma}]^2
$
and 
\begin{align*}
	\widehat{v}^2&=\frac{n_1+n_2}{n_1}\frac{1}{n_1}\sum_{i=1}^{n_1} \frac{(1+\exp(\widehat{\alpha}+X_{i\cdot}^\top\widehat{\beta}))^2}{\exp(\widehat{\alpha}+X_{i\cdot}^\top\widehat{\beta})} (\widehat{\gamma}^\top X_{i\cdot})^2+\frac{n_1+n_2}{n_2}\frac{1}{n_2}\sum_{i=1}^{n_2} \frac{(1+\exp(\widehat{\eta}+Z_{i\cdot}^\top\widehat{\gamma}))^2}{\exp(\widehat{\eta}+Z_{i\cdot}^\top\widehat{\gamma})} (\widehat{\beta}^\top Z_{i\cdot})^2\\
	&\quad+\frac{1}{n_1+n_2}\bigg[\sum_{i=1}^{n_1}(\widehat{\beta}X_{i\cdot}X_{i\cdot}^\top\widehat{\gamma}-\widehat{\beta}\widehat{\Sigma}\widehat{\gamma})^2+\sum_{i=1}^{n_2}(\widehat{\beta}Z_{i\cdot}Z_{i\cdot}^\top\widehat{\gamma}-\widehat{\beta}\widehat{\Sigma}\widehat{\gamma})^2\bigg]\\
	&\equiv \frac{n_1+n_2}{n_1}V_1+\frac{n_1+n_2}{n_2}V_2+V_3
\end{align*}
It suffices to show
\beq \label{v1.bnd}
\frac{|V_1-\E \eta_i(X,\widehat{\beta}')(\widehat{\gamma}^\top X_{i\cdot})^2|}{v^2}\to_P0,
\eeq
\beq \label{v3.bnd}
\frac{1}{v^2}\bigg|\frac{1}{n_1}\sum_{i=1}^{n_1}(\widehat{\beta}X_{i\cdot}X_{i\cdot}^\top\widehat{\gamma}-\widehat{\beta}\widehat{\Sigma}\widehat{\gamma})^2-\E [ \widehat{\beta}^\top (X_{i\cdot}X_{i\cdot}^\top -\Sigma)\widehat{\gamma}]^2\bigg|\to_P 0.
\eeq
To see (\ref{v1.bnd}), note that
$
|V_1-\E \eta_i(X,\widehat{\beta}')(\widehat{\gamma}^\top X_{i\cdot})^2|\le \big|V_1-\E\frac{(1+\exp(\widehat{\alpha}+X_{i\cdot}^\top\widehat{\beta}))^2}{\exp(\widehat{\alpha}+X_{i\cdot}^\top\widehat{\beta})} (\widehat{\gamma}^\top X_{i\cdot})^2 \big|+\big|\E\big[\frac{(1+\exp(\widehat{\alpha}+X_{i\cdot}^\top\widehat{\beta}))^2}{\exp(\widehat{\alpha}+X_{i\cdot}^\top\widehat{\beta})}-\eta_i(X,\widehat{\beta}')\big] (\widehat{\gamma}^\top X_{i\cdot})^2\big|.
$
On the one hand, by Lemma \ref{ratio.lem}, we have
\begin{align*}
	&\bigg|\E\bigg[\frac{(1+\exp(\widehat{\alpha}+X_{i\cdot}^\top\widehat{\beta}))^2}{\exp(\widehat{\alpha}+X_{i\cdot}^\top\widehat{\beta})}-\eta_i(X,\widehat{\beta}')\bigg] (\widehat{\gamma}^\top X_{i\cdot})^2\bigg|\\
	&\lesssim \bigg|\E[\exp(|{X'_{i\cdot}}^\top (\beta'-\widehat{\beta'})|)-1]\frac{(1+\exp(\widehat{\alpha}+X_{i\cdot}^\top\widehat{\beta}))^2}{\exp(\widehat{\alpha}+X_{i\cdot}^\top\widehat{\beta})} (\widehat{\gamma}^\top X_{i\cdot})^2\bigg|\\
	&\lesssim \sqrt{\E [\exp(|{X'_{i\cdot}}^\top (\beta'-\widehat{\beta'})|)-1]^2}\sqrt{\E \frac{(1+\exp(\widehat{\alpha}+X_{i\cdot}^\top\widehat{\beta}))^4}{\exp(2\widehat{\alpha}+2X_{i\cdot}^\top\widehat{\beta})} (\widehat{\gamma}^\top X_{i\cdot})^4}
\end{align*}
Now since  $e^x-1\le 2x$ for all $x\in(0,1)$, with probability at least $1-p^{-c}$,
\begin{align*}
	&\E [\exp(|{X'_{i\cdot}}^\top(\beta'-\widehat{\beta}')|)-1]^2\\
	&=\E [\exp(|{X'_{i\cdot}}^\top(\beta'-\widehat{\beta}')|)-1]^2\cdot 1\{|{X'_{i\cdot}}^\top(\beta'-\widehat{\beta}')|\le 1\}+\E [\exp(|{X'_{i\cdot}}^\top(\beta'-\widehat{\beta}')|)-1]^2\cdot 1\{|{X'_{i\cdot}}^\top(\beta'-\widehat{\beta}')|> 1\}\\
	&\lesssim \E [{X'_{i\cdot}}^\top(\beta'-\widehat{\beta}')]^2+\E \exp(2|{X'_{i\cdot}}^\top(\beta'-\widehat{\beta}'|))\cdot 1\{|{X'_{i\cdot}}^\top(\beta'-\widehat{\beta}')|> 1\}\\
	&\lesssim \frac{k\log p}{n}+\sqrt{\E \exp(4|{X'_{i\cdot}}^\top(\beta'-\widehat{\beta}'|))}\sqrt{P(|{X'_{i\cdot}}^\top(\beta'-\widehat{\beta}')|> 1)}\\
	&\lesssim \frac{k\log p}{n}+\frac{1}{p^c}.
\end{align*}
In addition, by the similar argument as in (\ref{ub.eta.eq}), we have
$
\sqrt{\E \frac{(1+\exp(\widehat{\alpha}+X_{i\cdot}^\top\widehat{\beta}))^4}{\exp(2\widehat{\alpha}+2X_{i\cdot}^\top\widehat{\beta})} (\widehat{\gamma}^\top X_{i\cdot})^4}\lesssim \|\gamma\|_2^2.
$
Hence, we have
$
\big|\E\big[\frac{(1+\exp(\widehat{\alpha}+X_{i\cdot}^\top\widehat{\beta}))^2}{\exp(\widehat{\alpha}+X_{i\cdot}^\top\widehat{\beta})}-\eta_i(X,\widehat{\beta}')\big] (\widehat{\gamma}^\top X_{i\cdot})^2\big|\lesssim \|\gamma\|_2^2\sqrt{\frac{k\log p}{n}}.
$
On the other hand, conditional on $\mathcal{D}_1$, by LLN,
$
\frac{V_1}{\|\widehat{\gamma}\|_2^2}-\E\frac{(1+\exp(\widehat{\alpha}+X_{i\cdot}^\top\widehat{\beta}))^2}{\exp(\widehat{\alpha}+X_{i\cdot}^\top\widehat{\beta})} \frac{(\widehat{\gamma}^\top X_{i\cdot})^2}{\|\widehat{\gamma}\|_2^2}\to 0, a.s..
$
Combining the above results, we have shown (\ref{v1.bnd}).

To see (\ref{v3.bnd}), let
$
S=\bigg|\frac{1}{n_1}\sum_{i=1}^{n_1}(\widehat{\beta}^\top X_{i\cdot}X_{i\cdot}^\top\widehat{\gamma}-\widehat{\beta}\widehat{\Sigma}\widehat{\gamma})^2-\E [\widehat{\beta}^\top (X_{i\cdot}X_{i\cdot}^\top -\Sigma)\widehat{\gamma}]^2\bigg|.
$
The following lemma concerns the tail probability of $S$.

\bel \label{eq.lem}
Under the conditions of Theorem 6, it holds that
\beq \label{eq1}
P\bigg( \frac{S}{v^2} \ge C\frac{(\|\beta\|_2^2+\frac{k\log p}{n})(\|\gamma\|_2^2+\frac{k\log p}{n})}{\|\beta\|_2^2+\|\gamma\|_2^2+\|\beta\|_2^2\|\gamma\|_2^2}\frac{\log^{5/2} n}{\sqrt{n}} \bigg)\lesssim n^{-c}.
\eeq
\eel

Lemma \ref{eq.lem} along with the fact that $L(\beta,\gamma)\gg \sqrt{\frac{k\log p}{n}}$ implies (\ref{v3.bnd}). This completes the proof of (\ref{v.consistent}).

By Theorem 4, we have, for $\delta_n \asymp (1+(\|\beta\|_2+\|\gamma\|_2)\sqrt{\log n})k\log p/n$,
$
P_\theta( \beta^\top\Sigma\gamma\in {\text{CI}_\alpha(\beta^\top\Sigma\gamma,\mathcal{D})})=P_\theta( |\widehat{\beta^\top\Sigma\gamma}-\beta^\top\Sigma\gamma|\le \widehat{\rho})=P_\theta( |A_n+B_n|\le \widehat{\rho}) \ge P_\theta( |B_n|\le \widehat{\rho}-\delta_n,|A_n|\le \delta_n)\ge 1-P_\theta( |B_n|\ge \widehat{\rho}-\delta_n)-P(|A_n|\ge \delta_n),
$
where by (\ref{T2-T4}), we have $\lim_{n,p\to\infty }P(|A_n|\ge \delta_n)= 0$. 
As for $P_\theta( |B_n|\ge \widehat{\rho}-\delta_n)$, if $L(\beta,\gamma)\gg \sqrt{\frac{k\log p}{n}}$, by (\ref{v.consistent}) there exists some sequence $\delta'_n\to 0$ such that $\lim_{n,p\to\infty}P_\theta(|\hat{v}-v|z_{\alpha/2}/v>\delta'_n)=0$, so 
$
P_\theta( |B_n|\ge \widehat{\rho}-\delta_n) \le P_\theta( \sqrt{n_1+n_2}|B_n|/v\ge z_{\alpha/2}-\delta'_n-\sqrt{n_1+n_2}\delta_n/v)+P_\theta(|\hat{v}-v|z_{\alpha/2}/v>\delta'_n).
$
By (\ref{clt.eq}) and the fact that, for $	k\ll \min\{\frac{n}{\log p\log n}, \frac{(\|\gamma\|_2+\|\beta\|_2+\|\beta\|_2\|\gamma\|_2)\sqrt{n}}{[1+(\|\beta\|_2+\|\gamma\|_2)\sqrt{\log n}]\log p}  \}$, $|\delta'_n+\sqrt{n_1+n_2}\delta_n/v|\to 0$, we have $\limsup_{n,p\to\infty}P_\theta( \sqrt{n_1+n_2}|B_n|/v\ge z_{\alpha/2}-\delta'_n-\sqrt{n_1+n_2}\delta_n/v|\mathcal{D}_1)\le\alpha$,
so that 
\begin{align*}
	&\limsup_{n,p\to\infty}P_\theta( |B_n|\ge \widehat{\rho}-\delta_n) \\
	&\le \limsup_{n,p\to\infty}\int P_\theta( \sqrt{n_1+n_2}|B_n|/v\ge z_{\alpha/2}-\delta'_n-\sqrt{n_1+n_2}\delta_n/v|\mathcal{D}_1)d P_{\mathcal{D}_1}\\
	&\le\int \limsup_{n,p\to\infty}P_\theta( \sqrt{n_1+n_2}|B_n|/v\ge z_{\alpha/2}-\delta'_n-\sqrt{n_1+n_2}\delta_n/v|\mathcal{D}_1)d P_{\mathcal{D}_1}\\
	&\le \alpha.
\end{align*}
This proves the first statement of the theorem. The second statement follows directly from the definition of $\widehat{\rho}$. Specifically, note that with high probability $ v\asymp \|\beta\|_2+\|\gamma\|_2+\|\beta\|_2\|\gamma\|_2$, and
$
\big||\widehat{\rho}|-\frac{1}{\sqrt{n}} v\big|\lesssim\frac{1}{\sqrt{n}}|v-\widehat{v}|.
$
By (\ref{v.consistent}), with probability at least $1-p^{-c}$,
$
|v-\widehat{v}|\ll \|\beta\|_2+\|\gamma\|_2+\|\beta\|_2\|\gamma\|_2.
$
This completes the proof of the length of the confidence interval.

\paragraph{Part II.} Similar to the proof in Part I, we first show that
\beq \label{v.consistent.2}
\bigg| \frac{\widehat{v}^2_R}{v_R^2}-1\bigg|\to_P 0.
\eeq
By Lemma \ref{v.bnd.lem}, we have $v_R^2\asymp 1+1/{\|\beta\|_2^2}+1/\|\gamma\|_2^2$. Then it suffices to show 
$
|\widehat{v}^2_R-v_R^2|/( 1+1/{\|\beta\|_2^2}+1/\|\gamma\|_2^2)\to_P 0.
$
Now  if (\ref{v.consistent}) and (\ref{demon.conv}) hold, we have
$
|\widehat{v}^2_R-v_R^2|\lesssim \frac{v^2}{{\beta^\top\Sigma\beta}{\gamma^\top\Sigma\gamma}}\big| \frac{{\beta^\top\Sigma\beta}{\gamma^\top\Sigma\gamma}}{{\widehat{\beta^\top\Sigma\beta}\widehat{\gamma^\top\Sigma\gamma}}} -1 \big|+\frac{v^2}{{\beta^\top\Sigma\beta}{\gamma^\top\Sigma\gamma}}\big|\frac{v^2}{\widehat{v}^2}-1  \big|
\ll 1+ 1/{\|\beta\|_2^2}+1/\|\gamma\|_2^2.
$
In other words, whenever (\ref{v.consistent}) and (\ref{demon.conv}) hold, (\ref{v.consistent.2}) holds. Taking into account the range of $R(\beta,\gamma,\Sigma)$, the results concerning the asymptotic coverage and the length of the CI follows from the similar argument as in the proof of Part I.

\section{Proofs of Lower Bound Results}

\subsection{Proof of Theorem 2}

The proof of the lower bound relies on the construction of two parameter spaces which are different but not ``testable" by any statistical procedure, which we refer as null and alternative.

We first construct $\HH_0$ and $\HH_1$ and show that $\HH_0\cup\HH_1\subset \Theta(k)$; then we will control the distribution distance $\chi^2(f_{\pi_{\HH_1}},f_{\pi_{\HH_0}} )$, where $f_{\pi_{\HH_1}}$ is the distribution with parameter $(\beta,\Sigma)$ that has a prior $\pi_{\HH_1}$ over $\HH_1$ and $f_{\pi_{\HH_0}}$ is the distribution with parameter $(\beta^*,\gamma^*,\Sigma^*)\in \HH_0$; lastly, we calculate the distance $\beta^\top\Sigma\gamma$ where $(\beta,\gamma,\Sigma) \in \HH_1$ and $(\beta^*,\gamma^*,\Sigma^*) \in \HH_0$. Then we can apply Theorem 2.15 of \cite{tsybakov2009introduction}, which is a general lower bound for testing two fuzzy hypotheses.
In our theorem, the lower bound consists of two pieces. The rest of our proof is separated into two parts, with the first parts  corresponding to the high-dimensional rate $\min\{ k\log p/n,L_n^2\}$  and the second part concerning the parametric rate $\min\{L_n/\sqrt{n},L_n^2\}$. 

In the following, we assume without loss of generality that $X_i=Z_i$ for $i=1,...,n$, corresponding to the observational scenario (II). The analysis under the scenario (I) can be obtained with minor adaptation in the proof.

\paragraph{The High-Dimensional Rate}

We first obtain the rate $\min\{ k\log p/n,L_n^2\}$. 

\emph{Case I. $ k\log p/n\lesssim L^2_n$.}
We prove the high-dimensional rate ${k\log p}/{n}$ following the steps as we stated earlier.

\underline{Step 1: Construction of $\HH_0$ and $\HH_1$.} The null space is taken as $\HH_0 = \{(0,0,{I_p})\}$, which consists of only one point.
To construct $\HH_1$, we define $\ell(M,n)$ as the set of all the $n$-element subsets of $M$. 
We define the alternative parameter space $\mathcal{H}_1= \{ (\beta_I,\gamma_I,\Sigma):\gamma_I=\beta_I, \|\beta_I\|_0 = k, \beta_{Ij} = \rho1\{j\in I\},  \Sigma={I}_p, \text{ for } I \in \ell([1:p],k) \}$,
so that  $\mathcal{H}_1$ contains all $(\beta_I,\gamma_I,{I}_p)$ for some special $k$-sparse $\beta_I=\gamma_i$ with nonzero components of magnitude $\rho$ (indexed by $I$). As a result, for any $(\beta,\gamma,\Sigma)\in \HH_1$ and $(\beta',\gamma',\Sigma')\in \HH_0$, it holds that
$
|\beta^\top\Sigma\gamma-\beta'^\top\Sigma'\gamma'|=\|\beta\|_2^2={k\rho^2}.
$
Apparently, $\HH_1 \subseteq \Theta(k,L_n)$.

\underline{Step 2: Control $\chi^2(f_{\pi_{\HH_1}},f_{\pi_{\HH_0}} )$.} 
Let $\pi$ denote the uniform prior on $I$ over $\ell([1:p],k)$. This prior induces a prior distribution $\pi_{\HH_1}$ over the parameter space $\HH_1$. Since for any $(\beta,\gamma,\Sigma)\in \HH_1$, we have $\beta=\gamma$. Thus $w_i\equiv y_i$ for all $i=1,...,n$ and we identify the joint distribution of $(y_i, w_i,X_i)$ with that of $(y_i,y_i,X_i)$, parametrized by $(\beta,\Sigma)$. Specifically,
$
p(X_i,y_i;\beta,\Sigma) =\frac{1}{\sqrt{(2\pi)^p| \Sigma| }}\exp\big\{ -\frac{1}{2}X_i^\top\Sigma^{-1} X_i \big\} \frac{\exp(y_iX_i^\top \beta)}{1+\exp(X_i^\top\beta)}.
$
For $(0,0,{I})\in\mathcal{H}_0$, the corresponding joint distribution is
$
g_0 = \prod_{i=1}^n p(X_i,y_i;0,{I_p}) =\frac{1}{(2\pi)^{np/2}} \prod_{i=1}^n  \frac{1}{2}e^{- \|X_i\|_2^2/2}.
$
Similarly, the marginal distribution of the samples with parameter in  $\mathcal{H}_1$ is denoted as
$
g_1 = \int_{\HH_1}  \prod_{i=1}^np(X_i,y_i;\beta,\Sigma) \pi_{\HH_1} =\frac{1}{{p \choose k}^n}  \sum_{(\beta,\Sigma)\in \mathcal{H}_1}  \prod_{i=1}^n p(X_i,y_i;\beta,\Sigma).
$
Therefore we have
$
\chi^2(g_1,g_0) = \int\frac{g_1^2}{g_0}-1 = \frac{1}{{p \choose k}^{2n}}\sum_{(\beta,\Sigma)\in \mathcal{H}_1} \sum_{(\beta',\Sigma') \in \mathcal{H}_1} \prod_{i=1}^n   \int \frac{ p(X_i,y_i;\beta,\Sigma)p(X_i,y_i;\beta',\Sigma')}{p(X_i,y_i;0,{ I_p})} -1.
$
The following lemma, obtained by \cite{ma2020global}, provides an important equation that significantly simplify our analysis of the Chi-square divergence.

\bel
Suppose $X_i\sim N(0, \Sigma)$ and let $p_f(X_i,y_i;\beta,\Sigma)$ be the joint density function of $(X_i,y_i)$. Then for any $(\beta,\Sigma)$ and $(\beta',\Sigma')$, it holds that
\begin{align} \label{key.eq}
	&\int \frac{p_f(X_i,y_i;\beta,\Sigma)p_f(X_i,y_i;\beta',\Sigma')}{p_f(X_i,y_i;0,{ I})} = \frac{\det(\Omega)\det(\Omega')}{\det(\Omega+\Omega'-I)}[1+\E \tanh(Z^\top\beta/2)\tanh(Z^\top\beta'/2)],
\end{align}
where $Z\sim N(0,(\Omega+\Omega'-{ I})^{-1})$, $\Omega=\Sigma^{-1}$ and $\Omega'=(\Sigma)^{-1}$.
\eel

Our next lemma, proved by \cite{ma2020global}, concerns some moment inequalities that useful to control the right-hand side of the above equation.

\bel[\citealt{ma2020global}]\label{mom.lem}
For a bivariate vector $(X,Y)\sim N(0,\Sigma)$ with $\Sigma=\sigma^2\begin{bmatrix}1&\rho\\ \rho&1 \end{bmatrix}$ for some $\sigma^2\le 1$ and $\rho\in(0,1)$, it holds that
$
\E \tanh(X/2)\tanh(Y/2)\le C\sigma^2\rho,
$
for some universal constant $C>0$.
\eel

By construction, we have $X^\top \beta\sim N(0,\|\beta\|_2^2)$ and $X^\top \beta'\sim N(0,\|\beta'\|_2^2)$, 
and $\text{Cov}(X^\top \beta,X^\top \beta')=\beta^\top\beta'=\delta^\top\delta'$.
As a result, by Lemma \ref{mom.lem}, we have
$
\E_{\Sigma,\Sigma'} h(X;\beta,\beta')\le 1+\delta^\top\delta'=1+j\rho^2
$
where $j=|\supp(\delta)\cap \supp(\delta')|=|I\cap I'|$ is the number of intersected components between $\delta$ and $\delta'$.
Hence $\chi^2(g,f)\le c\E (1+J\rho^2)^n-1$,
where $J$ follows a hypergeometric distribution $P(J=j)=\frac{{k \choose j}{p-k \choose k-j}}{{p\choose k}}, j=0,1,...,k-1.$
Then $\chi^2(g,f)\le c\E \exp(n\log(1+\rho^2 J))-1\le c\E e^{n\rho^2 J}-1.$
As shown on page 173 of \cite{aldous1985exchangeability}, $J$ has the same distribution as the random variable $\E(Z|\mathcal{B}_n)$ where $Z$ is a binomial random variable of parameters $(k,k/p)$ and $\mathcal{B}_n$ some suitable $\sigma$-algebra. Thus by Jensen's inequality we have $\E e^{CnJ\rho^2} \le \big( 1-\frac{k}{p}+\frac{k}{p}e^{n\rho^2} \big)^{k}$.
Hence, let $\rho^2 = \frac{1}{n}\log \big(1+\frac{p}{c_1k^2} \big)$
for some constant $c_1>0$,
we have $\chi^2(g,f) \le c_2$ for some small constant $c_2>0$.

\underline{Step 3. Obtain the Lower Bound.} Now by Theorem 2.15 of \cite{tsybakov2009introduction}, for $\xi=\beta^\top\Sigma\gamma$,
\begin{align} 
	\inf_{\widehat{\xi}}\sup_{\theta\in\Theta_E(k,L_n)} P\bigg(|\widehat{\xi}-\xi|\gtrsim k\frac{\log p}{n}\bigg)\ge c.\label{lower.eq}
\end{align}
Thus we have proven the lower bound ${k\log p}/{n}$.

\emph{Case II. $L^2_n\lesssim k\log p/n$.} We consider $\mathcal{H}_0 = \{ (0,0,{I})\},$  and $\mathcal{H}_1= \{ (\beta_I,\gamma_I,\Sigma):\gamma_I=\beta_I, \|\beta_I\|_0 = k, \beta_{Ij} = \rho1\{j\in I\},  \Sigma={\bf I}_p, \text{ for } I \in \ell([1:p],k) \}$, where $\rho=L_n/\sqrt{k}$. As a result, for any $(\beta,\gamma,\Sigma)\in \HH_1$ and $(\beta',\gamma',\Sigma')\in \HH_0$, it holds that
$
|\beta^\top\Sigma\gamma-\beta'^\top\Sigma'\gamma'|={k\rho^2}=L_n^2.
$
Apparently, $\HH_1 \subseteq \Theta(k,L_n)$. With the same argument as in Case I, we obtain for $\xi=\beta^\top\Sigma\gamma$,
\begin{align} 
	\inf_{\widehat{\xi}}\sup_{\theta\in\Theta_E(k,L_n)} P\bigg(|\widehat{\xi}-\xi|\gtrsim L_n^2\bigg)\ge c,\label{lower.eq2}
\end{align}
for some constant $c>0$.

\paragraph{The Parametric Rate.}
In this part, we obtain the rate $\min\{\frac{L_n}{\sqrt{n}},L_n^2\}+\frac{L_n^2}{\sqrt{n}}$, which along with the rate $\min\{k\log p/n,L_n^2\}$ from the previous part implies the final minimax lower bound.

\emph{Case I. $n^{-1/2}\lesssim L_n\lesssim 1$.}
To prove the parametric rate lower bound, we define $\mathcal{H}_0 = \{ (0,\gamma^*,{I_p})\},$ where $\gamma^*=(L_n,0,...,0)^\top$, and $\mathcal{H}_1 =\{ (\beta,\gamma,{ I_p}): \gamma=\gamma^*,\beta_1 =1/\sqrt{n}, \beta_i=0, \forall i\ne 1\}$.
Denote
$
g= \frac{1}{{(2\pi)^{np/2} }}\prod_{i=1}^n\exp\bigg\{ -\frac{1}{2}X_i^\top X_i \bigg\} \frac{\exp(y_iX_i^\top \beta)}{1+\exp(X_i^\top\beta)}\frac{\exp(w_iX_i^\top \gamma)}{1+\exp(X_i^\top\gamma)},
$
which is the joint density with parameter $(\beta,\gamma,I)\in\mathcal{H}_1$. Also denote $g_0$ as before as the joint density with $(\beta,\gamma,I)\in\mathcal{H}_0$. The $\chi^2$-divergence between distribution $g$ and $g_0$ is $\chi^2 = \int \frac{g^2}{g_0} -1\le (\E \Delta(X_i;\beta))^n-1$,
with $\Delta(X_i;\beta) = 2+4f(X_i^\top\beta)(f(X_i^\top\beta)-1)= 1+\tanh^2(X_i^\top\beta/2)$. By Lemma 6 of \cite{cai2020statistical}, we have $\Delta(X_i;\beta)\le 1+(X_i^\top\beta)^2$, so that
$
\E\Delta(X_i;\beta)\le 1+\sum_{j=1}^p\beta_j^2=1+\frac{1}{n}.
$
This implies that $\chi^2\le (1+\frac{1}{n})^n-1\le C$.
By Theorem 2.2 of \cite{tsybakov2009introduction}, we obtain
$
\inf_{\widehat{\xi}}\sup_{\theta\in\Theta_E(k,L_n)} P\bigg(|\widehat{\xi}-\xi|\gtrsim \frac{L_n}{\sqrt{n}}\bigg)\ge c,
$
for some constant $c>0$.

\emph{Case II. $ L_n\lesssim n^{-1/2}$.} In this case, we define $\mathcal{H}_0 = \{ (0,0,{I})\},$ and $\mathcal{H}_1 =\{ (\beta,\gamma,{ I}): \gamma=(L_n,0,...,0)^\top,\beta_1 =L_n, \beta_i=0, \forall i\ne 1\}$. It then follows form the similar argument as the proof in Case I that
$
\inf_{\widehat{\xi}}\sup_{\theta\in\Theta_E(k,L_n)} P\bigg(|\widehat{\xi}-\xi|\gtrsim L_n^2\bigg)\ge c,
$
for some constant $c>0$.

\emph{Case III. $L_n\gtrsim 1$.}
To prove the parametric rate lower bound, we define $\mathcal{H}_0 = \{ (e_1,e_1 ,{I_p})\},$ and $\mathcal{H}_1 =\{ (e_1,e_1,{ I_p+\frac{e_1e_1^\top}{\sqrt{n}L^2_n}})\}$. Then for $(\beta,\gamma,\Sigma)\in \HH_1$ and $(\beta',\gamma',\Sigma')\in \HH_0$, it holds that
$
|\beta^\top\Sigma\gamma-\beta'^\top\Sigma'\gamma'|=\frac{L_n^2}{\sqrt{n}}.
$
Denote
$
g=  \frac{1}{{(2\pi)^{np/2} }}\prod_{i=1}^n\exp\big\{ -\frac{1}{2}X_i^\top \Sigma^{-1}X_i \big\} \frac{\exp(y_iX_i^\top \beta)}{1+\exp(X_i^\top\beta)}
$
as the joint density with parameter $(\beta,\gamma,\Sigma)\in\mathcal{H}_1$. Also denote $g_0$  as the joint density with $(\beta,\gamma,\Sigma)\in\mathcal{H}_0$. By Lemma 7 of \cite{cai2018semi}, the $\chi^2$-divergence between distribution $g$ and $g_0$ satisfies
$
\chi^2(g,g_0) = \int \frac{g^2}{g_0} -1=\big( 1-\frac{1}{n}\big)^{-\frac{n}{2}}-1.
$
For sufficiently large $n$, we have $\chi^2(g,g_0) \le c$.
By Theorem 2.2 of \cite{tsybakov2009introduction}, we obtain
$
\inf_{\widehat{\xi}}\sup_{\theta\in\Theta_E(k,L_n)} P\big(|\widehat{\xi}-\xi|\gtrsim \frac{L^2_n}{\sqrt{n}}\big)\ge c,
$
for some constant $c>0$.

\section{Proofs of Auxiliary Lemmas}

\subsection{Proof of Lemma \ref{v.bnd.lem}}

For simplicity we write $\E$ in place of $\E_{\mathcal{D}_2}$. In addition, we write $(X'_{i\cdot},\beta',\widehat{\beta}')$ as $(X_{i\cdot},\beta,\widehat{\beta})$ since the arguments  are the same for  finite intercepts.

\bel \label{ratio.lem}
For all $1\le i\le n_1$, it holds that
\[
\exp(-|X_{i\cdot}^{\top}(\beta-\widehat{\beta})|)\le \frac{(1+\exp(X_{i\cdot}^{\top}\widehat{\beta}))^2\exp(X_{i\cdot}^{\top}{\beta})}{(1+\exp(X_{i\cdot}^{\top}{\beta}))^2\exp(X_{i\cdot}^{\top}\widehat{\beta})}\le \exp(|X_{i\cdot}^{\top}(\beta-\widehat{\beta})|).
\]
\eel
\begin{proof}
	The lemma can be proven using the following inequality from Example 8 of  \cite{huang2012estimation},
	$
	e^{|t|}\ge\frac{e^{t+\theta}(1+e^{\theta})^2}{e^{\theta}(1+e^{\theta+t})^2}\ge e^{-|t|}.
	$
\end{proof}

Since by Lemma \ref{lasso.bnd.lem}, with probability at least $1-p^{-c}$,
$
\|{\gamma}\|_2-\sqrt{\frac{k\log p}{n}}\lesssim\|\widehat{\gamma}\|_2\lesssim \|{\gamma}\|_2+\sqrt{\frac{k\log p}{n}},
$
we have, for $\|\gamma\|_2\gg \sqrt{k\log p/n}$, under the same event $\|{\gamma}\|_2\asymp \|\widehat{\gamma}\|_2$. Same results hold for $\beta$.
Note that
$
\frac{(1+\exp(X_{i\cdot}^{\top}\widehat{\beta}))^2}{\exp(X_{i\cdot}^{\top}\widehat{\beta})}= \exp(X_{i\cdot}^{\top}\widehat{\beta})+\exp(-X_{i\cdot}^{\top}\widehat{\beta})+2.
$
By Lemmas \ref{ratio.lem} and \ref{lasso.bnd.lem}, as long as $k\lesssim \frac{n}{\log p}$, with probability at least $1-p^{-c}$,
\begin{align}
	\E \eta_i(\widehat{\beta})(\widehat{\gamma}^\top X_{i\cdot})^2&\lesssim \E (e^{X_{i\cdot}^\top\widehat{\beta}}+e^{-X_{i\cdot}^\top\widehat{\beta}}+1)(\widehat{\gamma}^\top X_{i\cdot})^2 \exp(|X_{i\cdot}^\top(\beta-\widehat{\beta})|)\nonumber\\
	&\le \sqrt{\E (e^{X_{i\cdot}^\top\widehat{\beta}}+e^{-X_{i\cdot}^\top\widehat{\beta}}+1)^2(\widehat{\gamma}^\top X_{i\cdot})^4} \sqrt{\E \exp(2|X_{i\cdot}^\top(\beta-\widehat{\beta})|)}\nonumber\\
	&\le [\E(e^{X_{i\cdot}^\top\widehat{\beta}}+e^{-X_{i\cdot}^\top\widehat{\beta}}+1)^4]^{1/4}[\E(\widehat{\gamma}^\top X_{i\cdot})^8]^{1/4}\nonumber\\
	&\lesssim \|\widehat{\gamma}\|_2^2 \label{ub.eta.eq}
\end{align}
For the lower bound, by Lemma \ref{ratio.lem}, it follows that
$
\E \eta_i(\widehat{\beta})(\widehat{\gamma}^\top X_{i\cdot})^2\gtrsim \E\exp(-|X_{i\cdot}^\top(\beta-\widehat{\beta})|)(\widehat{\gamma}^\top X_{i\cdot})^2.
$
Note that $\E(\widehat{\gamma}^\top X_{i\cdot})^2\ge c \|\widehat{\gamma}\|_2^2$. It suffices to show that
$
\E [1-\exp(-|X_{i\cdot}^\top(\beta-\widehat{\beta})|)](\widehat{\gamma}^\top X_{i\cdot})^2\le \|\widehat{\gamma}\|_2^2c/2.
$
To see this, by Cauchy-Schwartz inequality,  with probability at least $1-p^{-c}$,
\begin{align*}
	&\E [1-\exp(-|X_{i\cdot}^\top(\beta-\widehat{\beta})|)](\widehat{\gamma}^\top X_{i\cdot})^2\le \sqrt{\E [1-\exp(-|X_{i\cdot}^\top(\beta-\widehat{\beta})|)]^2} \sqrt{\E(\widehat{\gamma}^\top X_{i\cdot})^4}\\
	&\lesssim \|\widehat{\gamma}\|_2^2\sqrt{\E |X_{i\cdot}^\top (\beta-\widehat{\beta})|^2}\lesssim \|\widehat{\gamma}\|_2^2\sqrt{\frac{k\log p}{n}}\le \|\widehat{\gamma}\|_2^2c/2
\end{align*}
for some $k\lesssim \frac{n}{\log p}$, 
where the second last inequality follows from Lemma \ref{lasso.bnd.lem}. 
For the upper bound for $\E[\widehat{ \beta}^\top(X_{i\cdot}X_{i\cdot}^\top-\Sigma)\widehat{\gamma}]^2$, we have
\beq \label{2nd.mom2}
\E (\widehat{\beta}^\top X_{i\cdot})^2(\widehat{\gamma}^\top X_{i\cdot})^2\le \sqrt{\E (\widehat{\beta}^\top X_{i\cdot})^4}\sqrt{\E (\widehat{\gamma}^\top X_{i\cdot})^4}\lesssim \|\widehat{\beta}\|_2^2\|\widehat{\gamma}\|_2^2,
\eeq
where the last inequality follows again from the subguassian property of the random design. The lower bound follows directly from assumption (A2).

\subsection{Proof of Lemma \ref{eq.lem}}

The following analysis is conditional on $\mathcal{D}_1$. Define $A_i=\widehat{\beta}^\top X_{i\cdot}$ and $B_i=\widehat{\gamma}^\top X_{i\cdot}$. Then we have
$
\E ( \widehat{\beta}^\top (X_{i\cdot}X_{i\cdot}^\top -\Sigma)\widehat{\gamma})^2=\E(A_iB_i-\E A_iB_i)^2,
$ and
$
\frac{1}{n}\sum_{i=1}^n({\widehat{\beta}}^\top X_{i\cdot}X_{i\cdot}^\top{\widehat{\gamma}}-{\widehat{\beta}}\widehat{\Sigma}{\widehat{\gamma}})^2=\frac{1}{n}\sum_{i=1}^n \big(A_iB_i-\frac{1}{n}\sum_{i=1}^n A_iB_i\big)^2.
$
We have
\begin{align*}
	&\frac{1}{n}\sum_{i=1}^n({\widehat{\beta}}^\top X_{i\cdot}X_{i\cdot}^\top{\widehat{\gamma}}-{\widehat{\beta}}\widehat{\Sigma}{\widehat{\gamma}})^2-\E ( \widehat{\beta}^\top (X_{i\cdot}X_{i\cdot}^\top -\Sigma)\widehat{\gamma})^2\\
	&=\frac{1}{n}\sum_{i=1}^n(A_i^2B_i^2-\E A_i^2B_i^2)-2\E A_iB_i\cdot \frac{1}{n}\sum_{i=1}^n(A_iB_i-\E A_iB_i)-\bigg( \frac{1}{n}\sum_{i=1}^n(A_iB_i-\E A_iB_i)\bigg)^2.
\end{align*}
Note that $\E A_iB_i\asymp \|\widehat{\beta}\|_2\|\widehat{\gamma}\|_2$. It suffices to obtain upper bounds for $\frac{1}{n}\sum_{i=1}^n(A_i^2B_i^2-\E A_i^2B_i^2)$ and $\frac{1}{n}\sum_{i=1}^n(A_iB_i-\E A_iB_i)$. By concentration inequality for sub-exponential random variables, we have
\beq \label{AB}
P\bigg( \frac{1}{n}\sum_{i=1}^n(A_iB_i-\E A_iB_i)\ge \|\widehat{\beta}\|_2\|\widehat{\gamma}\|_2\sqrt{\frac{\log n}{n}} \bigg)\le n^{-c}.
\eeq
By concentration inequality for sub-exponential random variables, we have
$
\sum_{i=1}^n P(|A_iB_i|\ge C\sqrt{\log n})\le n^{-c}.
$
Now define $\bar{C}_i=A_iB_i1\{|A_iB_i|\le C\sqrt{\log n}\}$ and $\tilde{C}_i=A_iB_i1\{|A_iB_i|\ge C\sqrt{\log n}\}$. We have
$
\frac{1}{n}\sum_{i=1}^n(A_i^2B_i^2-\E A_i^2B_i^2)=\frac{1}{n}\sum_{i=1}^n(\bar{C}^2_i-\E \bar{C}^2_i)+\frac{1}{n}\sum_{i=1}^n(\tilde{C}_i^2-\E\tilde{C}_i^2).
$
To control the first term, we apply the following lemma obtained in \cite{cai2011adaptive}.

\bel \label{conc.lem}
Let $\xi_1,...,\xi_n$ be independent centred random variables. Suppose that there exists some $\eta>0$ and $M_n$ such that $\sum_{i=1}^n \E \xi_i^2\exp(\eta|\xi_i|)\le M_n^2$. Then for $0<t\le M_n$,
$
P(\sum_{i=1}^n\xi_i\ge C_\eta M_n t)\le e^{-t^2}
$
where $C_\eta=\eta+\eta^{-1}$.
\eel

Taking $\eta\asymp \frac{1}{\log^2 n}$, we have
$
\sum_{i=1}^n(\bar{C}^2_i-\E \bar{C}^2_i)^2 \exp(\eta|\bar{C}^2_i-\E \bar{C}^2_i|)\le C\sum_{i=1}^n \E(\bar{C}^2_i-\E \bar{C}^2_i)^2 \lesssim \|\widehat{\gamma}\|_2^4\|\widehat{\beta}\|_2^4 n.
$
By Lemma \ref{conc.lem} with $M_n\asymp \|\widehat{\gamma}\|_2^2\|\widehat{\beta}\|_2^2\sqrt{n}$ and $t=\sqrt{\log n}$, we have
$
P\big( \frac{1}{n}\sum_{i=1}^n (\bar{C}^2_i-\E \bar{C}^2_i)\ge C \|\widehat{\gamma}\|_2^2\|\widehat{\beta}\|_2^2\frac{\log^{5/2}n}{\sqrt{n}}\big) \lesssim n^{-c}.
$
For the expectation $\E\tilde{C}_i^2$, we have
$
\E\tilde{C}_i^2\le \sqrt{\E[ A_i^4B_i^4 ]P(|A_iB_i|\ge C\sqrt{\log n})}\lesssim \|\widehat{\beta}\|_2^2\|\widehat{\gamma}\|_2^2n^{-c}.
$
Combining the above inequalities, 
$
P\big( \frac{1}{n}\sum_{i=1}^n(A_i^2B_i^2-\E A_i^2B_i^2)\ge C\|\widehat{\beta}\|_2^2\|\widehat{\gamma}\|_2^2\frac{\log^{5/2}}{\sqrt{n}} \big)\le \sum_{i=1}^nP(|A_iB_i|\ge C\sqrt{\log n})
+\big( \frac{1}{n}\sum_{i=1}^n (\bar{C}^2_i-\E \bar{C}^2_i)\ge C \|\widehat{\gamma}\|_2^2\|\widehat{\beta}\|_2^2\frac{\log^{5/2}n}{\sqrt{n}} \big)+P\big(\E\tilde{C}_i^2\ge C \|\widehat{\gamma}\|_2^2\|\widehat{\beta}\|_2^2\frac{\log^{5/2}n}{\sqrt{n}} \big)\lesssim n^{-c}.
$
Thus, the above inequality along with (\ref{AB}) and the upper bounds for $\|\widehat{\beta}\|_2\|\widehat{\gamma}\|_2$ implies (\ref{eq1}).

\section{Supplementary Tables and Figures} \label{supp.sec}

\setcounter{table}{0}
\renewcommand{\thetable}{S\arabic{table}}

\subsection{Additional Details of Simulation Setup}

\paragraph{Covariance matrix.} In Section 6 of our main paper, we carried out simulations that compare different methods. In one of the settings, the design covariates were generated from a multivariate Gaussian distribution, whose covariance matrix is a blockwise diagonal matrix of 10 identical unit diagonal Toeplitz matrices as follows
\[
\Sigma_M=\begin{bmatrix}
	1& \frac{3(p-2)}{10(p-1)}&\frac{3(p-3)}{10(p-1)}& ...& \frac{3}{10(p-1)} & 0\\
	\frac{3(p-2)}{10(p-1)} & 1 & \frac{3(p-2)}{10(p-1)}&...&\frac{6}{10(p-1)} & \frac{3}{10(p-1)}\\
	\vdots &&\ddots&&&\\
	0& \frac{3}{10(p-1)} &  \frac{6}{10(p-1)} &...&\frac{3(p-2)}{10(p-1)}&1
\end{bmatrix}.
\]
\paragraph{Selection of tuning parameter.}  Although our theoretical analysis ensured the asymptotic validity of our inference procedure whenever $\lambda=C\sqrt{\log p/n}$ for some constant $C>0$, there is currently no theoretical results on how to determine the optimal value for $C$ under finite samples.  In our numerical implementations, we used cross-validation to determine $C$. We found that $C\approx 0.12$ was oftentimes suggested by cross-validation, and led to desirable and stable numerical performance over a wide range of settings, including all our  simulation scenarios. Thus we set $C=0.12$ for convenience and to speed up computation.  

\subsection{Simulation for Parameter Estimation} 

Table \ref{table:S02} and \ref{table:S01} below show the square roots of the empirical mean square errors for each estimator based on 500 rounds of simulations for each setting. It can be seen that each of the proposed estimators outperformed the alternative estimators in all the settings. In particular, in Table \ref{table:S02}, plg and lpj were shown to badly estimate $\beta^\top\Sigma\gamma$ and $\beta^\top\Sigma\beta$ individually, but not so much for $R$. To understand this, since $R$ is the ratio between $\beta^\top\Sigma\gamma$ and $\sqrt{\beta^\top\Sigma\beta\gamma^\top\Sigma\gamma}$, large errors in the estimators of  $\beta^\top\Sigma\gamma$ and $\sqrt{\beta^\top\Sigma\beta\gamma^\top\Sigma\gamma}$  may not necessarily lead to a very large error in the final estimator for $R$, after taking the ratio. For example, if we define 
\[
E_1 = \widehat{\beta^\top\Sigma\gamma}-\beta^\top\Sigma\gamma,\qquad E_2=\sqrt{\widehat{\beta^\top\Sigma\beta}\widehat{\gamma^\top\Sigma\gamma}}-\sqrt{\beta^\top\Sigma\beta\gamma^\top\Sigma\gamma},
\]
a simple calculation yields
\begin{align*}
	|\widehat R-R|=\bigg|\frac{\widehat{\beta^\top\Sigma\gamma}}{\sqrt{\widehat{\beta^\top\Sigma\beta}\widehat{\gamma^\top\Sigma\gamma}}}-R\bigg|\le \bigg|\frac{E_1}{E_2}-R\bigg|
\end{align*}
Apparently, the error of $\widehat R$ can still be small even if $E_1$ and $E_2$ are individually large, as long as $E_1/E_2$ is close to $R$.

\begin{table}
	\centering
	\caption{Estimation errors with block-diagonal covariance $\Sigma=\Sigma_B$ and $k=25$. pro: proposed estimators; plg: simple plug-in estimators; lpj:  component-wise projected Lasso estimators; rpj:  the component-wise projected Ridge estimators.} 
		\begin{tabular}{c|cccc|cccc|cccc}
			\hline
			\multirow{2}{1em}{$p$}&\multicolumn{4}{c|}{$\beta^\top\Sigma\gamma$} &\multicolumn{4}{c|}{$\beta^\top\Sigma\beta$} &\multicolumn{4}{c}{$R$}\\
			\cline{2-13}
			&   pro & plg & lpj &  rpj & pro & plg &  lpj &  rpj &   pro & plg & lpj &  rpj \\  
			\hline
			&\multicolumn{12}{c}{$n=200$}\\
			700& 1.4 &17.7  &7.3 &2.5  &2.4 &55.4 &43.2  &7.5  &0.14 &0.18  &0.20 &0.23\\
			800 & 1.3 &22.0  &8.9 &2.3  &2.1 &67.8 &59.8  &7.5  &0.15  &0.20  &0.21 &0.23\\
			900&  1.1 &26.8 &12.4  &2.2  &2.5 &76.6 &72.6 &7.3  &0.12  &0.19 &0.19  &0.23\\
			1000 & 1.2 &29.5 &10.9  &2.1 &2.5 &87.8 &88.3  &7.4  &0.13  &0.21  &0.19 &0.24\\
			&\multicolumn{12}{c}{$n=300$}\\
			700&1.0 &14.7  &3.5  &2.0 &1.7 &43.2 &20.3  &6.7 &0.11  &0.19 & 0.18  &0.19\\
			800 &  1.0 &18.8  &4.6  &1.8  &2.4 &47.9 &26.0 &6.9  &0.12  &0.21  &0.17  &0.22\\
			900&   1.0 &17.3  &4.7  &2.0  &2.8 &53.8 &36.3  &7.7  &0.12  &0.19  &0.17  &0.19\\
			1000 & 0.9 &23.5  &6.7  &2.1  &2.0 &61.4 &37.6  &6.6  &0.11  &0.19  &0.18  &0.21\\
			&\multicolumn{12}{c}{$n=400$}\\
			700& 0.8 &10.0  &1.4  &1.4  &1.3 &35.1 &11.6 &5.9  &0.10  &0.19  &0.14  &0.14\\
			800 &   0.8 &13.0  &2.9  &1.8  &2.4 &40.5 &17.2 &7.1  &0.10  &0.19  &0.15  &0.15\\
			900&  0.8 &17.2 & 2.1  &1.7  &3.0 &45.5 &19.0  &6.1  &0.10  &0.21  &0.17  &0.17\\
			1000 & 1.4& 21.2  &3.5  &2.6  &2.4 &54.1 &21.9  &6.7 & 0.12  &0.21 & 0.19  &0.17\\
			%
			\hline
		\end{tabular}
		\label{table:S02}
	\end{table}

	\begin{table}
		\caption{Estimation errors with exchangeable covariance matrix $\Sigma=\Sigma_E$ and sparsity parameter $k=25$. \label{table:S01}}
		\centering
		\begin{tabular}{c|cccc|cccc|cccc}
			\hline
			\multirow{2}{1em}{$p$}&\multicolumn{4}{c|}{$\beta^\top\Sigma\gamma$} &\multicolumn{4}{c|}{$\beta^\top\Sigma\beta$} &\multicolumn{4}{c}{$R$}\\
			\cline{2-13}
			&   pro & plg &  lpj &  rpj  & pro & plg &  lpj &  rpj  &   pro & plg & lpj &  rpj  \\  
			\hline
			&\multicolumn{12}{c}{$n=200$}\\
			700& 0.8&  1.3&  72.7&  1.9&  1.7&  7.2& 171.7&  6.7&  0.09&  0.15&  0.26&   0.21\\
			800 &  1.1&  1.4& 163.7&  1.9&  1.3& 6.6& 273.9&  7.2&  0.13&  0.18&   0.30&   0.25\\
			900&  1.1&  1.6& 195.2& 2.2&  1.7&  6.9& 326.8&   7.1&  0.13&  0.17&  0.31&   0.24\\
			1000 & 1.1& 1.9 &  135.2&   1.8&   1.5& 7.0&  440.3&   7.0&   0.11&   0.17&  0.28& 0.23\\
			&\multicolumn{12}{c}{$n=300$}\\
			700& 0.8   & 1.1 & 14.1  & 1.7  &1.4&  6.1  &49.9  & 5.8 & 0.10  & 0.14  & 0.20 &  0.19\\
			800 &  0.8 & 1.2 & 22.6  & 1.9  & 1.2 & 6.0 &64.1  & 6.0  & 0.10  & 0.13 & 0.19  & 0.21\\
			900&  0.9& 1.4 & 13.6  & 1.5  & 1.1& 6.0 &135.7  &6.4   &0.10  &0.14  & 0.17  & 0.17\\
			1000 & 0.8& 1.3  &76.8  &2.0  & 1.4 & 6.4 &159.1  & 6.7  & 0.08  & 0.14   &0.19   & 0.21\\
			&\multicolumn{12}{c}{$n=400$}\\
			700& 0.6 & 1.2 &5.4  & 1.5  & 1.6&  5.8  &22.5  & 6.4  &0.07  & 0.09  & 0.15   &0.15\\
			800 &  0.6 & 1.2  &6.4  & 1.8  &1.3& 6.0 & 32.2  & 5.8  & 0.07  & 0.12  & 0.16 &  0.16\\
			900&   0.7 & 1.0  &12.3   &1.5  & 1.1 & 6.1  &69.3  & 6.0  &0.08  & 0.13  & 0.15  & 0.16\\
			1000 &  0.7 & 1.0 &17.8  & 1.5  & 0.9 & 6.1 &118.3  & 5.9  & 0.09  & 0.13  & 0.15  & 0.15\\
			\hline
		\end{tabular}
	\end{table}

	\subsection{Simulations for Confidence Intervals} 
	
	Table  \ref{table:t02} evaluates the empirical performance of the proposed CIs and the plg-based CIs under exchangeable covariance structure $\Sigma=\Sigma_E$ under various settings described in Section 6 of the main paper.   We observe that in both Table  \ref{table:t02} and Table 4 of the main paper, when the sample size increases from 300 to 500, the empirical coverage probability of the proposed CIs for the genetic covariance and correlation parameters seems to  inflate and become slightly larger than the nominal level. This is likely due to the limitation of our empirically determined tuning parameter, which as a practical proxy necessarily differs from the underlying theoretically optimal value, and therefore may lead to results that slightly deviate from our theoretical prediction. Nevertheless, even though the proposed method became a little more conservative, the length of the CIs became shorter under larger $n$, and its advantage over the alternative methods was notable.
	
	Tables \ref{table:t1} and \ref{table:t2} show the averaged coverage probabilities and lengths of the rpj-based bootstrap confidence intervals introduced in Section 6.2 of the main paper, based on 500 rounds of simulation for each setting.

	\begin{table}
		\caption{Coverage  and length of the CIs with exchangeable covariance $\Sigma=\Sigma_E$, $\alpha=0.05$ and sparsity $k=25$. Coverage is denoted as cov (\%) and length is denoted as len.	\label{table:t02}}
		\centering
		\begin{tabular}{c|cccc|cccc|cccc}
			\hline
			\multirow{3}{1em}{$p$}&\multicolumn{4}{c|}{$\beta^\top\Sigma\gamma$} &\multicolumn{4}{c|}{$\beta^\top\Sigma\beta$} &\multicolumn{4}{c}{$R$}\\
			\cline{2-13}
			& \multicolumn{2}{c|}{ pro} &\multicolumn{2}{c|}{boot(plg)} &\multicolumn{2}{c|}{ pro} &\multicolumn{2}{c|}{boot(plg)} &\multicolumn{2}{c|}{ pro} &\multicolumn{2}{c}{boot(plg)} \\
			\cline{2-13}
			&   cov & len &cov& len&  cov& len  &   cov& len &   cov& len&   cov& len \\  
			\hline
			&\multicolumn{12}{c}{$n=300$}\\
			700 &   95.2 &  5.37&  54.2& 2.07 & 92.0&  7.02  & 8.4& 2.52 & 97.0 & 0.31 &72.2& 0.38 \\
			800&  96.4 & 5.34&  49.8 &2.11 &94.6&  6.67 &8.8& 2.66& 95.2 & 0.30 &71.2 & 0.38 \\
			900 &  93.4& 5.33 & 52.0& 1.98&  92.8&  6.94  &11.2& 2.61 & 93.4 & 0.32 &75.5& 0.37\\
			1000 &  95.2 & 4.73& 45.6 & 1.99 &91.8&  6.46 & 7.4& 2.53 &94.6 & 0.29  &70.2& 0.37 \\
			&\multicolumn{12}{c}{$n=400$}\\
			700 &   96.6 &  5.28&  63.6& 2.35 & 94.4 &  6.75 &21.4& 2.87  & 97.0 & 0.30& 77.6& 0.39\\
			800 &  97.4 &  4.99& 55.2& 2.32&  92.4& 6.69   & 13.4 & 2.82 & 97.0& 0.28 &73.0& 0.38 \\
			900 &  97.0&  5.16&  50.8& 2.22 & 90.6 &  6.89   &13.6& 2.87 & 97.6& 0.31 &70.4& 0.38 \\
			1000 &  95.6 & 4.87  &59.6&2.27 & 91.6&  6.49 &19.0 & 3.07   & 95.4& 0.28  &80.0& 0.38\\
			&\multicolumn{12}{c}{$n=500$}\\
			700 &   98.2 & 5.03& 54.0& 2.62 & 95.6 & 6.42  &32.4& 3.40   & 99.6 & 0.27& 77.6& 0.39\\
			800 &  98.4 &  4.67 & 59.2& 2.55& 94.4& 6.55& 32.0& 3.40& 97.2& 0.28 & 77.6& 0.39\\
			900&  99.2&  4.76 & 64.4& 2.46 & 91.4 & 6.37 &27.0& 3.10 & 99.0& 0.28 & 76.8& 0.38 \\
			1000 &  96.4 & 5.15 &64.0& 2.47& 93.6&  6.71  &26.8& 3.28 & 98.2 & 0.29 &76.6& 0.39\\
			\hline
		\end{tabular}
	\end{table}

	\begin{table}[h!]
		\centering
		\caption{Coverage  and length of the rpj-based bootstrap CIs with $\Sigma=\Sigma_B$,  $\alpha=0.05$ and $k=25$. Coverage is denoted as cov (\%) and length is denoted as len.}
		\begin{tabular}{c|c|cc|cc|cc}
			\hline
			\multirow{3}{1em}{$n$}&\multirow{3}{1em}{$p$}&\multicolumn{2}{c|}{$\beta^\top\Sigma\gamma$} &\multicolumn{2}{c|}{$\beta^\top\Sigma\beta$} &\multicolumn{2}{c}{$R$}\\
			\cline{3-8}
			&&   cov & len &cov& len&  cov& len   \\  
			\hline
			&600&   7.6&    0.78&    18.8&    6.74&    80.8&    0.51 \\
			&700 & 6.6&    0.80&   18.8&   6.29&   76.0&    0.50 \\
			300&800&  10.0&    0.74&    19.6&    6.13&    85.6&    0.50\\
			&900 & 8.6&    0.71&    17.6&    6.03&   71.8&    0.48 \\
			&1000 &  6.2&    0.68&    15.8&    6.04&   84.8&    0.47 \\
			\cline{1-8}
			&600&  10.2&   0.84&  13.8&   6.40&   87.2&   0.53\\
			&700 &  8.2&    0.64&    13.6&    6.07&    80.0&    0.52\\
			400&800 &  9.2&    0.71&    9.0&    6.04&    85.8&   0.51\\
			&900 & 10.4&    0.76&    7.4&    6.07&    76.6&    0.49\\
			&1000 & 12.8&    0.66&    10.6&    6.01&   78.4&    0.51\\
			\cline{1-8}
			&600&    8.2&    0.87&    20.4&    7.17&    83.6&    0.52\\
			&700 &  10.6&    0.82&    10.2&    6.64&    84.2&    0.51 \\
			500&800 & 11.4&    0.53&   5.2&   5.24&    84.6&    0.49\\
			&900& 10.8&    0.74&   11.6&    5.73&    79.6&    0.50\\
			&1000 & 13.6&    0.60&    6.4&    5.72&    81.8&    0.49\\
			\hline
		\end{tabular}	\label{table:t2}
	\end{table}

	\begin{table}[h!]
		\centering
		\caption{Coverage  and length of the rpj-based bootstrap CIs with $\Sigma=\Sigma_E$, $\alpha=0.05$ and $k=25$. Coverage is denoted as cov (\%) and length is denoted as len.}
		\begin{tabular}{c|c|cc|cc|cc}
			\hline
			\multirow{3}{1em}{$n$}&\multirow{3}{1em}{$p$}&\multicolumn{2}{c|}{$\beta^\top\Sigma\gamma$} &\multicolumn{2}{c|}{$\beta^\top\Sigma\beta$} &\multicolumn{2}{c}{$R$}\\
			\cline{3-8}
			&&   cov & len &cov& len&  cov& len   \\  
			\hline
			&600&    18.6&    0.84&   17.2&    6.41&    78.8&    0.52\\
			&700 &  13.0&    0.78&   17.8&    6.46&    78.6&    0.51 \\
			300&800&  16.4&    0.75&    15.0&    6.26&    81.0&    0.53 \\
			&900 & 10.8&    0.69&   11.2&    5.74&    81.0&    0.52\\
			&1000 &  13.2&    0.88&    8.6&    5.74&    84.4&    0.52 \\
			\cline{1-8}
			&600&   18.2&    0.86&    17.6&    6.71&    81.4&    0.56\\
			&700 &  14.0&    0.83&    12.6&   6.53&    80.2&    0.53\\
			400&800 &  11.8&    0.70&    7.8&    5.47&    75.6&    0.54 \\
			&900 &  11.2    &0.72&    12.0&    5.77&    83.0&    0.52 \\
			&1000 &  21.2&    0.91&    10.2&    5.86&    81.2&    0.53\\
			\cline{1-8}
			&600&    14.0   & 0.86&    10.4&    6.23&    86.6&    0.55 \\
			&700 &   12.4 &   0.78&    9.8&    5.95&    82.4&    0.54\\
			500&800 & 13.4&    0.87  &  10.6&    6.00&   92.8 &   0.53\\
			&900&  9.6&    0.67&    9.4&    5.43&    83.2&    0.52 \\
			&1000 & 19.0&    0.86&    4.2&    5.71&    85.2&    0.55\\
			\hline
		\end{tabular}	\label{table:t1}
	\end{table}

	\subsection{Simulation for Hypotheses Testing}

	Same as the previous simulations, the random covairates are drawn from a multivariate Gaussian distribution with a covariance matrix $\Sigma_B$ or $\Sigma_E$. For the regression coefficients, given the support $\mathcal{S}$, we generate $\beta_j$ and $\gamma_j$ randomly from $[-1,1]$ for all $j\in \mathcal{S}$ such that $|\beta^\top \Sigma\gamma|>3$. We compare the empirical type I errors and statistical powers of our proposed tests and the bootstrap tests based on the plg estimators, which can be obtained by inverting the bootstrap confidence intervals from the main paper. Specifically, the type I errors are evaluated under the null hypotheses where the functionals take their true values, while the statistical powers are calculated based on the null hypotheses where $B_0=Q_0=R_0=0$. From Table \ref{table:t3} and Table \ref{table:t03}, we see that,  across all the settings,  the proposed tests have type I error around the nominal level 0.05, whereas the bootstrap tests have much inflated type I errors, which, due to the fundamental bias of the plg estimators, leads to higher statistical powers. The statistical power of the proposed tests increases as sample size grows, and their performance is stable across the two covariance structures.

	\begin{table}
		\centering
		\caption{Type I errors and powers with $\Sigma=\Sigma_B$, $\alpha=0$.$05$ and sparsity $k=25$. pro: proposed tests; boot: the plg-based bootstrap tests.} 	\label{table:t3}
		\begin{tabular}{c|cccc|cccc|cccc}
			\hline
			\multirow{3}{1em}{$p$}&\multicolumn{4}{c|}{$\beta^\top\Sigma\gamma$} &\multicolumn{4}{c|}{$\beta^\top\Sigma\beta$} &\multicolumn{4}{c}{$R$}\\
			\cline{2-13}
			& \multicolumn{2}{c|}{ type I error} &\multicolumn{2}{c|}{power} &\multicolumn{2}{c|}{type I error} &\multicolumn{2}{c|}{power} &\multicolumn{2}{c|}{ type I error} &\multicolumn{2}{c}{power} \\
			\cline{2-13}
			&   pro & boot &pro & boot&  pro & boot &   pro & boot & pro & boot&   pro & boot \\  
			&\multicolumn{12}{c}{$n=300$}\\
			600&   0.05 & 0.93  &0.42& 0.85 &  0.04& 0.95 &0.92& 1.00 & 0.03 & 0.36 &0.42& 0.85 \\
			700 &   0.07 & 0.89 &0.42& 0.85 & 0.06& 0.92 &0.93& 1.00 &0.03& 0.39 &0.43& 0.85\\
			800&  0.07& 0.96 &0.47& 0.84& 0.09&  0.91 &0.92&1.00  & 0.04& 0.34&0.47&0.84  \\
			900 &  0.08&  0.96 &0.51& 0.79 & 0.07&  0.96 &0.92& 1.00   & 0.04& 0.40& 0.51&0.79\\
			&\multicolumn{12}{c}{$n=400$}\\
			600&    0.05 & 0.81 &0.52&0.83&  0.04 & 0.87  &0.95& 1.00 & 0.02 & 0.35& 0.52& 0.83\\
			700 &   0.04 &  0.84&0.52& 0.84& 0.06&  0.86 &0.89& 1.00  & 0.01 & 0.36 &0.53&0.84\\
			800 &  0.05 & 0.87& 0.51& 0.87& 0.06&  0.86  &0.92& 1.00 & 0.03& 0.41& 0.50&0.88 \\
			900 &  0.07& 0.82 & 0.54 & 0.83& 0.07&  0.90 & 0.93& 1.00 & 0.04 & 0.37 & 0.54&0.83\\
			&\multicolumn{12}{c}{$n=500$}\\
			600&    0.03  & 0.77&  0.57& 0.84& 0.04 & 0.75  &0.89& 1.00 & 0.01 & 0.38 &0.57& 0.83\\
			700&   0.02& 0.76& 0.54& 0.80& 0.05 &  0.82 &0.88& 1.00  & 0.01& 0.36& 0.54& 0.81\\
			800&  0.03&  0.81 & 0.60& 0.85 & 0.04& 0.78  &0.93& 1.00   & 0.02& 0.37 &0.60& 0.85 \\
			900 &  0.03&  0.85 & 0.61 & 0.81 &0.04 &  0.82   & 0.92& 1.00 & 0.02& 0.31& 0.61 & 0.81 \\
			\hline
		\end{tabular}
	\end{table}
	
	\begin{table}
		\caption{Type I errors and powers with exchangeable covariance $\Sigma=\Sigma_E$, $\alpha=0.05$ and sparsity $k=25$.	\label{table:t03}}
		\centering
		\begin{tabular}{c|cccc|cccc|cccc}
			\hline
			\multirow{3}{1em}{${p}$}&\multicolumn{4}{c|}{$\beta^\top\Sigma\gamma$} &\multicolumn{4}{c|}{$\beta^\top\Sigma\beta$} &\multicolumn{4}{c}{$R$}\\
			\cline{2-13}
			& \multicolumn{2}{c|}{ type I error} &\multicolumn{2}{c|}{power} &\multicolumn{2}{c|}{type I error} &\multicolumn{2}{c|}{power} &\multicolumn{2}{c|}{ type I error} &\multicolumn{2}{c}{power} \\
			\cline{2-13}
			&   pro & boot &pro & boot&  pro & boot &   pro & boot & pro & boot&   pro & boot \\  
			\hline
			&\multicolumn{12}{c}{$n=300$}\\
			600&   0.11 & 0.72  &0.44& 0.84 &  0.08& 0.88 &0.91& 1.00 & 0.06 & 0.30 &0.44& 0.84 \\
			700 &   0.12 & 0.71 &0.47& 0.86 & 0.10& 0.92  &0.96& 1.00 &0.06& 0.27 &0.48& 0.86\\
			800&  0.14& 0.71 &0.51& 0.82& 0.12&  0.86 &0.97&1.00  & 0.07& 0.28&0.51&0.82  \\
			900 &  0.14&  0.70 &0.47& 0.83 & 0.11&  0.91 &0.93& 1.00   & 0.08& 0.24& 0.47&0.83\\
			&\multicolumn{12}{c}{$n=400$}\\
			600&    0.05 & 0.57 &0.54&0.87&  0.05 & 0.78  &0.94& 1.00 & 0.03 & 0.23& 0.54& 0.87\\
			700 &   0.07 &  0.56&0.51& 0.85& 0.09&  0.77 &0.92& 1.00  & 0.03 & 0.27 &0.51&0.85\\
			800 &  0.08 & 0.56& 0.57& 0.81& 0.05&  0.79  &0.94& 1.00 & 0.03& 0.28& 0.57&0.81 \\
			900 &  0.11& 0.59 & 0.58 & 0.80& 0.10&  0.88 & 0.96& 1.00 & 0.07 & 0.30 & 0.84&0.79\\
			&\multicolumn{12}{c}{$n=500$}\\
			600&    0.04  & 0.45&  0.57& 0.86& 0.03 & 0.59  &0.96& 1.00 & 0.01 & 0.22 &0.57& 0.86\\
			700&   0.04& 0.47& 0.58& 0.81& 0.04 &  0.61 &0.94& 1.00  & 0.03& 0.28& 0.58& 0.81\\
			800&  0.05&  0.50 & 0.62& 0.83 & 0.02& 0.74  &0.96& 1.00   & 0.02& 0.25 &0.62& 0.83 \\
			900 &  0.06&  0.48 & 0.61 & 0.87 &0.04 &  0.72   & 0.94& 1.00 & 0.04& 0.19& 0.61 & 0.86 \\
			\hline
		\end{tabular}
	\end{table}

	\subsection{Additional simulations based on simulated genetic data}
	
\begin{figure}
		\centering
		\includegraphics[angle=0,width=7cm]{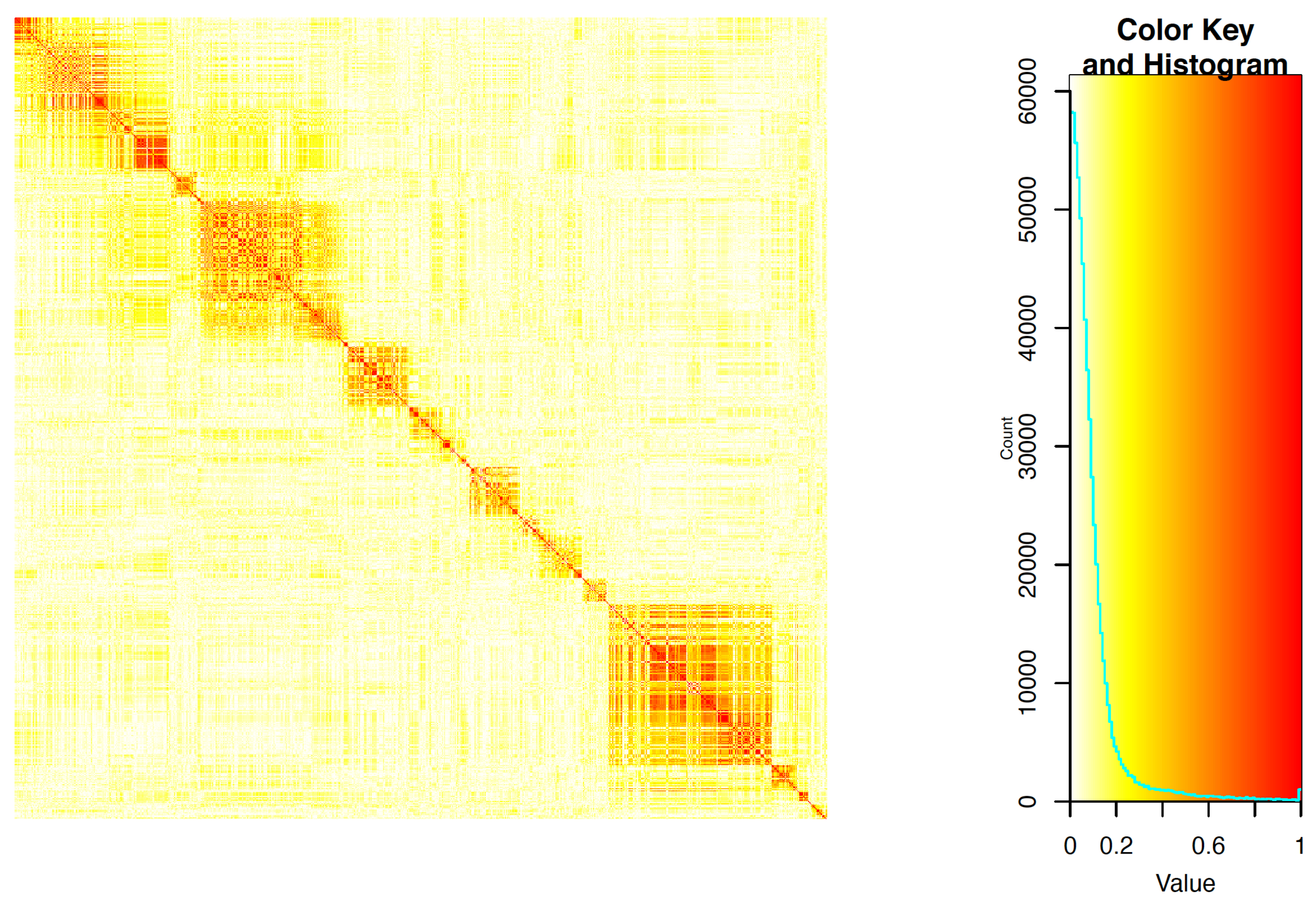}
		\label{cor.fig}
		\caption{Correlation of the simulated genotypes with $n=200$ and $p=600$.} 
	\end{figure}

	To justify our method applied to real genetic datasets where the binary traits may be associated to SNPs on multiple chromosomes, 
	we generated genotypes of $2n$ unrelated individuals containing $2p$ SNPs based on the aforementioned reference haplotype map, and the sequencing data over two different regions, one from chromosome 9 (GrCH37: bp 40,900 to bp 2,000,000) and the other from chromosome 10 (GrCH37: bp 7,000 to bp 2,000,000), of 503 European samples from the 1000 Genomes Project Phase 3, with $p$ SNPs from each region.  
	For the two binary traits, we let one trait be associated with 50 SNPs from the above regions with 25 SNPs for each chromosome, and let the other trait be associated only with 25 SNPs from chromosome 10, among which 12 SNPs are shared between two traits. The true effect sizes were generated in the same manner as in previous simulations, with the additional constraints that,  the absolute value of genetic covariance over 
	chromosome 10 is more than 3. After obtaining the simulated data, we apply the proposed test at a chromosomal basis to detect the genetic covariance/correlation between two traits.  In Table \ref{table:t9}, the proposed  tests under the null hypothesis $B_0=R_0=0$ were evaluated based on 500  simulations. Our results indicate the usefulness of our proposed methods in producing valid inferences from real genetic data. Figure \ref{cor.fig} shows the correlation matrix of the above generated genotypes ($n=200, p=600$) corresponding to chromosome 9, which indicates significant LD structure. 
	
	\begin{table}
		\centering
		\caption{\footnotesize Type I errors and powers for simulated genetic data with $\alpha=0$.$05$.} 	
			\begin{tabular}{c|cccc|cccc}
				\hline
				\multirow{3}{1em}{${p}$}&\multicolumn{4}{c|}{$\beta^\top\Sigma\gamma$} &\multicolumn{4}{c}{$R$}\\
				\cline{2-9}
				& \multicolumn{2}{c|}{type I error (chr9)} &\multicolumn{2}{c|}{power (chr10)} &\multicolumn{2}{c|}{type I error (chr9)} &\multicolumn{2}{c}{power (chr10)} \\
				\cline{2-9}
				&   $n=300$& 400 & 300 & 400 & 300 & 400&   300 & 400 \\  
				\hline
				500 &   0.02 & 0.03 &0.41& 0.50 & 0.02& 0.03  &0.41& 0.50 \\
				600&  0.02& 0.04 &0.47& 0.50& 0.02&  0.04 &0.47 & 0.50    \\
				700 &  0.02&  0.02 &0.49& 0.54 & 0.02&  0.02 &0.49& 0.54 \\
				800 &  0.02&  0.03 &0.45& 0.56 & 0.02&  0.03 &0.45& 0.56  \\
				\hline
			\end{tabular}
			\label{table:t9}
		\end{table}

		\paragraph{Numbers of SNPs per chromosome.} In Table \ref{table:s1}, we list the numbers of SNPs in each of the 22 autosomes that were used for our data analysis in the main paper.
		
		\begin{table}
			\centering
			\caption{Numbers of SNPs on each autosome}
			\begin{tabular}{c|c|c|c}
				\hline
				Chromosome& \# of SNPs & 	Chromosome& \# of SNPs  \\
				\hline
				1&36760&	12&23560\\
				2& 39339&	13&18040 \\
				3& 32977& 	14&16111\\
				4& 28852& 	15&14510\\
				5& 30078& 	16&14857\\
				6& 31575& 	17&12710\\
				7&26175& 	18&14692\\
				8&27248& 	19&8345\\
				9& 23387& 	20&12444\\
				10&25384& 			21&7217\\
				11&23732& 		21&7217
			\end{tabular}	\label{table:s1}
		\end{table}

		\subsection{Improved Power by Modeling Binary Outcomes}
		
		To demonstrate the advantage of our logistic regression modelling of binary outcomes, and the potential limitations of  treating the binary outcomes as continuous variables, in this section, we carry out additional simulations to evaluate the performance of a linear model counterpart of our proposed inference procedure, where $\beta$ and $\gamma$ were defined as the coefficient vectors of two linear regressions, respectively. Specifically, we  followed the same argument in Section 2.2 of the paper, and by replacing the expit function $h(u)=\frac{e^u}{1+e^u}$ from the logistic regression by the identity map $h(u)=u$, we derived debiased estimators for the corresponding parameters of interest under the linear regression model. These debiased estimators under the linear regression model are given by $
		\widehat{\beta^{\top}\Sigma\gamma}=\widehat{\beta}^{\top} \widehat{\Sigma}\widehat{\gamma}-\widehat{\gamma}^{\top}\frac{1}{n_1}\sum_{i=1}^{n_1}(\widehat{\alpha}+X_{i\cdot}^\top\widehat{\beta}-y_i)X_{i\cdot}-\widehat{\beta}^{\top}\frac{1}{n_2}\sum_{i=1}^{n_2}(\widehat{\zeta}+Z_{i\cdot}^\top\widehat{\gamma}-w_i)Z_{i\cdot}$, 
		$\widehat{\beta^{\top}\Sigma\beta}=\widehat{\beta}^{\top} \widehat{\Sigma}\widehat{\beta}-2\widehat{\beta}^{\top}\frac{1}{n_1}\sum_{i=1}^{n_1}(\widehat{\alpha}+X_{i\cdot}^\top\widehat{\beta}-y_i)X_{i\cdot}$, $\widehat{\gamma^{\top}\Sigma\gamma}=\widehat{\gamma}^{\top} \widehat{\Sigma}\widehat{\gamma}-2\widehat{\gamma}^{\top}\frac{1}{n_2}\sum_{i=1}^{n_2}(\widehat{\zeta}+Z_{i\cdot}^\top\widehat{\gamma}-w_i)Z_{i\cdot}$, 
		and $\widehat R$ defined as in (2.7) of the main text. Their asymptotic variances can be estimated by $
		\widehat{v}^2=\frac{n_1+n_2}{n^2_1}\sum_{i=1}^{n_1}  (\widehat{\gamma}^\top X_{i\cdot})^2+\frac{n_1+n_2}{n^2_2}\sum_{i=1}^{n_2}  (\widehat{\beta}^\top Z_{i\cdot})^2
		+\frac{1}{n_1+n_2}\big\{\sum_{i=1}^{n_1}(\widehat{\beta}X_{i\cdot}X_{i\cdot}^\top\widehat{\gamma}-\widehat{\beta}\widehat{\Sigma}\widehat{\gamma})^2+\sum_{i=1}^{n_2}(\widehat{\beta}Z_{i\cdot}Z_{i\cdot}^\top\widehat{\gamma}-\widehat{\beta}\widehat{\Sigma}\widehat{\gamma})^2\big\}$, $\widehat{v}^2_\beta=\frac{4(n_1+n_2)}{n^2_1}\sum_{i=1}^{n_1}  (\widehat{\beta}^\top X_{i\cdot})^2
		+\frac{1}{n_1+n_2}\bigg\{\sum_{i=1}^{n_1}(\widehat{\beta}X_{i\cdot}X_{i\cdot}^\top\widehat{\beta}-\widehat{\beta}\widehat{\Sigma}\widehat{\beta})^2+\sum_{i=1}^{n_2}(\widehat{\beta}Z_{i\cdot}Z_{i\cdot}^\top\widehat{\beta}-\widehat{\beta}\widehat{\Sigma}\widehat{\beta})^2\bigg\}$, and $\widehat{v}_R^2= \frac{\widehat{v}^2}{\widehat{\beta^\top\Sigma\beta}\widehat{\gamma^\top\Sigma\gamma}}$, respectively. Under the simulation setup concerning hypotheses testing described in Section S4.4, we applied the statistical tests based on the above estimators and their asymptotic variance estimators, by treating the binary outcomes as continuous variables. Although strictly speaking, applying the above linear-model based method to data with binary outcomes was not valid due to model mispecification, we made this comparison mainly to illustrate the limitation of such a practice. Comparing with Table \ref{table:t3}, for parameters $\beta^\top\Sigma\gamma$ and $R$, the linear-model based procedure had desirable type I errors, but had lower powers than the logistic-model based procedure; for the parameter $\beta^\top\Sigma\beta$, the linear model based procedure was not valid as it failed to maintain the $\alpha=0.05$ type I error. These demonstrate the merit of our more careful treatment of the binary outcomes.

		\begin{table}[h!]
			\centering
			\caption{Type I error and power of the linear-model based CIs with $\Sigma=\Sigma_B$,  $\alpha=0.05$ and $k=25$.}
			\begin{tabular}{c|c|cc|cc|cc}
				\hline
				\multirow{3}{1em}{$n$}&\multirow{3}{1em}{$p$}&\multicolumn{2}{c|}{$\beta^\top\Sigma\gamma$} &\multicolumn{2}{c|}{$\beta^\top\Sigma\beta$} &\multicolumn{2}{c}{$R$}\\
				\cline{3-8}
				&&   type I error & power & type I error & power &  type I error & power   \\  
				\hline
				&700&   0.02&    0.14&    1.00 &    1.00&   0.02&    0.14 \\
				300	&800 & 0.02&    0.15&   1.00&   1.00&  0.01&    0.15 \\
				&900& 0.06&    0.13&   1.00&    1.00&    0.02&   0.13\\
				\cline{1-8}
				&700 &  0.05&    0.30&    1.00&    1.00&   0.01&    0.30\\
				400&800 &  0.09&    0.26&    1.00&    1.00&   0.01&  0.26\\
				&900 & 0.06&    0.31&    1.00&    1.00&    0.00&    0.31\\
				\cline{1-8}
				&700 &  0.10&    0.45&    1.00&    1.00&    0.01&     0.45\\
				500&800 & 0.06&    0.37&   1.00&   1.00&   0.00&   0.37\\
				&900& 0.05&    0.42&   1.00&    1.00&    0.01&   0.42\\
				\hline
			\end{tabular}	\label{table:t4}
		\end{table}

		\section{Inference with Data from Overlapped Samples} \label{scenario2.sec}
		
		In the main paper, our discussions have been focusing on the observations under Scenario (I). In fact, by slightly modifying the methods described in Sections 2 and 3, a  similar set of inference procedures can be developed based on observations under Scenario (II).
		
		Specifically, we still denote the logistic Lasso estimators as $(\widehat{\alpha},\widehat{\beta})$ and $(\widehat{\zeta},\widehat{\gamma})$. 
		Let $\widetilde{\Sigma}=\frac{1}{n_1+n_2-m}[\sum_{i=1}^{m}X_{i\cdot}X_{i\cdot}^\top+\sum_{i=m+1}^{n_1}X_{i\cdot}X_{i\cdot}^\top+\sum_{i=m+1}^{n_2}Z_{i\cdot}Z_{i\cdot}^\top]$. In light of the arguments in the main paper, we define the bias-corrected genetic covariance estimator
		\beq
		\begin{aligned}
			\widetilde{\beta^{\top}\Sigma\gamma}&=\widehat{\beta}^{\top} \widetilde{\Sigma}\widehat{\gamma}-\widehat{\gamma}^{\top}\frac{1}{n_1}\sum_{i=1}^{n_1}\frac{(1+e^{\widehat{\alpha}+X_{i\cdot}^{\top}\widehat{\beta}})^2}{e^{\widehat{\alpha}+X_{i\cdot}^{\top}\widehat{\beta}}}\{h(\widehat{\alpha}+X_{i\cdot}^\top\widehat{\beta})-y_i\}X_{i\cdot}\\
			&\quad-\widehat{\beta}^{\top}\frac{1}{n_2}\sum_{i=1}^{n_2}\frac{(1+e^{\widehat{\zeta}+Z_{i\cdot}^{\top}\widehat{\gamma}})^2}{e^{\widehat{\zeta}+Z_{i\cdot}^{\top}\widehat{\gamma}}}\{h(\widehat{\zeta}+Z_{i\cdot}^\top\widehat{\gamma})-w_i\}Z_{i\cdot}.\label{eq: weighted FDE2}
		\end{aligned}
		\eeq
		The estimators $\widetilde{\beta^\top\Sigma\beta}$ and $\widetilde{\gamma^\top\Sigma\gamma}$ can be defined similarly the main paper, with $\widehat{\Sigma}$ replaced by $\widetilde{\Sigma}$ and $Z_{i\cdot}$ replaced by $X_{i\cdot}$. Hence, the bias-corrected estimator for the genetic correlation $R$ can be defined as
		\beq \label{R.hat2}
		\widetilde{R} = \left\{ \begin{array}{ll}
			\frac{\widetilde{\beta^\top\Sigma\gamma}}{\surd({\widetilde{\beta^\top\Sigma\beta}\widetilde{\gamma^\top\Sigma\gamma}})}, & \textrm{if $(\widetilde{\beta^\top\Sigma\gamma})^2<{\widetilde{\beta^\top\Sigma\beta}\widetilde{\gamma^\top\Sigma\gamma}}$}\\
			0, & \textrm{if ${\widetilde{\beta^\top\Sigma\beta}\widetilde{\gamma^\top\Sigma\gamma}}=0$}\\
			\text{sign}( {\widetilde{\beta^\top\Sigma\gamma}}),& \textrm{otherwise}
		\end{array} \right. .
		\eeq
		The asymptotic variance of 
		$\widetilde{\beta^\top\Sigma\gamma}$
		is given by
		\begin{align*}
			v^2&=E \big\{ \widehat{\beta}^\top (X_{i\cdot}X_{i\cdot}^\top -\Sigma)\widehat{\gamma}\big\}^2+\bigg(\frac{n_1+n_2-m}{m}+\frac{n_1+n_2-m}{n_1-m}\bigg)E\{ \eta_i^{(X)}(\widehat{\gamma}^\top X_{i\cdot})^2\}\\
			&\quad+\bigg(\frac{n_1+n_2-m}{m}+\frac{n_1+n_2-m}{n_2-m}\bigg)E\{ \eta_i^{(Z)}(\widehat{\beta}^\top Z_{i\cdot})^2\},
		\end{align*}
		which can be estimated by
		\[
		\widetilde{v}^2=\Delta_1+\bigg(\frac{n_1+n_2-m}{m}+\frac{n_1+n_2-m}{n_1-m}\bigg)\Delta_2+\bigg(\frac{n_1+n_2-m}{m}+\frac{n_1+n_2-m}{n_2-m}\bigg)\Delta_3,
		\]
		where
		\[
		\Delta_1=\frac{1}{n_1+n_2-m}[\sum_{i=1}^m (\widehat{\beta}X_{i\cdot}X_{i\cdot}^\top\widehat{\gamma}-\widehat{\beta}\widetilde{\Sigma}\widehat{\gamma})^2+\sum_{i=m+1}^{n_1} (\widehat{\beta}X_{i\cdot}X_{i\cdot}^\top\widehat{\gamma}-\widehat{\beta}\widetilde{\Sigma}\widehat{\gamma})^2+\sum_{i=m+1}^{n_2} (\widehat{\beta}Z_{i\cdot}Z_{i\cdot}^\top\widehat{\gamma}-\widehat{\beta}\widetilde{\Sigma}\widehat{\gamma})^2],
		\]
		\[
		\Delta_2=\frac{1}{n_1}\sum_{i=1}^{n_1} \frac{(1+e^{\widehat{\alpha}+X_{i\cdot}^\top\widehat{\beta}})^2(\widehat{\gamma}^\top X_{i\cdot})^2}{e^{\widehat{\alpha}+X_{i\cdot}^\top\widehat{\beta}}},\qquad
		\Delta_3=\frac{1}{n_2}\sum_{i=1}^{n_2} \frac{(1+e^{\widehat{\eta}+Z_{i\cdot}^\top\widehat{\gamma}})^2(\widehat{\beta}^\top Z_{i\cdot})^2}{e^{\widehat{\eta}+Z_{i\cdot}^\top\widehat{\gamma}}}.
		\]
		Similarly,  the asymptotic variance of $\widetilde{\beta^\top\Sigma\beta}$ is given by 
		$$
		v^2_\beta=\frac{4(n_1+n_2-m)}{n_1}E\{ \eta_i^{(X)} (\widehat{\beta}^\top X_{i\cdot})^2\}+E\{\widehat{\beta}^\top(X_{i\cdot}X_{i\cdot}^\top-\Sigma)\widehat{\beta}\}^2,
		$$
		which can be estimated by
		$$
		\widetilde{v}^2_\beta=\frac{4(n_1+n_2-m)}{n^2_1}\sum_{i=1}^{n_1} \frac{(1+e^{\widehat{\alpha}+X_{i\cdot}^\top\widehat{\beta}})^2}{e^{\widehat{\alpha}+X_{i\cdot}^\top\widehat{\beta}}} (\widehat{\beta}^\top X_{i\cdot})^2
		+\Delta_1.
		$$
		Finally, the asymptotic variance of $\tilde{R}$ can be estimated by $\widetilde{v}_R^2=\frac{\widetilde{v}^2}{\widetilde{\beta^{\top}\Sigma\beta}\widetilde{\gamma^{\top}\Sigma\gamma}}.$
		Confidence intervals and statistical tests can be constructed based on the above estimators and their variance estimates. For example, an $(1-\alpha)$-level confidence interval for $R$ can be constructed as ${\text{CI}^*_{\alpha}(R,\mathcal{D})}=\big[ \widetilde{R}-\widetilde{\rho}_R,  \widetilde{R}+\widetilde{\rho}_R\big]$ with $\widetilde{\rho}_R=\frac{z_{\alpha/2}\widetilde{v}_R}{\sqrt{n}}$, while for the null hypothesis $H_0: R=R_0$ and the test statistic  $T={\sqrt{n}(\widetilde{R}-R_0)}/{\widetilde{v}_R}$, we reject $H_0$ whenever $|T|>z_{\alpha/2}$. 
		
		Theoretically, by slightly modifying the proofs, all the theoretical properties obtained under Scenario (I)  still hold, only with condition (A1) replaced by the following condition.
		
		\noindent {\bf (A1')}  Each element in $\{Z_{i\cdot}\}_{i=1}^{m}\cup \{X_{i\cdot}\}_{i=m+1}^{n_1}\cup \{Z_{i\cdot}\}_{i=m+1}^{n_2}$  is a centred $i.i.d.$ sub-Gaussian vector with covariance matrix $\Sigma\in\R^{p\times p}$ satisfying $ M^{-1}\le \lambda_{\min}(\Sigma)\le \lambda_{\max}(\Sigma)\le M$ for some constant $M>1$.
		
		\section{Comparison with Guo et al. (2019) and Other Extensions}
		
		Like the current paper, \cite{guo2019optimal} also considered quantification of genetic relatedness. However, there are three main differences between the two papers: 
		\begin{enumerate}
			\item Continuous vs. binary outcomes: \cite{guo2019optimal}  considered a pair of linear regression models with continuous outcomes, whereas the current paper considered a pair of logistic regression models with binary outcomes. The main challenge of the current study comes from the nonlinearity of the regression function, making the derivation of the bias-correction term and its theoretical analysis much more challenging compared to the linear regression setting.
			\item Distinct parameters of interest: the main focus of \cite{guo2019optimal}  was the inner-product parameter $\beta^\top\gamma$ and its normalized version $\frac{\beta^\top\gamma}{\|\beta\|_2\|\gamma\|_2}$, whereas the current paper focused on the bilinear form $\beta^\top\Sigma\gamma$, and its normalized version $R=\frac{\beta^\top\Sigma\gamma}{\sqrt{\beta^\top\Sigma\beta\gamma^\top\Sigma\gamma}}$. The two pairs of parameter only coincides when the population covariance matrix is identity ($\Sigma=I$), which rarely happens in practice. This seemingly small discrepancy  in the parameters of interest makes the corresponding inferential procedures and their implementations very different. For example, in \cite{guo2019optimal} , the bias-correction term relies on a so-called projection vector, which have to be obtained by solving a high-dimensional convex optimization. In contrast, in the current study the bias-correction term admits a closed-form expression, which can be evaluated immediately given the data and the initial Lasso estimator.
			\item Results on statistical inference: the most important difference between the two papers is that, \cite{guo2019optimal}  only considered point estimation of the parameters of interest, whereas the current paper proposed estimator, confidence intervals and statistical test for the parameters of interest. The construction of the CIs and statistical tests requires establishing the asymptotic normality of the proposed estimators, which is technically much  more challenging than obtaining the rate of the convergence of the estimators. 
		\end{enumerate}
		Interestingly, the rates of convergence for the estimators proposed in the two papers share the similar forms (compare Theorem 1 of \cite{guo2019optimal} with Theorem 1 of the current paper). However, as we pointed out in the above Item 3, there is no statistical test or CI proposed in \cite{guo2019optimal} for their genetic relatedness parameters. Due to the lack of statistically valid inference procedures under the linear regression setting considered in \cite{guo2019optimal}, it is in general very difficult to compare the two methods in terms of their statistical powers for detecting genetic correlations. Nevertheless, it demonstrate the advantage of the current paper, whose results and the proposed method are potentially more useful in practice.

		Finally, if we have one binary trait and one continuous trait, and our goal is again to make inference about their genetic covariance $\beta^\top\Sigma\gamma$, the methodology developed in this paper can still be applied, to derive a similar bias-corrected estimator, along with its theoretical properties such as asymptotic normality. More specifically, one can follow the argument in Section 2.2, and make appropriate modifications by adapting to the form of linear regression (such as replacing the expit function $f(u)=\frac{e^u}{e^u+1}$ by the identity map $f(u)=u$), to obtain the final bias-correction term for the initial plug-in estimator $\widehat\beta^\top\widehat\Sigma\widehat\gamma$. The thus obtained estimator should have similar theoretical properties as discussed in the current paper. Nevertheless, working out the explicit forms of the final debiased estimator as well as its theoretical properties still require involved nontrivial calculations. Since this is beyond the scope of the current paper, we leave this interesting question for future investigation.

\end{document}